\documentclass[twocolumn]{aastex631}
\graphicspath{{./}}

\begin{document}

\title{Nearby SNR: a possible common origin to multi-messenger anomalies in spectra, ratios and anisotropy of cosmic rays}

\correspondingauthor{Yi-Qing Guo,Wei Liu,Xiao-Jun Bi}
\email{guoyq@ihep.ac.cn, liuwei@ihep.ac.cn, bixj@ihep.ac.cn}

\author{Bing-Qiang Qiao}
\affiliation{Key Laboratory of Particle Astrophysics, Institute of
	High Energy Physics, Chinese Academy of Sciences, Beijing 100049, China}

\author{Yi-Qing Guo}
\affiliation{Key Laboratory of Particle Astrophysics, Institute of High Energy Physics, Chinese Academy of Sciences, Beijing 100049, China}
\affiliation{University of Chinese Academy of Sciences, Beijing 100049, China}


\author{Wei Liu}
\affiliation{Key Laboratory of Particle Astrophysics, Institute of
	High Energy Physics, Chinese Academy of Sciences, Beijing 100049, China}

\author{Xiao-Jun Bi}
\affiliation{Key Laboratory of Particle Astrophysics, Institute of High Energy Physics, Chinese Academy of Sciences, Beijing 100049, China}
\affiliation{University of Chinese Academy of Sciences, Beijing 100049, China}



\begin{abstract}
The multi-messenger anomalies, including spectral hardening or excess for nuclei, leptons, ratios of $\bar p/p$ and B/C, and anisotropic reversal, were observed in past years. AMS-02 experiment also revealed different spectral break for positron and electron at 284 GeV and beyond TeV respectively.
It is natural to ask whether all those anomalies originate from one unified physical scenario. In this work, the spatially-dependent propagation (SDP) with a nearby SNR source is adopted to reproduce above mentioned anomalies. There possibly exists dense molecular cloud(DMC) around SNRs and the secondary particles can be produced by pp-collision or fragmentation between the accelerated primary cosmic rays and DMC. As a result, the spectral hardening for primary, secondary particles and ratios of $B/C$ and $\bar p/p$ can be well reproduced. Due to the energy loss at source age of 330 kyrs, the characteristic spectral break-off for primary electron is at about 1 TeV hinted from the measurements. The secondary positron and electron from charged pion take up $5\%$ energy from their mother particles, so the positron spectrum has a cut-off at $\sim$250 GeV. 
Therefore, the different spectral break for positron and electron together with other anomalies can be fulfilled in this unified physical scenario. More interesting is that we also obtain the featured structures as spectral break-off at 5 TV for secondary particles of Li, Be, B, which can be served to verify our model. We hope that those tagged structures can be observed by the new generation of space-borne experiment HERD in future. 
\end{abstract}



\section{Introduction} \label{sec:intro}

The origin of cosmic rays (CRs) has been a centurial mystery since its discovery.
The scientists have been always devoted to resolve this problem. With new
generation space-borne and ground-based experiments, CRs measurements are stepping into an era of high precision and a series of new phenomena in
spectra, ratio of secondary-to-primary and anisotropy, as multi-messenger anomalies, are revealed now. It may be an effective way to pinpoint the origin problem by joint study of multi-messenger information.

Firstly, the nuclei spectra have been measured with unprecedent precision and a fine structure of spectral hardening at 200 GV has been discovered by
ATIC-2, CREAM and PAMELA experiments \citep{2007BRASP..71..494P,2009BRASP..73..564P,2010ApJ...714L..89A,2017ApJ...839....5Y,2011Sci...332...69A}.  Lately, AMS-02 experiment also confirmed it \citep{2015PhRvL.114q1103A,2015PhRvL.115u1101A} and further revealed that other heavy nuclei including secondary particles have similar anomaly \citep{2018PhRvL.120b1101A,2017PhRvL.119y1101A,2018PhRvL.121e1103A,2020PhRvL.124u1102A,2021PhRvL.127b1101A,2021PhRvL.126d1104A}. 
More interesting is that the spectral break-off around $\sim$14 TeV was observed by CREAM, NUCLEON and DAMPE experiments \citep{2017ApJ...839....5Y,2017JCAP...07..020A,2018JETPL.108....5A,2019SciA....5.3793A}. Furthermore, recent spectral measurement of Helium showed that 
the drop-off starts from $\sim$34 TV, which supported the rigidity dependent cut-off \citep{2021PhRvL.126t1102A}. 
Several kinds of models have been proposed to explain the origin of spectral hardening, including the nearby source \citep{2013APh....50...33S,2017PhRvD..96b3006L,2019JCAP...10..010L,2019JCAP...12..007Q,2012ApJ...752...68V}, the combined effects from different group sources
and the spatially-dependent propagation(SDP)\citep{2016ApJ...819...54G,2018PhRvD..97f3008G,2018ApJ...869..176L,2021ApJ...911..151M}. Considering the break-off at the rigidity of 
14 TV in spectrum, it seems that the nearby source model becomes natural and accessible.
However, other observational clues are required to support this point of view.

Secondly, the spectra of positron and electron is another good choice to shed new light on this topic.
The famous spectral excess of positron above 20 GeV was discovered by PAMELA experiment \citep{2009Natur.458..607A}. Then the AMS-02 experiment confirmed this remarkable result \citep{2014PhRvL.113l1101A,2019PhRvL.122d1102A}. Just recently, a sharp drop-off at 284 GeV was observed by AMS-02 experiment with above 4$\sigma$ confidence level \citep{2019PhRvL.122d1102A}. {As for electrons, the measurement by the AMS-02 experiment showed that the energy spectrum could not be described by a single power-law form, the power index changes at about 42 GeV. Contrary to the positron flux, which has an exponential energy cutoff of about 810 GeV, at the 5$\sigma$ level the electron flux does not have an energy cutoff below 1.9 TeV \citep{2019PhRvL.122j1101A}.}
However, the drop-off around 1 TeV in the total spectrum of positron and electron was first reported by HESS collaboration \citep{2008PhRvL.101z1104A,2009A&A...508..561A} and validated by MAGIC \citep{2011ICRC....6...47B}, VERITAS \citep{2015ICRC...34..868S} experiments. The DAMPE experiment also performed the  direct measurement to this feature and announced that the break-off was at $\sim$0.9 TeV \citep{2017Natur.552...63D}. It is obvious that the spectra of positron and electron have extra-components at high energy similar to nuclei one. The nearby source is also {an alternative for interpreting the above feasures.}
One of the natural advantages of positron and electron is their fast cooling in the interstellar radiation field (ISRF), which can roughly decide the source distance and age. For example, the spectral cut-off
at $\sim$TeV requires that the age of nearby source is about 330 kyrs \citep{2009PhRvD..80f3003F,2016PTEP.2016b1E01K,2021JCAP...05..012Z,2022ApJ...930...82L}, which will set a 
much more strict constrain on its place. Where is the nearby source and its rough direction may point out a new bright road.

Lastly, the anisotropy of CRs is one of the best choices to fulfill this role. 
Thanks to unremitting efforts 
of ground-based experiments, the measurements of large-scale anisotropy has made great progress from
hundreds of GeV to several PeV \citep{2006Sci...314..439A,2017ApJ...836..153A,2016ApJ...826..220A,2013ApJ...765...55A,2015ApJ...809...90B}. 
It is obvious that the phase has reversed at 100 TeV and the direction roughly point to local  magnetic field and Galactic Center (GC) below and above 100 TeV respectively. 
Coincidentally, the amplitude 
has a dip structure at $\sim$100 TeV, starting from $\sim$10 TeV. 
The most importance thing is that
there exists a common transition energy scale between the structures of the energy spectra
and the anisotropies. The local source possibly plays a very important role to resolve the conjunct
problems of spectra and anisotropies. Furthermore, the direction of anisotropy can roughly
outline the position of such local source. In our recent work, we proposed a local source under
the SDP model to reproduce the co-evolution of the spectra and
anisotropies and found that the optimal candidate of local source is possible a SNR at Geminga’s birth
place \citep{2019JCAP...10..010L,2019JCAP...12..007Q}. 

Based on above discussions, the local source is necessary to understand the multi-messenger anomalies. 
However, the latest observations bring new challenges into this model, such as the different spectral break-off for
positron and electron and a series of results of nuclei spectra and ratios that were published by 
AMS-02 experiment recently. Therefore, a systematic study is useful to understand those new observations. More important is that the featured structure
is required and necessary to verify this model. In this work, a unified physical scenario as 
the SDP with a nearby source, Geminga SNR, is adopted to reproduce all the above anomalies.
Simultaneously, we obtain the tagged structure to examine our model.
The paper is organized as follows. Section 2 describes the model and method briefly, Section 3 presents all the calculated results
and Section 4 gives the conclusion.

\section{Model and Method Description} \label{sec:model}

The CRs in solar system come from two parts as the global one from the galactic background sources (bkg) and the local one from nearby SNRs (loc SNR). 
For the background sources, it is viable to assume that the spatial distribution of CRs from them arrives at steady state. Nevertheless for the nearby single SNR, the time-dependent transport of CRs after injection is requisite. 

Furthermore, the dense molecular cloud (DMC) plays a key role in the star formation, which means that there possibly exists DMC around SNRs \citep{2008A&A...481..401A,2012ApJ...749L..35U,2021ApJ...923..106Z}. In our model, we assume that the DMC or dense interstellar medium(DISM) exists around the nearby SNR. Therefore, the general physical picture can be sketched as three steps. Firstly, the nuclei and electrons can be accelerated to very high energy (VHE) together during the SNR explosion. Then the VHE CR nuclei and electrons undergo the interaction with DMC and ISRF and then produce secondary particles. Lastly, the primary and secondary particles will go through the interstellar space and experience a long travel in the galaxy. 
Certainly, a limited part will arrive at the earth and be observed by the various experiments.

\begin{figure}[htp]
	\centering
	\includegraphics[width=8.cm]{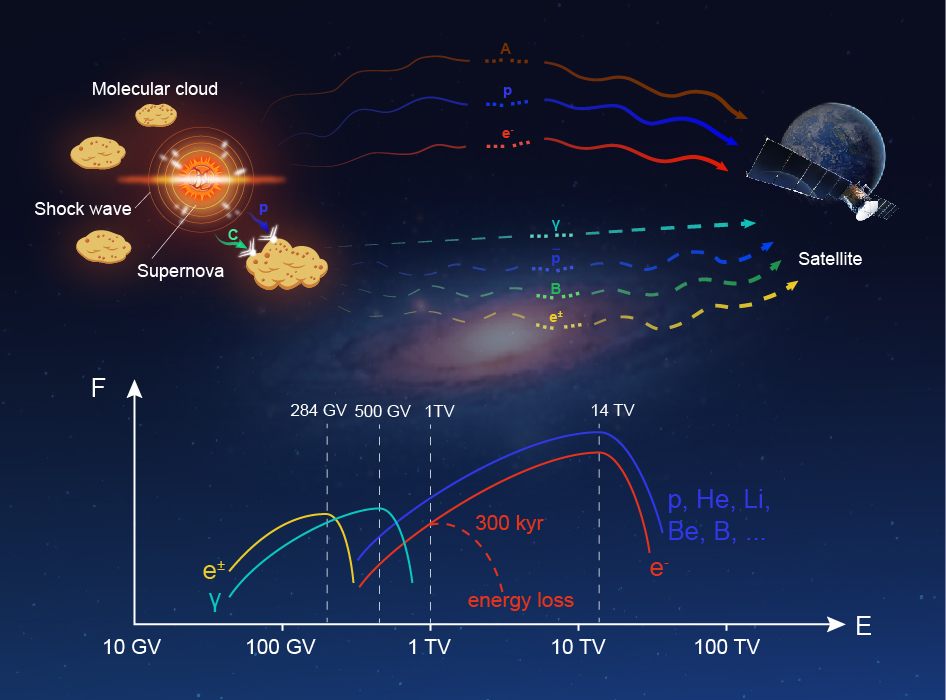}
	\caption{The cartoon illustration to describe the production mechanism of the multi-messenger anomalies: the spectral hardening and break-off at rigidity of 200 GV and 14 TV for both primary and secondary nuclei, the energy cut-off for positron and electron at around 300 GeV and 1 TeV respectively.}
	\label{fig:model}
\end{figure}

The results can be imaged as the cartoon illustration of Figure \ref{fig:model}.
Here, the spectral break-off of primary nuclei and electrons with exponential form is adopted to be 5 TV for the sake of reproducing the observed bump structure of proton at $\sim$14 TV by DAMPE
satellite experiment \citep{2019SciA....5.3793A}. {The primary electrons have to suffer the energy loss cooling by scattering off the ISRF with the time around 330 kyrs \citep{1994A&A...281L..41S,2005AJ....129.1993M,2007Ap&SS.308..225F}}, which leads to the sharp dropping of electron spectrum in the energy region of $\sim$TeV as observed by DAMPE, HESS, MAGIC and VERITAS \citep{2008PhRvL.101z1104A,2009A&A...508..561A,2011ICRC....6...47B,2015ICRC...34..868S,2017Natur.552...63D}. The anti-proton and positron will be produced in the pp-collision between VHE proton and DMC. Due to the spectral break-off of primary nuclei at the rigidity of $5$ TV, this causes the cut-off of positron, the secondary particle, at about $\sim$300 GeV. Simultaneously, the $\gamma$-rays from $\pi^0$ decay in the pp-collision is produced and has a spectral cut-off around 500 GeV.  In addition, the heavy nuclei
will undergo the fragmentation with the DISM or DMC and then the secondary nuclei such as $Li, Be, B$ are produced. The secondary nuclei from the fragmentation
inherit the property of their mother particle and have the same morphology of energy break-off around 5 TV,  which can be served to differentiate with other models \citep{2022PhRvD.105b3002Z,2018ApJ...869..176L}.

The detailed descriptions of cosmic ray propagation, Galactic background sources, and nearby SNRs are displayed in the following appendix A, B and C respectively.

\begin{table*}
	\caption{The parameters of transport spectrum with SDP model.}
	\centering
	\begin{tabular}{cccccccc}
	\hline
	\hline
	${D}_0~[\rm cm^{-2}s^{-1}]$ & $\delta_0$  & ${N}_{m}$  & $\xi$  & $n$ & $\eta$ & ${\cal \nu}_{\rm A}$~[km s$^{-1}$]   & ${z}_h$~[kpc] \\
	\hline
	$4.2 \times 10^{28}$ & $0.62$  &  $0.5$ & $0.1$ &  $4.0$  & $0.05$ & $6$ & $5$ \\
	\hline
	\hline
	\end{tabular}
	\label{table_1}
\end{table*}

\begin{table*}
	\begin{center}
		\begin{tabular}{|c|ccccc|ccc|}
			\hline
			& \multicolumn{5}{c|}{Background} & \multicolumn{3}{c|}{Local source} \\
			\hline
			Element & Normalization$^\dagger$ & ~~~$\nu_1$~~~  & ~~~${\cal R}_{\rm br}$~~~  & ~~~$\nu_2$~~~  & ~~~$\mathcal R_{c}$~~~ & ~~~$q_0$~~~~~ & ~~~~~$\nu'$~~~ & ~~~${\cal R}'_c$~~~ \\
			\hline
			& $[({\rm m}^2\cdot {\rm sr}\cdot {\rm s}\cdot {\rm GeV})^{-1}]$ & & [GV] & & [PV] & [GeV$^{-1}$] & &  [TV] \\
			\hline
			$e^-$   & $2.9\times 10^{-1}$    & 1.6 & 5.5 & 2.83   &  ...  & $7.20\times 10^{49}$  & 2.25 & 1 \\
			p   & $4.2\times 10^{-2}$    & 2.15 & 8 & 2.38   &  7  & $1.60\times 10^{52}$  & 2.19 & 15 \\
			He & $2.7\times 10^{-3}$   & 2.15 & 8 & 2.32     &  7  & $2.98\times 10^{51}$  & 2.05  &  15  \\
			C   & $9.8\times 10^{-5}$   & 2.15 & 8 & 2.33    &  7  & $1.66\times 10^{50}$    & 2.10 &  15  \\
			N   & $6.0\times 10^{-6}$  & 2.15 &  8 & 2.34    &  7  & $7.73\times 10^{48}$  & 1.95 &   15  \\
			O   & $1.4\times 10^{-4}$   & 2.15 & 8 & 2.37    &  7  & $1.87\times 10^{50}$ & 2.10  &   15  \\
			Ne & $1.7\times 10^{-5}$   & 2.15 & 8 & 2.38   &  7  & $2.22\times 10^{49}$ & 2.05  &   15 \\
			Na & $1.3\times 10^{-6}$   & 2.15 & 8 & 2.33   &  7  & $8.41\times 10^{47}$ & 2.01  &   15  \\
			Mg & $2.2\times 10^{-5}$   & 2.15 & 8 & 2.41     &  7  & $2.33\times 10^{49}$ & 2.05  &   15  \\
			Al & $2.2\times 10^{-6}$   & 2.15 & 8 & 2.40     &  7  & $1.60\times 10^{48}$ & 1.95  &   15  \\
			Si & $2.5\times 10^{-5}$    & 2.15 & 8 & 2.41   &  7  & $2.09\times 10^{49}$ & 2.05  &   15  \\
			Fe & $2.4\times 10^{-5}$    & 2.15 & 8 & 2.35    &  7  & $2.19\times 10^{49}$ & 2.05   &  15  \\
			\hline
		\end{tabular}\\
		$^\dagger${The normalization for CR nuclei and cosmic ray electrons (CREs) are set at kinetic energy per nucleon $E^p_{k} = 100$ GeV/n and $E^{e^-}_{k} = 10$ GeV/n respectively.}
	\end{center}
	\caption{Injection parameters of the background and local source.}
	\label{para_inj}
\end{table*}  

%
%
%

\section{Results} \label{sec:results}
Based on above discussions, the spectra of primary, secondary and all particles are calculated to reproduce the measurements. Simultaneously, the ratios between primary to primary, secondary to primary and secondary to secondary are presented. For the sake of completeness of this work, we also give the anisotropy for CR nuclei and electrons. 
In the model calculations, the parameters of propagation and inject spectra both for local source and galactic ones
are listed in Table \ref{table_1} and \ref{para_inj} respectively.

\subsection{Spectra} \label{subsec:spectra}
The CR spectra are the most important effects to understand their propagation in the galaxy.
Thanks to the new generations of spaced-borne experiments, the spectral measurements of CRs are stepping into precise era and revealed a series of new phenomena, such as the spectral hardening at 200 GV and break-off at $\sim$14 TV, the famous excess of positron and extra-component of electron at high energy. In this section, the unified model calculations are described to reproduce and understand the measurements.


\subsubsection{Primary Particles}
\begin{figure}[htp]
	\centering
	\includegraphics[width=4.cm]{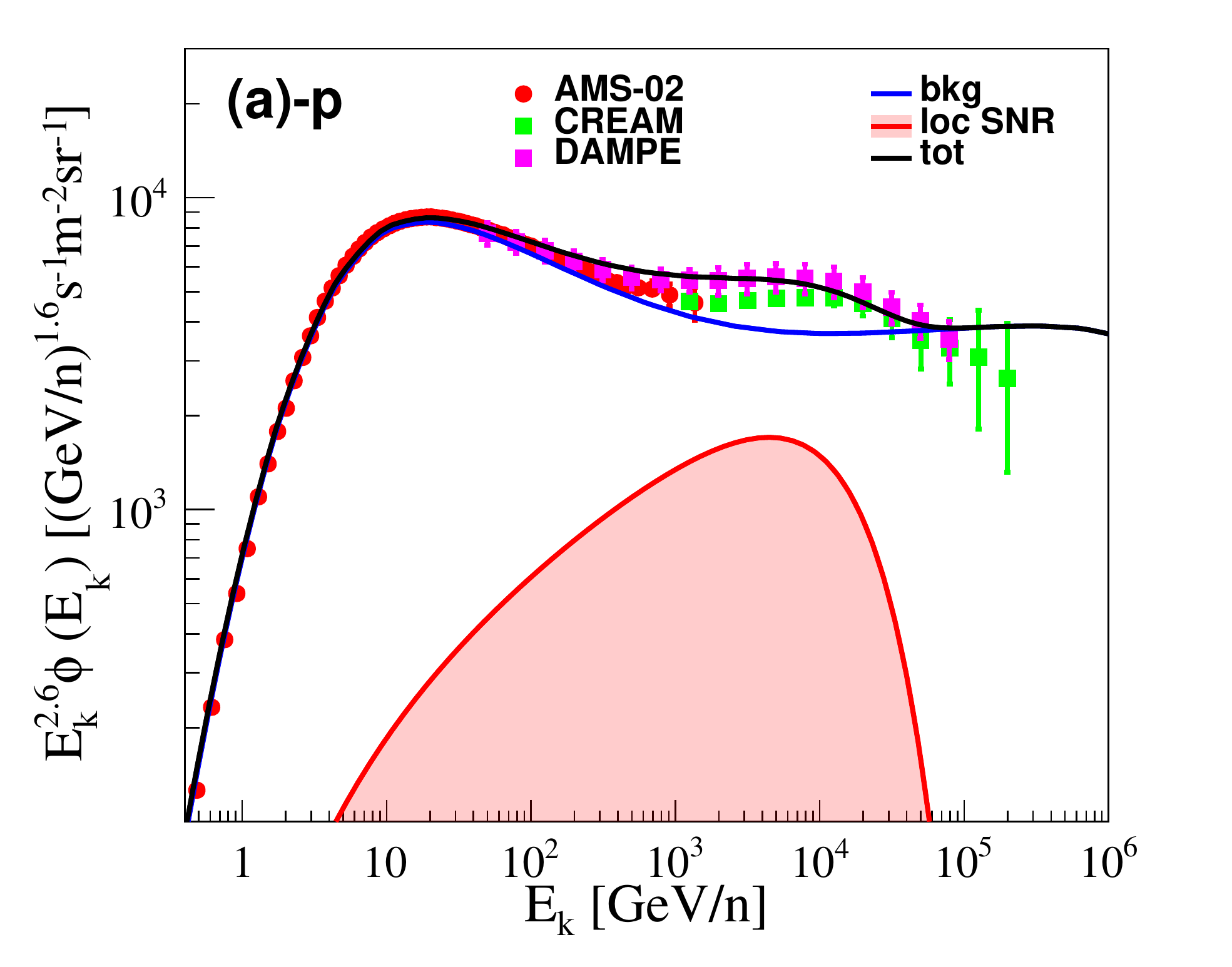}
	\includegraphics[width=4.cm]{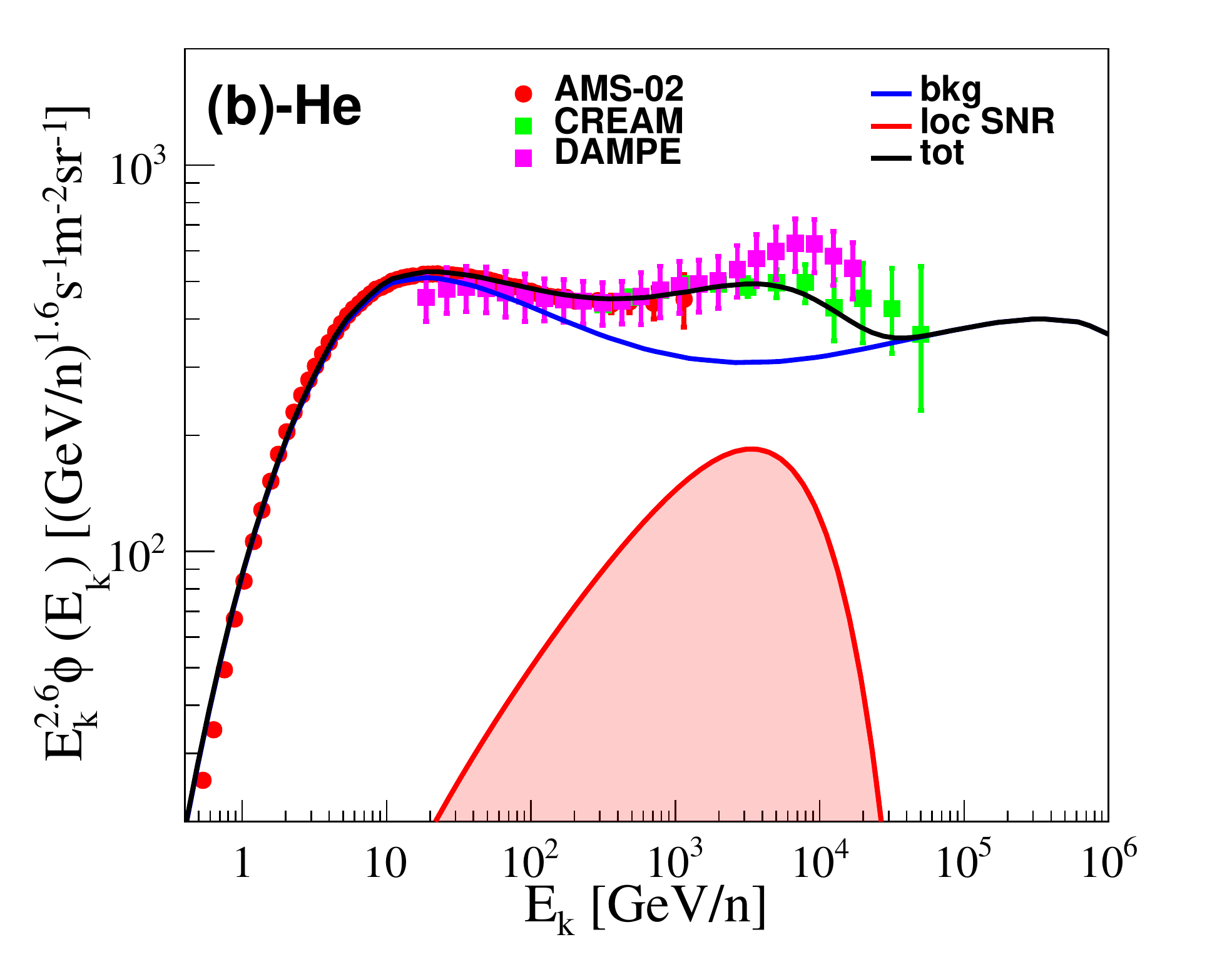}
	\caption{The comparison between model calculations and experimental measurements for proton and Helium spectra as (a) and (b) panel. Here the red shadow shape is from the contribution of local SNR (e.g. Geminga), the blue solid line comes from the background components and the black solid line is the total contribution. The data points are measured by AMS-02, CREAM and DAMPE experiments\citep{2015PhRvL.114q1103A,2017PhRvL.119y1101A,2017ApJ...839....5Y,2019SciA....5.3793A,2021PhRvL.126t1102A}.}
	\label{fig:pHespec}
\end{figure}

\begin{figure}[htp]
	\centering
	\includegraphics[width=4.cm]{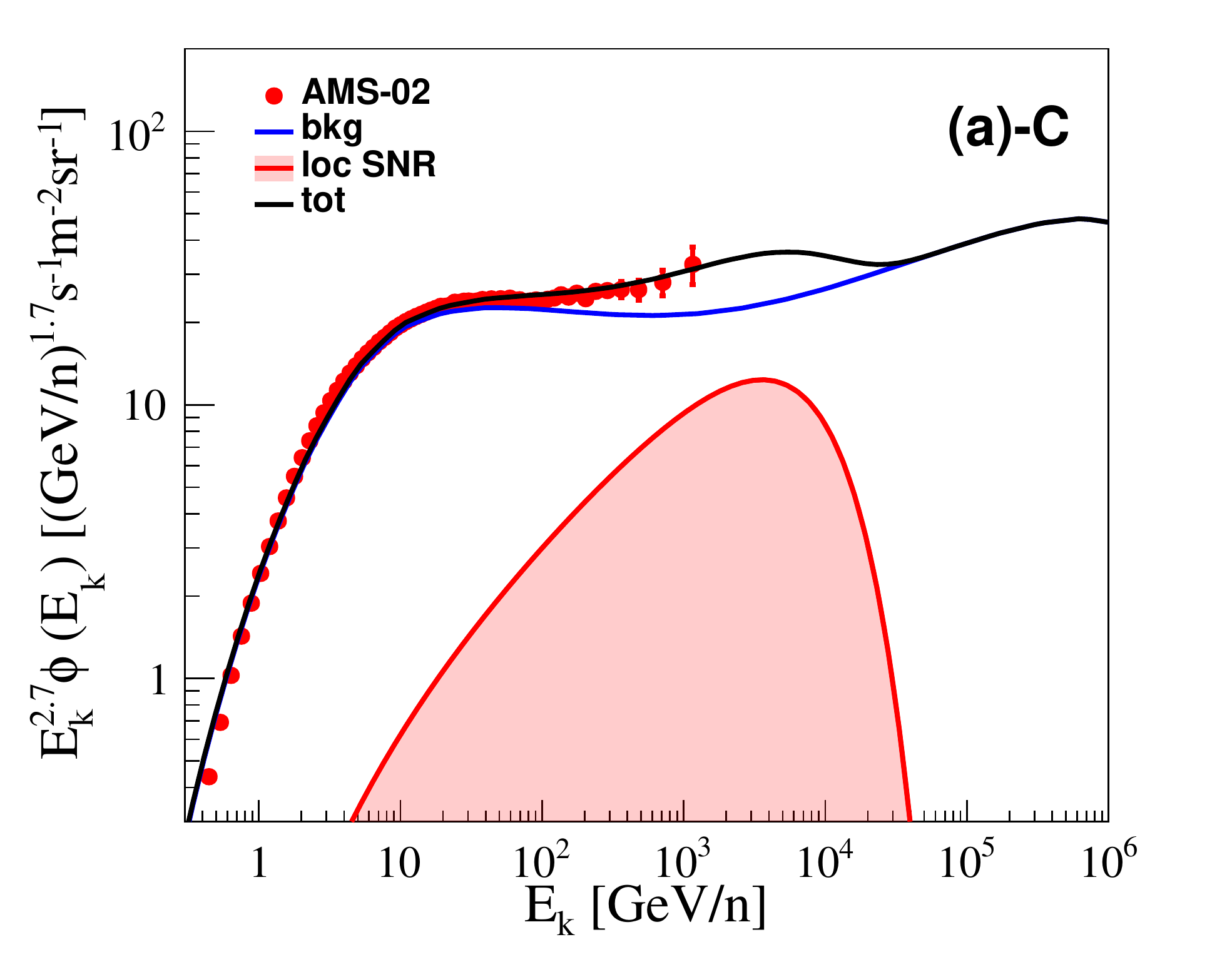}
	\includegraphics[width=4.cm]{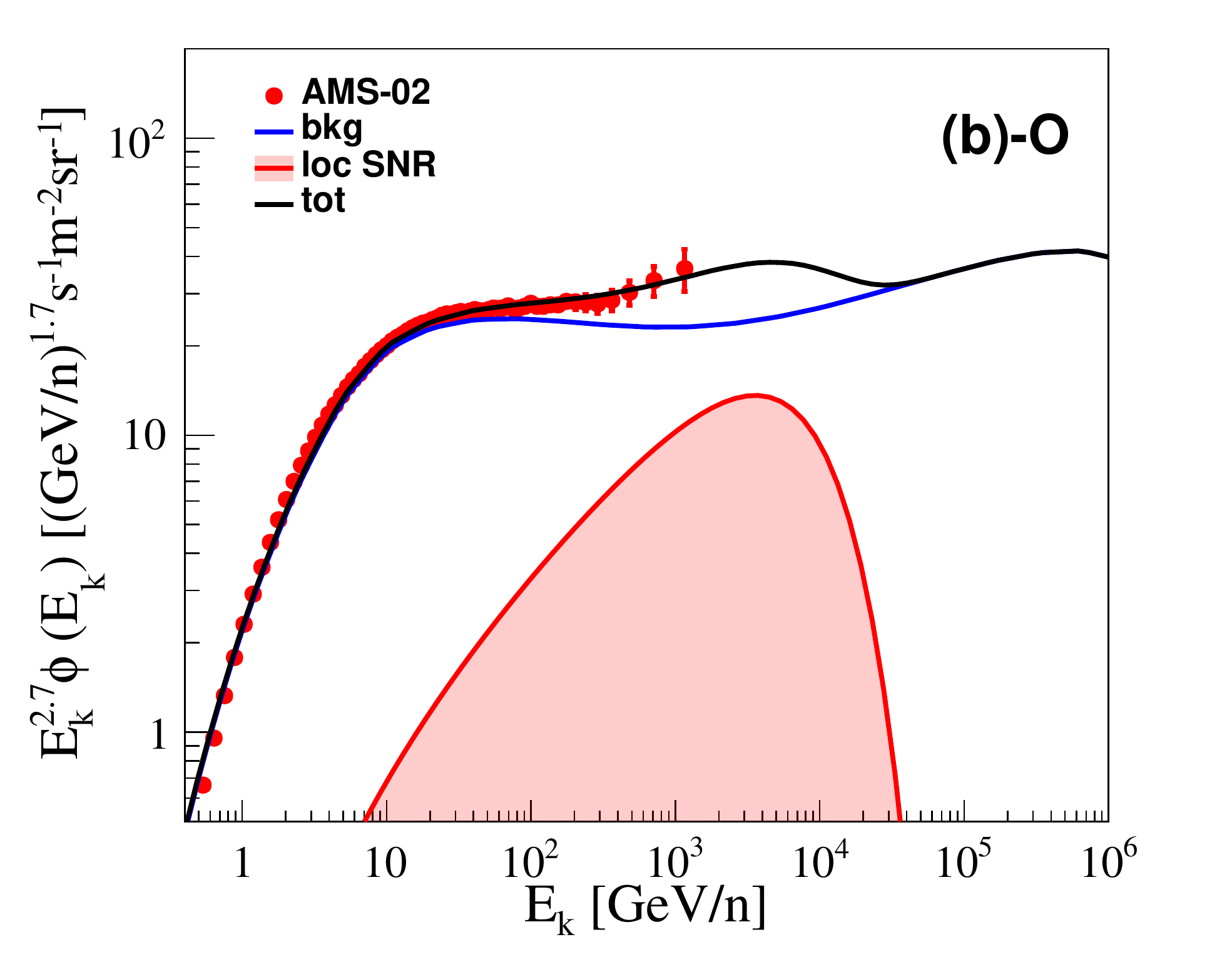}
	\includegraphics[width=4.cm]{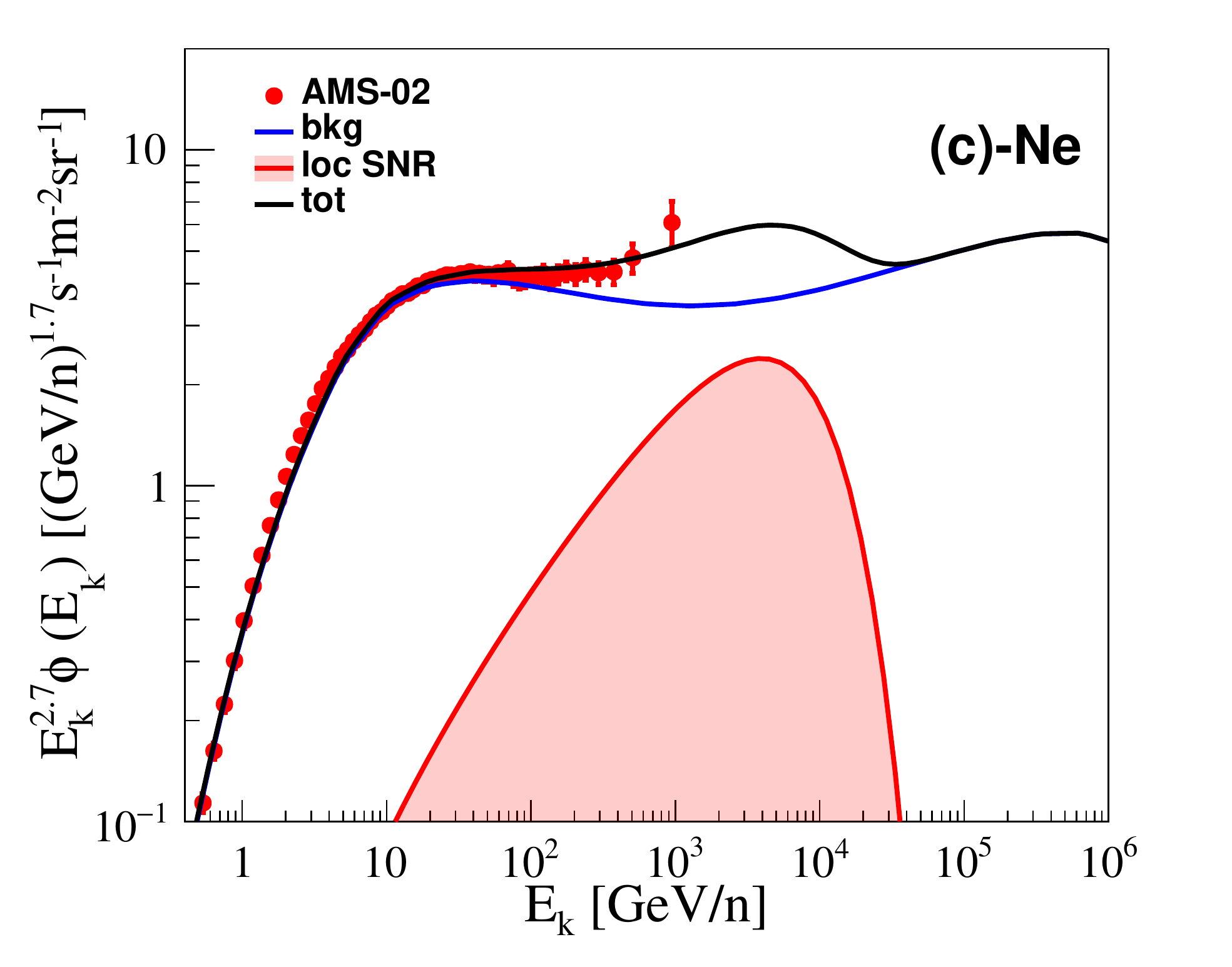}
	\includegraphics[width=4.cm]{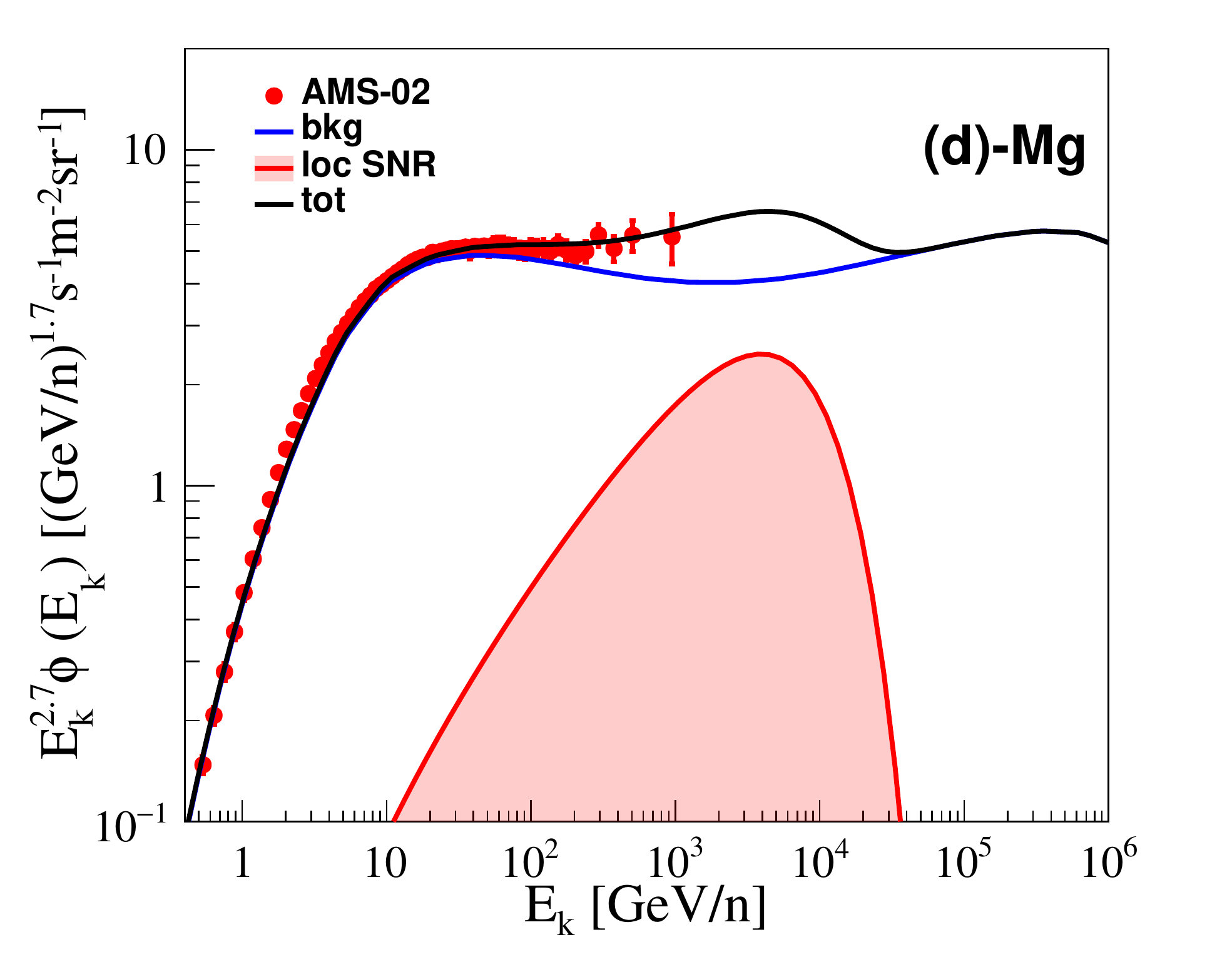}
	\includegraphics[width=4.cm]{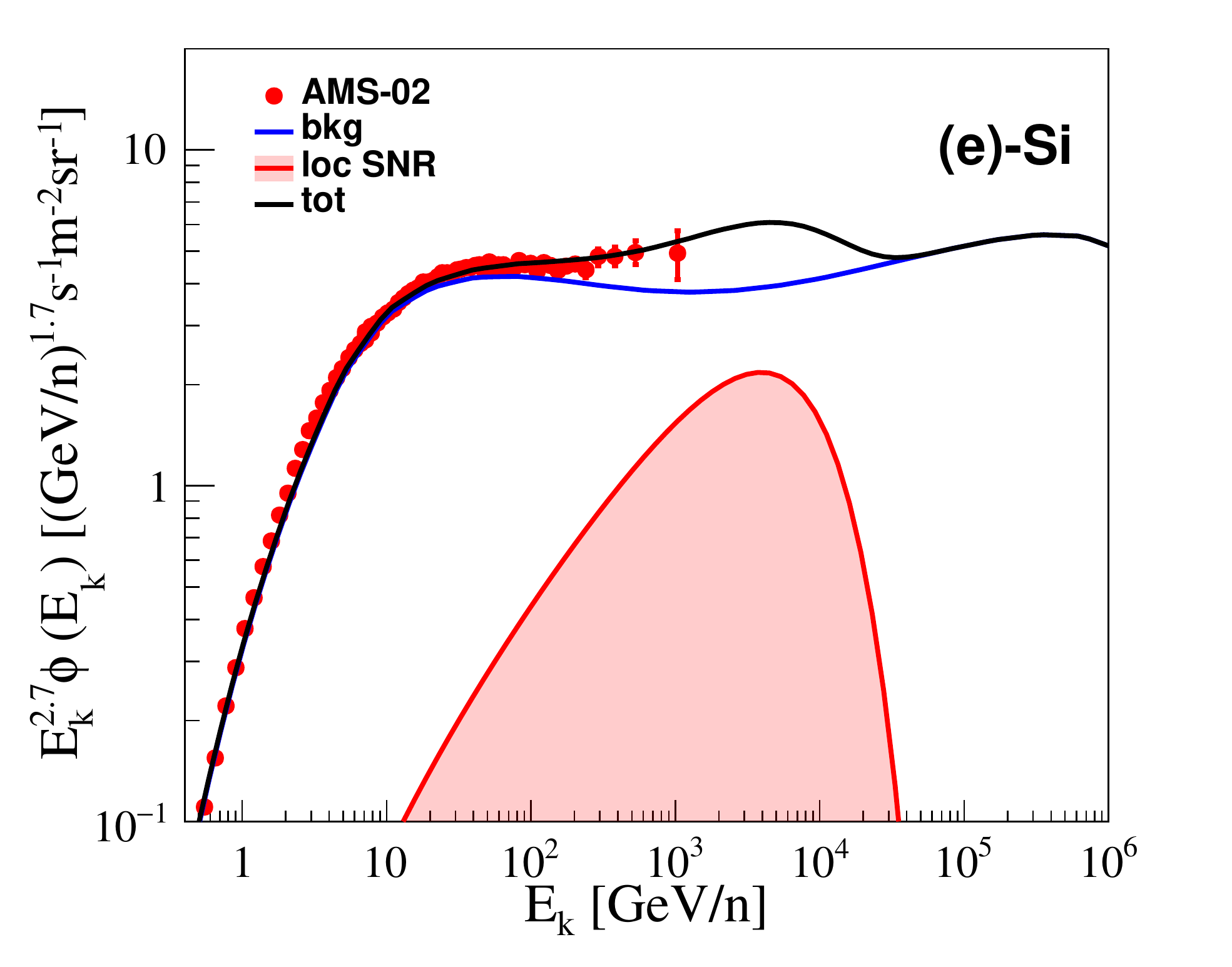}
	\includegraphics[width=4.cm]{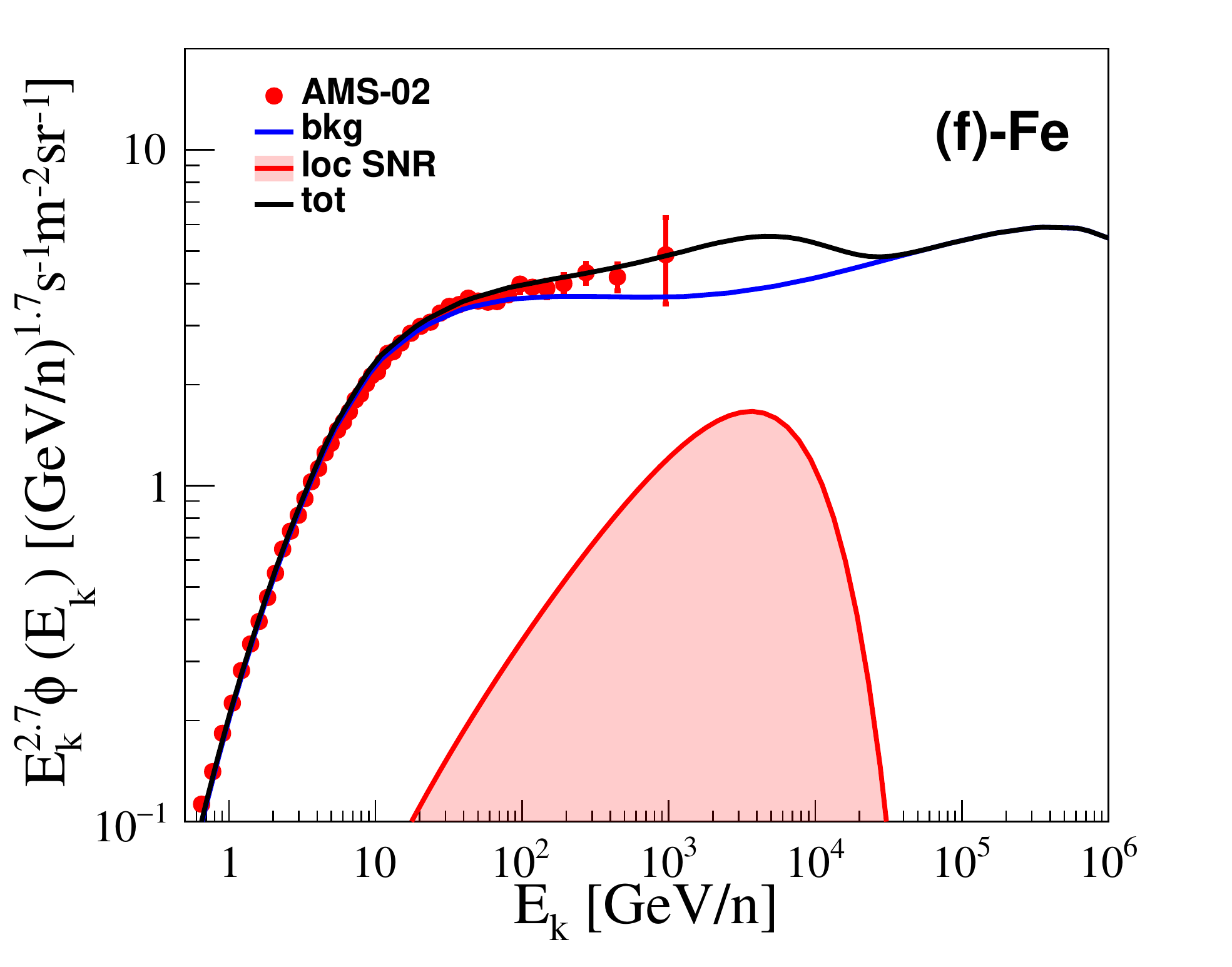}
	\caption{Similar to Figure \ref{fig:pHespec}, the individual species of $C, O, Ne, Mg, Si$ and $Fe$ from panel (a) to (f). The data points are from the AMS-02 experiment\citep{2017PhRvL.119y1101A,2020PhRvL.124u1102A,2021PhRvL.126d1104A}.}
	\label{fig:prispec}
\end{figure}

Figure \ref{fig:pHespec} and \ref{fig:prispec} show the measurements and model calculations for most of nuclei species individually. In the model calculations, the red solid line with shadow is the contribution from local SNR, the blue solid line is from galactic background sources and the black solid line is the sum of them. It is obvious that the spectral hardening can be reproduced well for all the species and its origin dominantly originates from the contribution of local source. To satisfy the energy break-off of proton at the rigidity of 14 TV observed by DAMPE experiment \citep{2019SciA....5.3793A}, 
the injection spectrum of local source is parameterized as
a cutoff power-law form, $q_{\rm inj}({\cal R})=q_0{\cal R}^{-\nu'}
\exp(-{\cal R}/{\cal R}'_{\rm c})$, where the normalization $q_0$ and spectral index $\nu'$ are determined
through fitting to the CR energy spectra. The parameter ${\cal R}'_{\rm c}$ is adopted to be 15 TV, which leads to the bend of CR spectra starting around the rigidity of 5 TV. For detailed parameter information about the spectra, please refer to Table \ref{para_inj}. Figure \ref{fig:pHespec} shows that the spectral break-off of proton is consistent well between model expection and data points from DAMPE and CREAM measurements \citep{2017ApJ...839....5Y,2019SciA....5.3793A}, but the Helium species has a little difference with DAMPE measurement and is roughly consistent with CREAM observations at several highest energy points. The reason is that the measured spectral break-off from DAMPE for Helium species is at the rigidity of $\sim$34 TV, which is a little higher than proton under the Z-dependent energy cut-off frame \citep{2021PhRvL.126t1102A}. 
To keep the uniformity of all the nuclei, we choose the 
cut-off rigidity at 15 TV, which doesn't affect the results of other heavier primary nuclei and the conclusions. Under this physical scenario, our model expects that all the heavier primary nuclei have the same rigidity break-off at 5 TV as shown in Figure \ref{fig:prispec}, which can be observed by the HERD experiment in future \citep{2022PhyS...97e4010K}.

\subsubsection{Secondary Particles}
\begin{figure}[htp]
	\centering
	\includegraphics[width=4.cm]{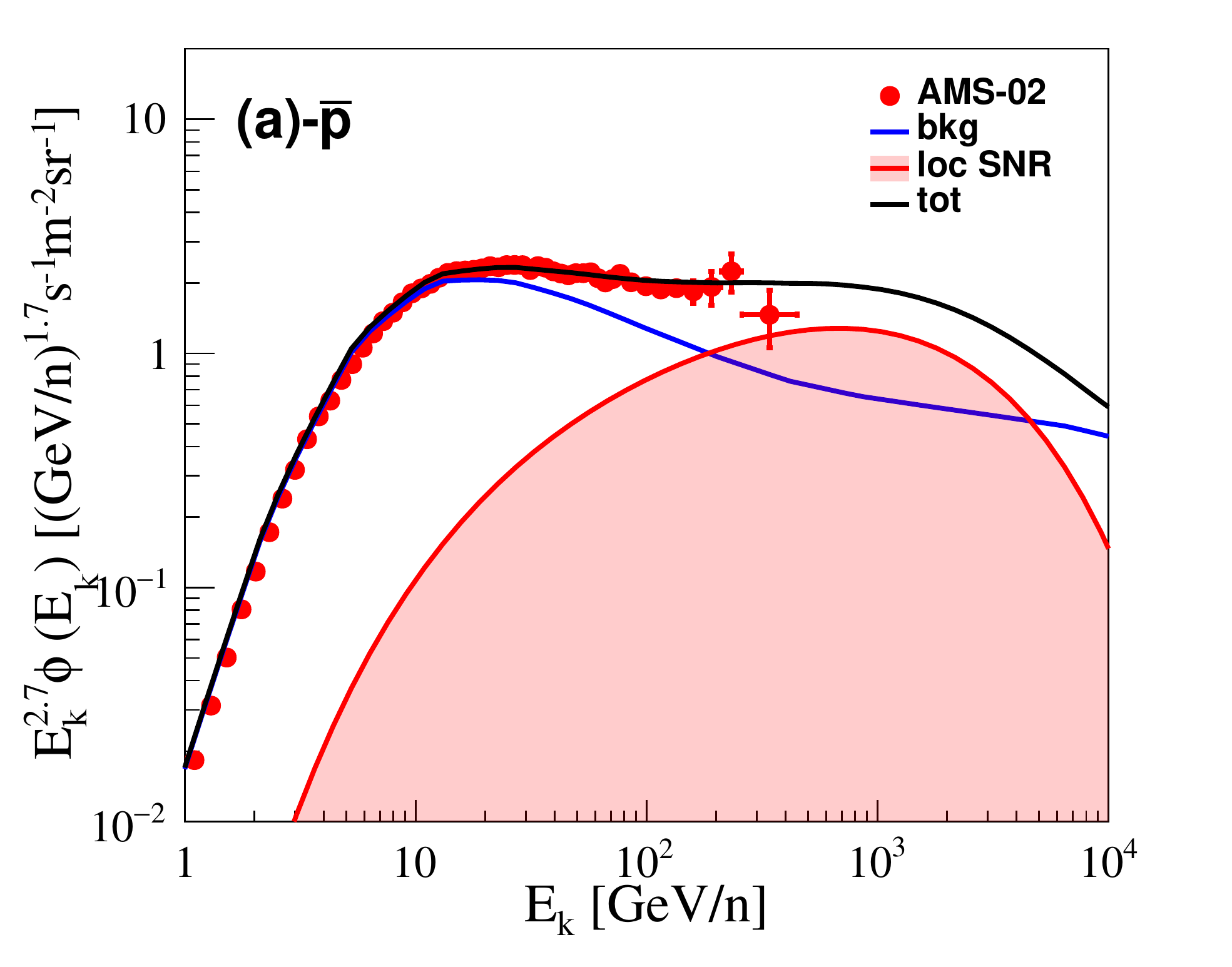}
	\includegraphics[width=4.cm]{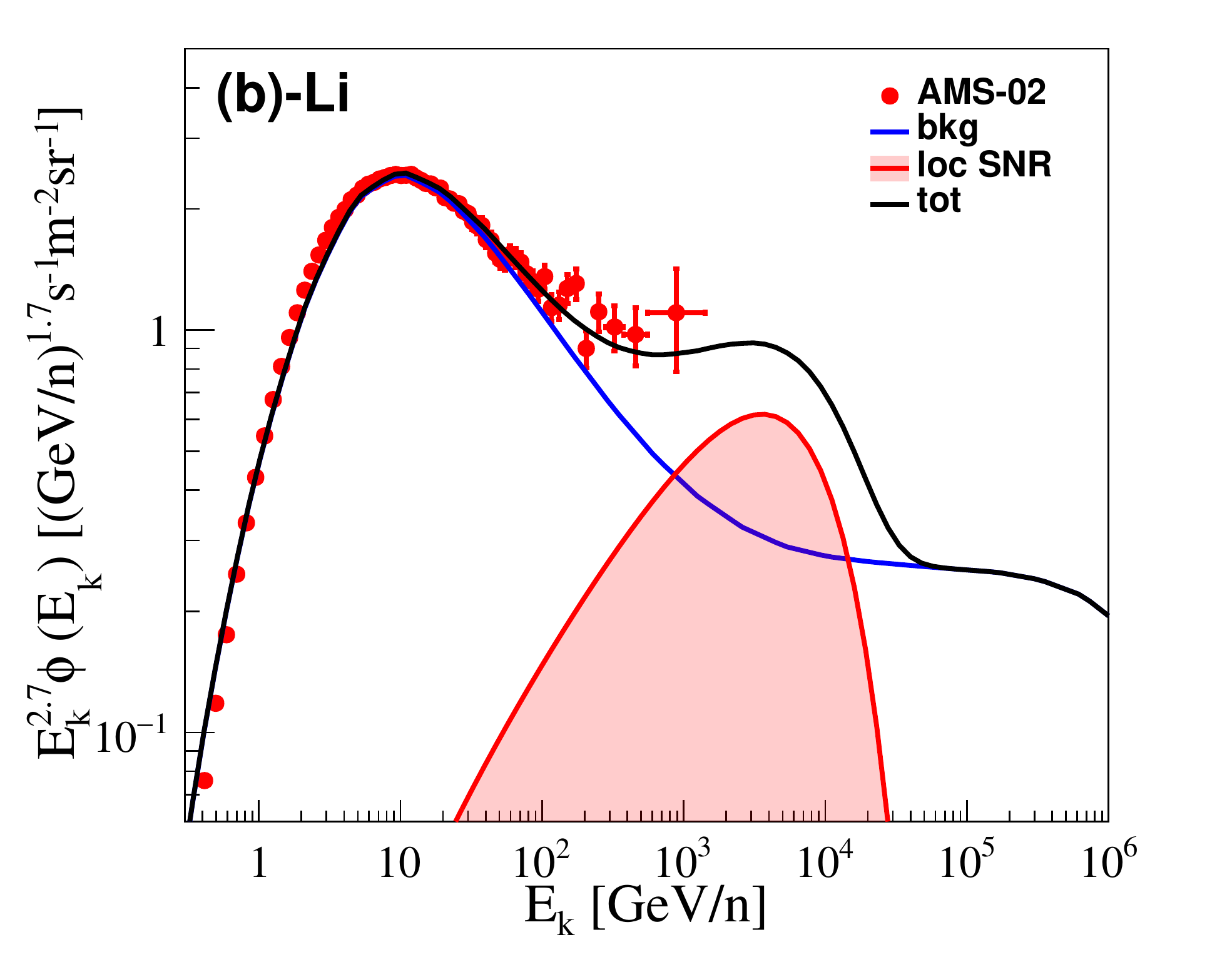}
	\includegraphics[width=4.cm]{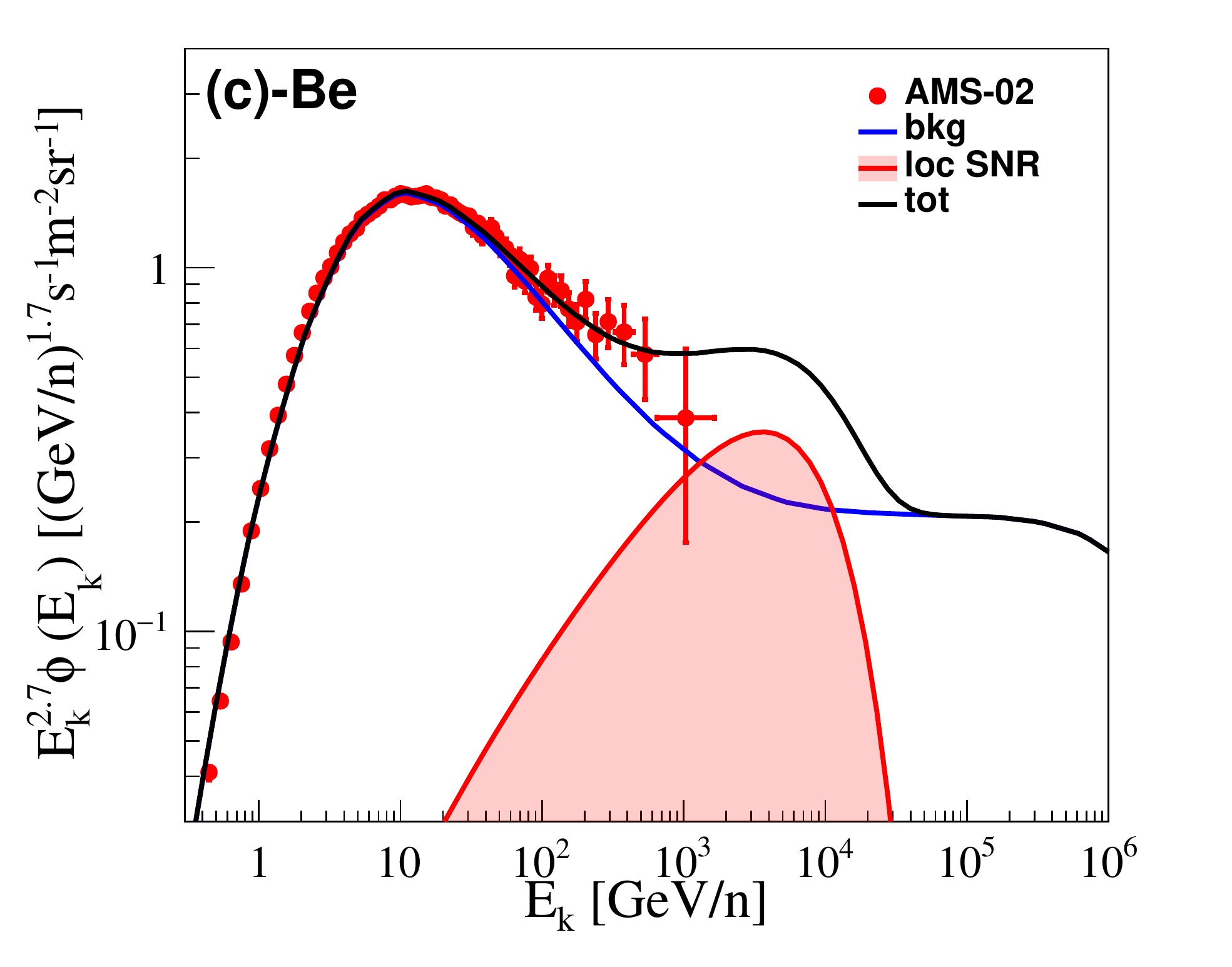}
	\includegraphics[width=4.cm]{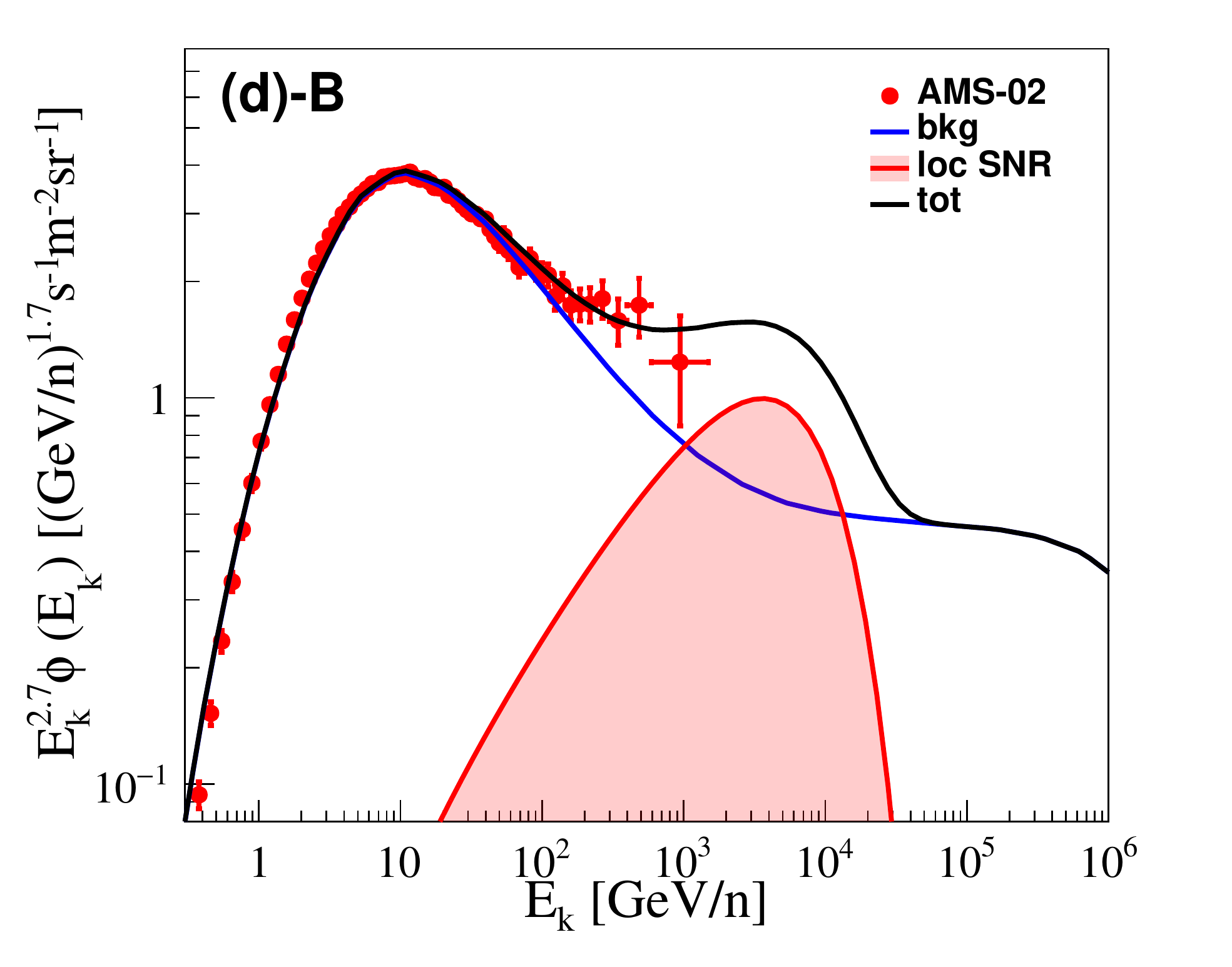}
	\caption{Similar to Figure \ref{fig:pHespec}, the secondary particles of $\bar p, Li, Be$ and $B$ from panel (a) to panel (d). The data points are from the AMS-02 experiment.\citep{2016PhRvL.117i1103A,2018PhRvL.120b1101A}.}
	\label{fig:secpspec}
\end{figure}

Following the primary species, the secondary particles spectra also have two components as the local and the global sources. 
Figure \ref{fig:secpspec} gives the spectra of $\bar p$, $Li, Be$ and $B$ from panel (a) to (d). 
The model calculations are well consistent with the observations from AMS-02 \citep{2016PhRvL.117i1103A,2018PhRvL.120b1101A}.
The hardening of $\bar p$ spectrum starts
from tens of GeV owing to 200 GV hardening of its mother proton. For the species of $Li, Be$ and $B$, they are produced through fragmentation of heavier nuclei, such as $C, O$ and so on. They keep the same behaviour as their mother particles with the hardening at 200 GV and break-off at around $\sim$10 TV. The typical energy break-off for secondary particles is pivotal to validate the dominant interactions around source regions for the nearby source, which are probe to unveil the origin of positron and $\bar p$ excess at high energy. We hope that the spectral bend around $\sim$TeV can be observed by HERD experiment in near future.

\subsubsection{Two components particles} \label{subsubsec:autonumber}
\begin{figure*}[htp]
	\centering
	\includegraphics[width=5.5cm]{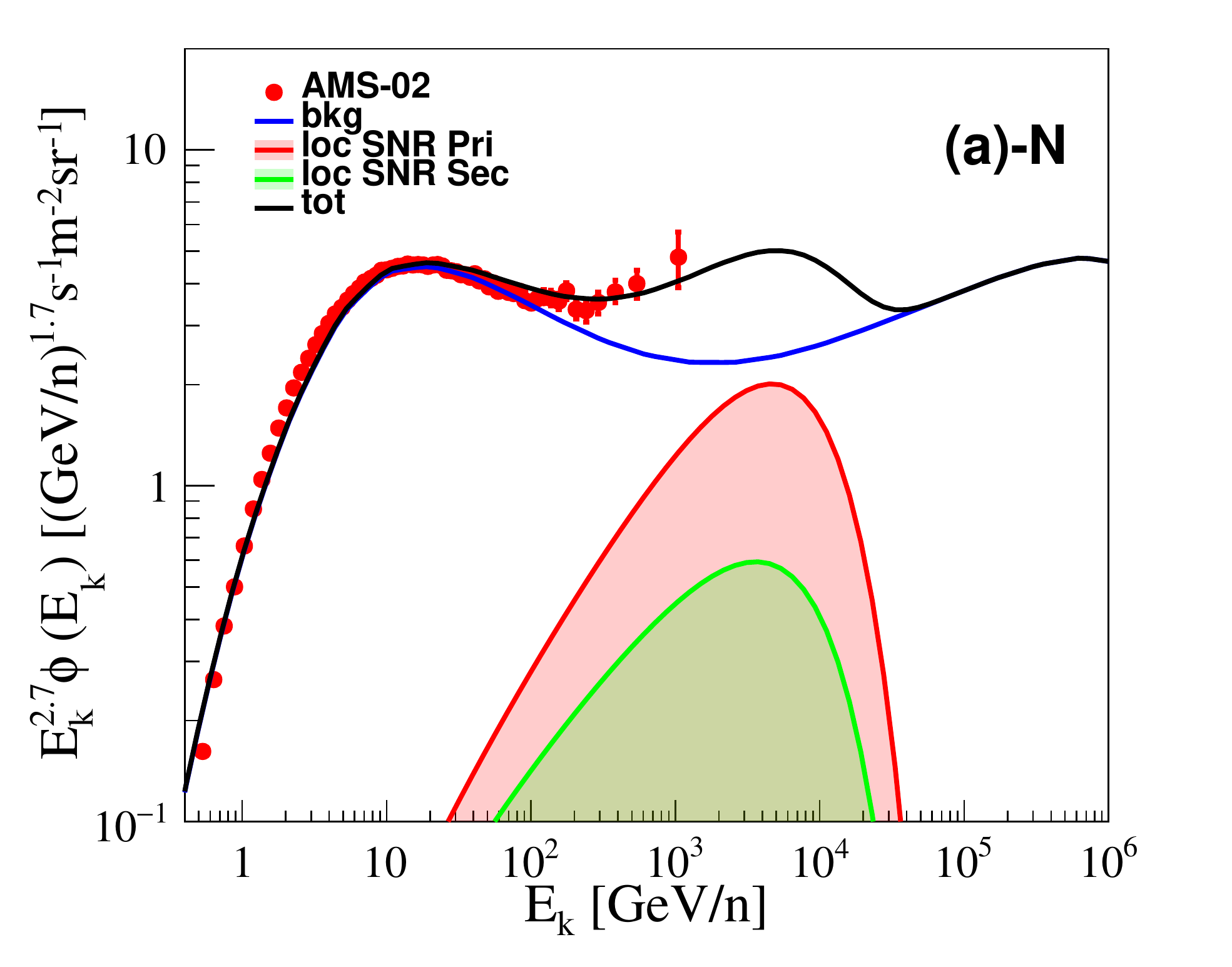}
	\includegraphics[width=5.5cm]{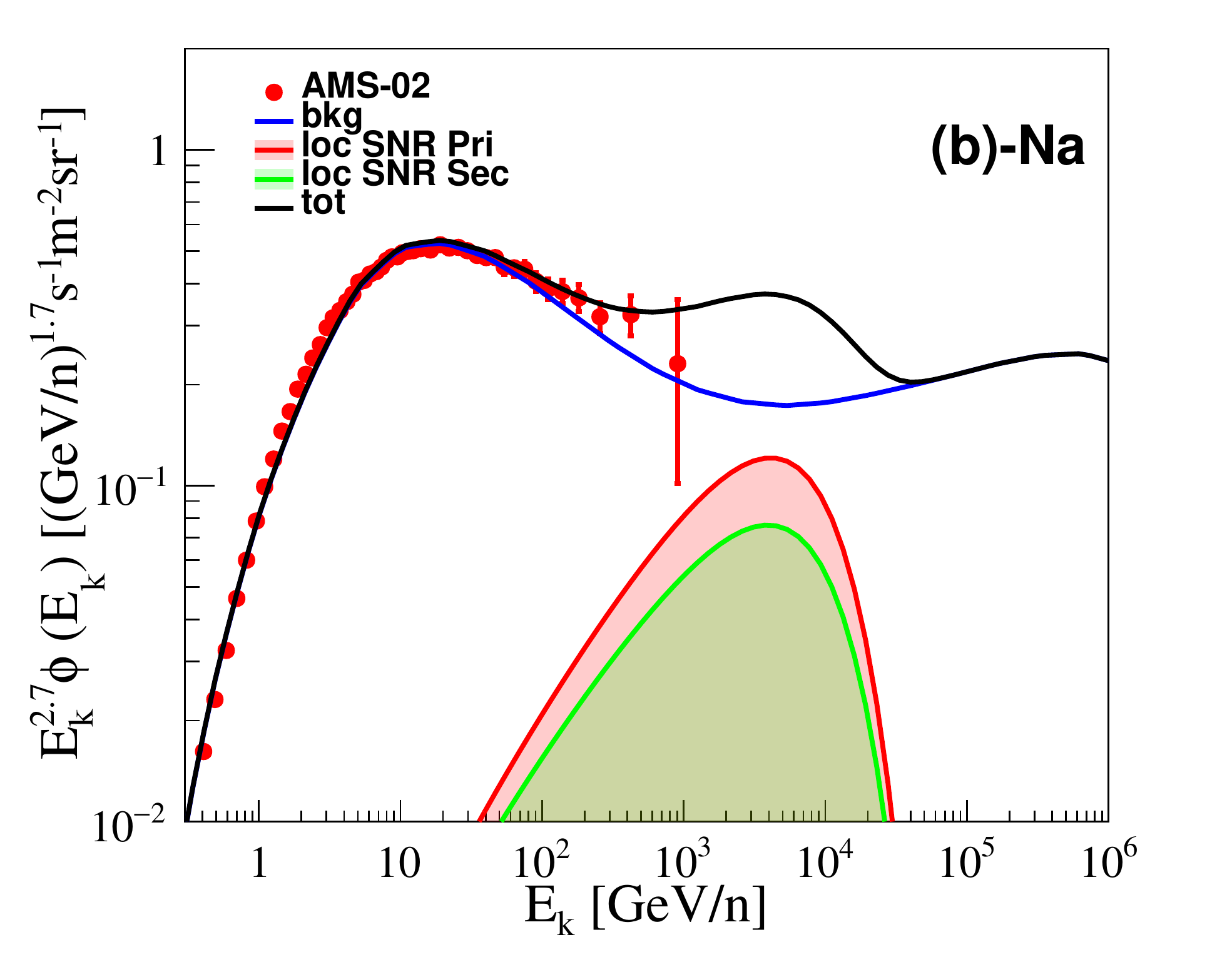}
	\includegraphics[width=5.5cm]{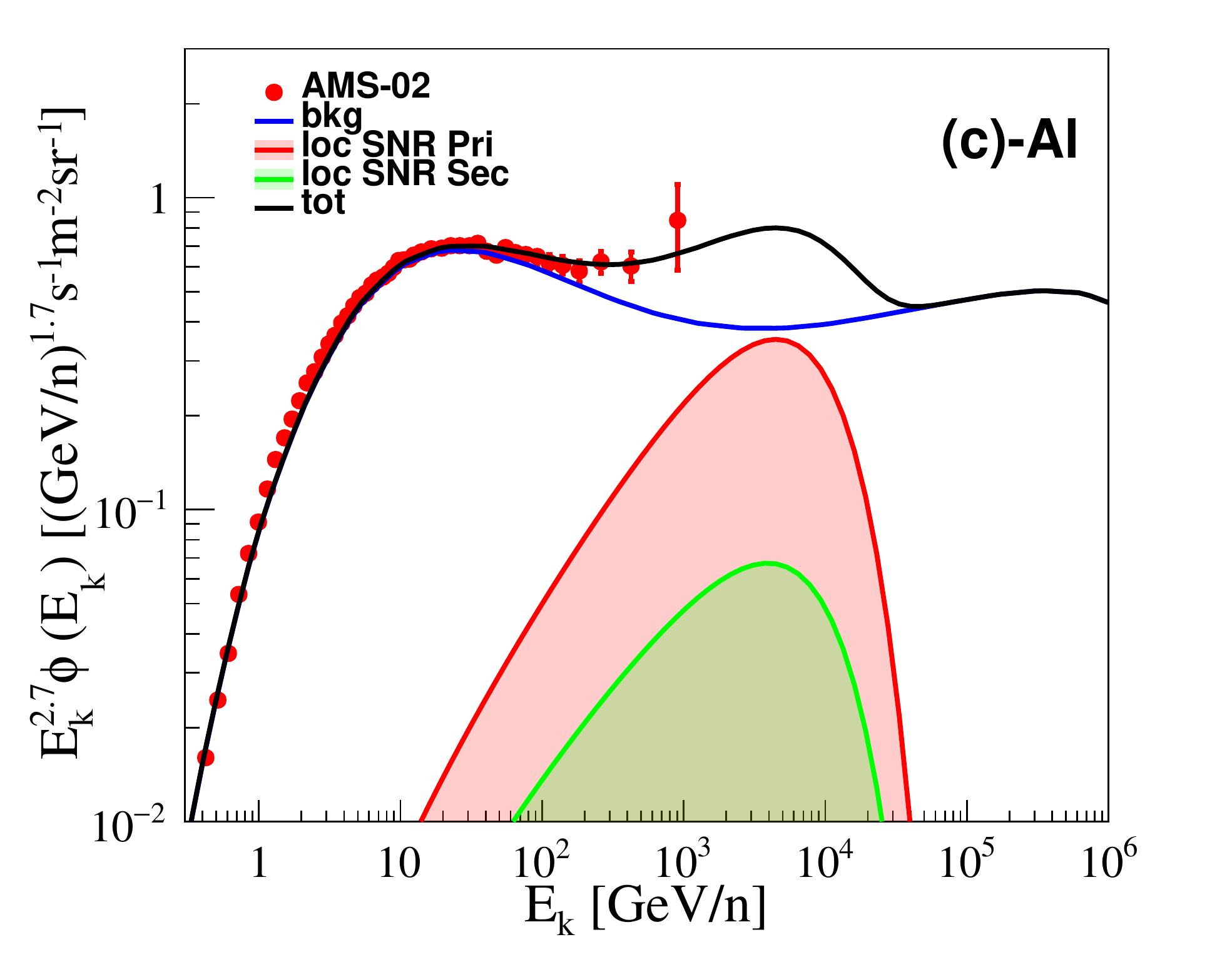}
	\caption{Similar to Figure \ref{fig:pHespec}, the individual species of $N, Na, Al$, including both primary and secondary components, from panel (a) to (c). The data points are from the AMS-02 experiment\citep{2018PhRvL.121e1103A,2021PhRvL.127b1101A}.}
	\label{fig:prisecspec}
\end{figure*}

There are also some special particles, such as $N, Na, Al$, including both the primary and secondary components. 
They are thought to be
produced both in galactic background sources, and by the collisions of heavier nuclei with the interstellar medium (ISM) ($\rm O\rightarrow \rm N+ \rm X$, $\rm Mg\rightarrow \rm Na+ \rm X$, $\rm Si\rightarrow \rm Na+ \rm X$, $\rm Si\rightarrow \rm Al+ \rm X$)\citep{2015ARA&A..53..199G,2013A&ARv..21...70B,2007ARNPS..57..285S}. As shown in Figure \ref{fig:prisecspec}, the red solid line with shadow is the contribution from primary component accelerated  directly by local SNR, the green solid line with shadow indicates the secondary component produced by the collisions of heavier nuclei with ISM, the blue solid line is from galactic background sources and the black solid line is the sum of them.
It can be seen that the model calculations are agreement with the observations, and their spectra also have a break-off structure similar to that of primary particles at $\sim5$ TV. 

\subsubsection{All particles} \label{subsubsec:all particles}
\begin{figure}[htp]
	\centering
	\includegraphics[width=8.5cm]{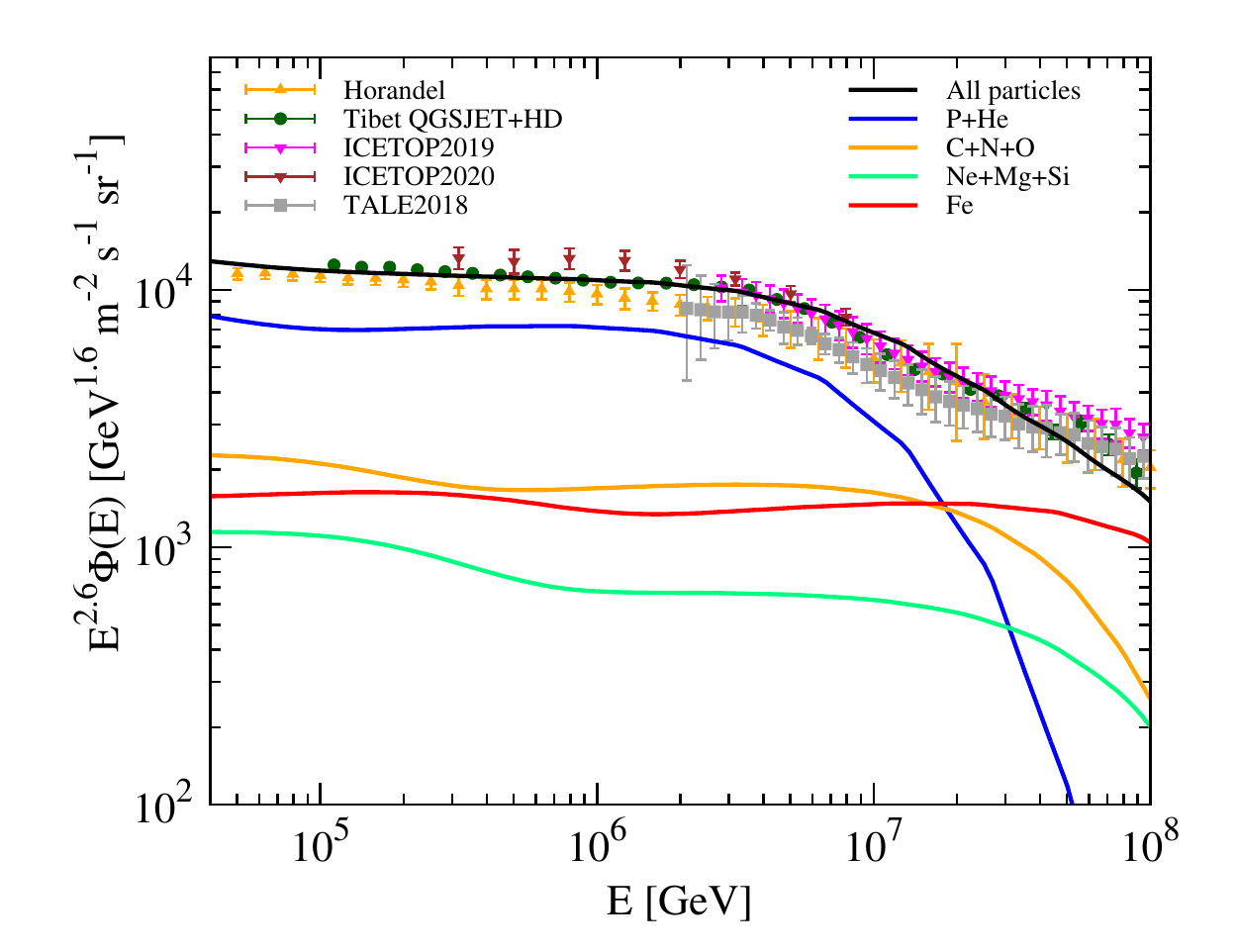}
	\caption{The observed and calculated all-particle spectra. In the model calculations, the solid blue, orange, green, red and black line represent the P+He, C+N+O, Ne+Mg+Si, Fe species and all-particle spectra respectively. The data points are measured by TALE \citep{2018ApJ...865...74A}, Tibet-AS$\gamma$ \citep{2008ApJ...678.1165A}, ICETOP \citep{2019PhRvD.100h2002A,2020PhRvD.102l2001A} experiments and from weighted by Horandel \citep{2003APh....19..193H}.}
	\label{fig:allp}
\end{figure}

The space-borne experiments have decisive advantages in the seperation ability for different species. However they have limited effective detector area, which lead to the lower statistical event numbers in high energy. On the contrary, the ground-based experiments have opposite properties. This makes that the  spectral measurements for individual species step into dilemma above tens of TeV. The all-particle spectra can make up this shortcoming to constrain the high energy contributions. Figure \ref{fig:allp} shows the model calculations and measurements for all-particle spectrum. In the calculations, four groups as $H+He, C+N+O, Ne+Mg+Si$ and $Fe$ are demonstrated in { blue, orange, green and red} solid lines. It is clear that the knee structure of the all-particle
spectrum can be properly reproduced by the background component assuming a Z-dependent
cutoff with $R_c \sim 7$ PV. In this case, the light species of protons and He nuclei dominantly composes the knee structure. This is because we try to fit the KASCADE spectra of proton and Helium, which was also favoured by the diffuse $\gamma$-ray measurement at ultra-high energy by AS$\gamma$ experiment \citep{2021PhRvL.126n1101A}.

\subsubsection{Positron and Electron}
\begin{figure}[htp]
	\centering
	\includegraphics[width=4.cm]{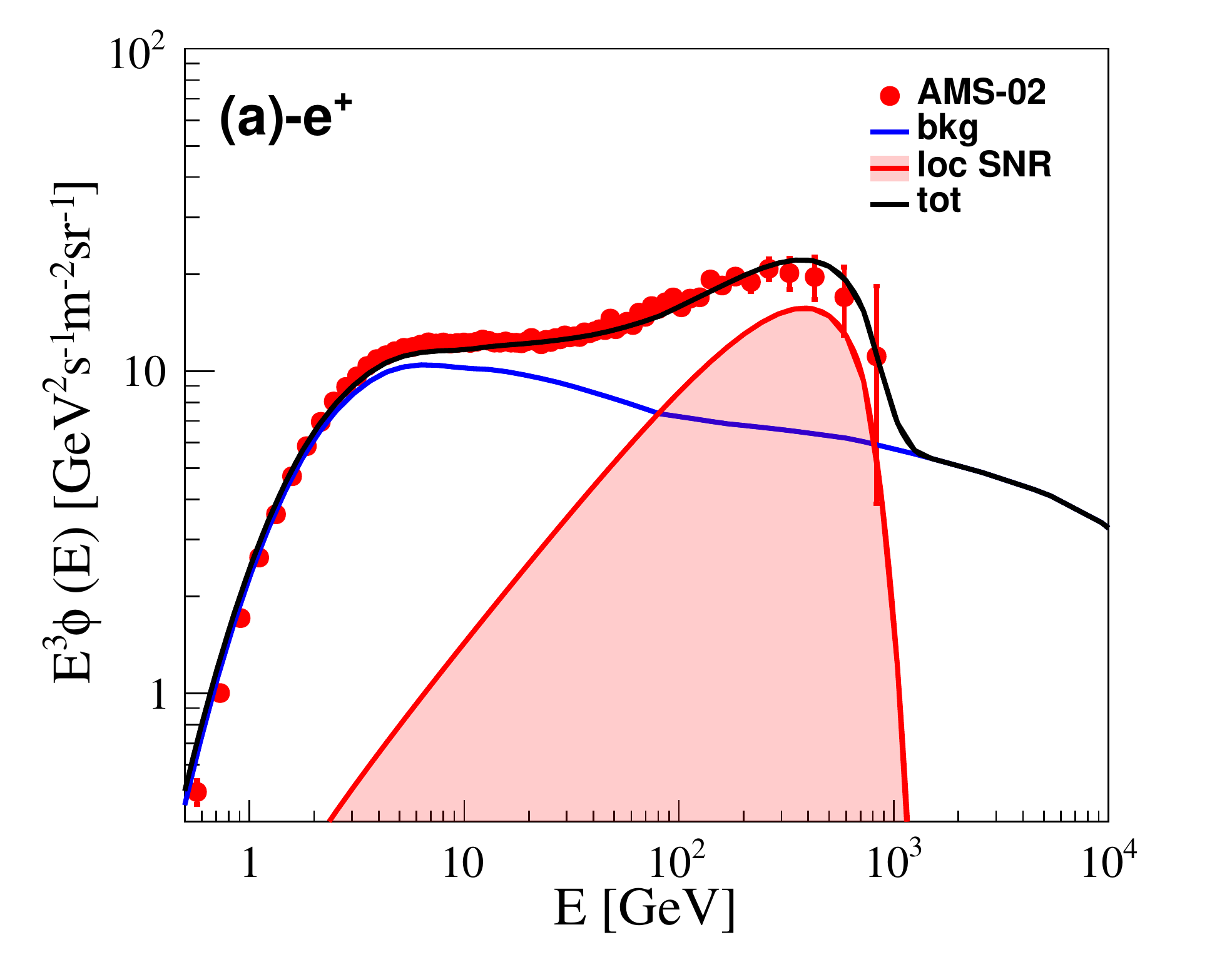}
	\includegraphics[width=4.cm]{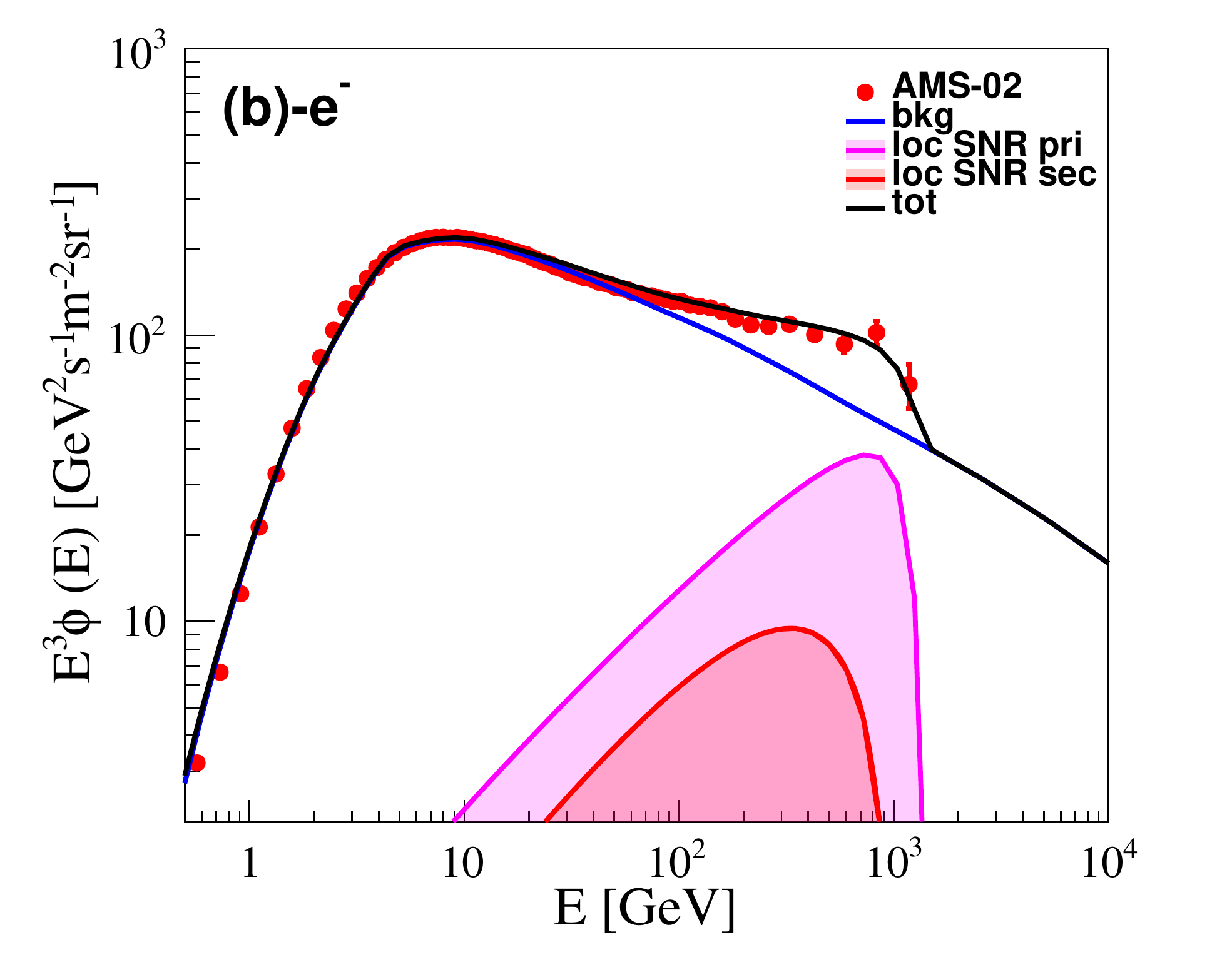}
	\includegraphics[width=4.cm]{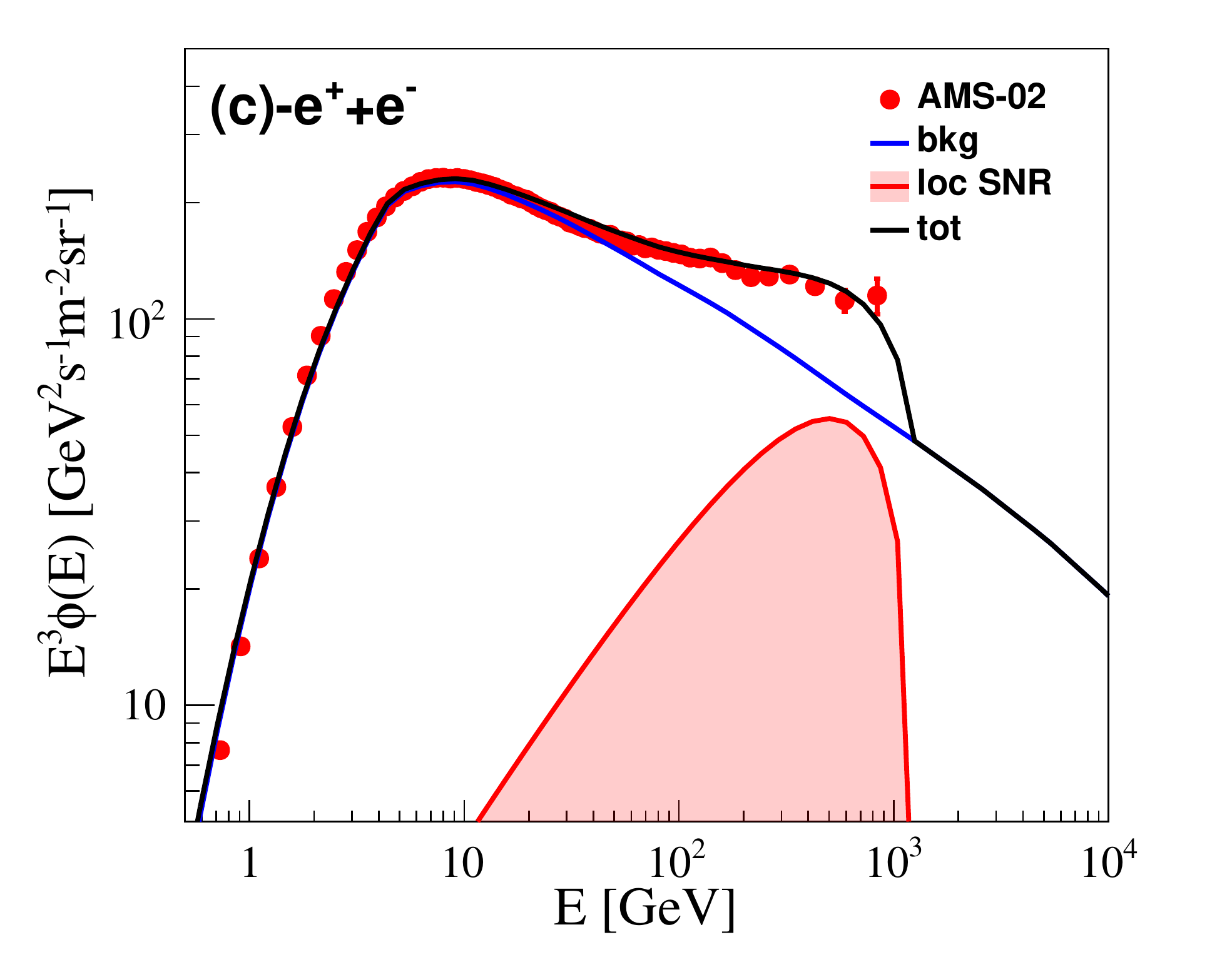}
	\includegraphics[width=4.cm]{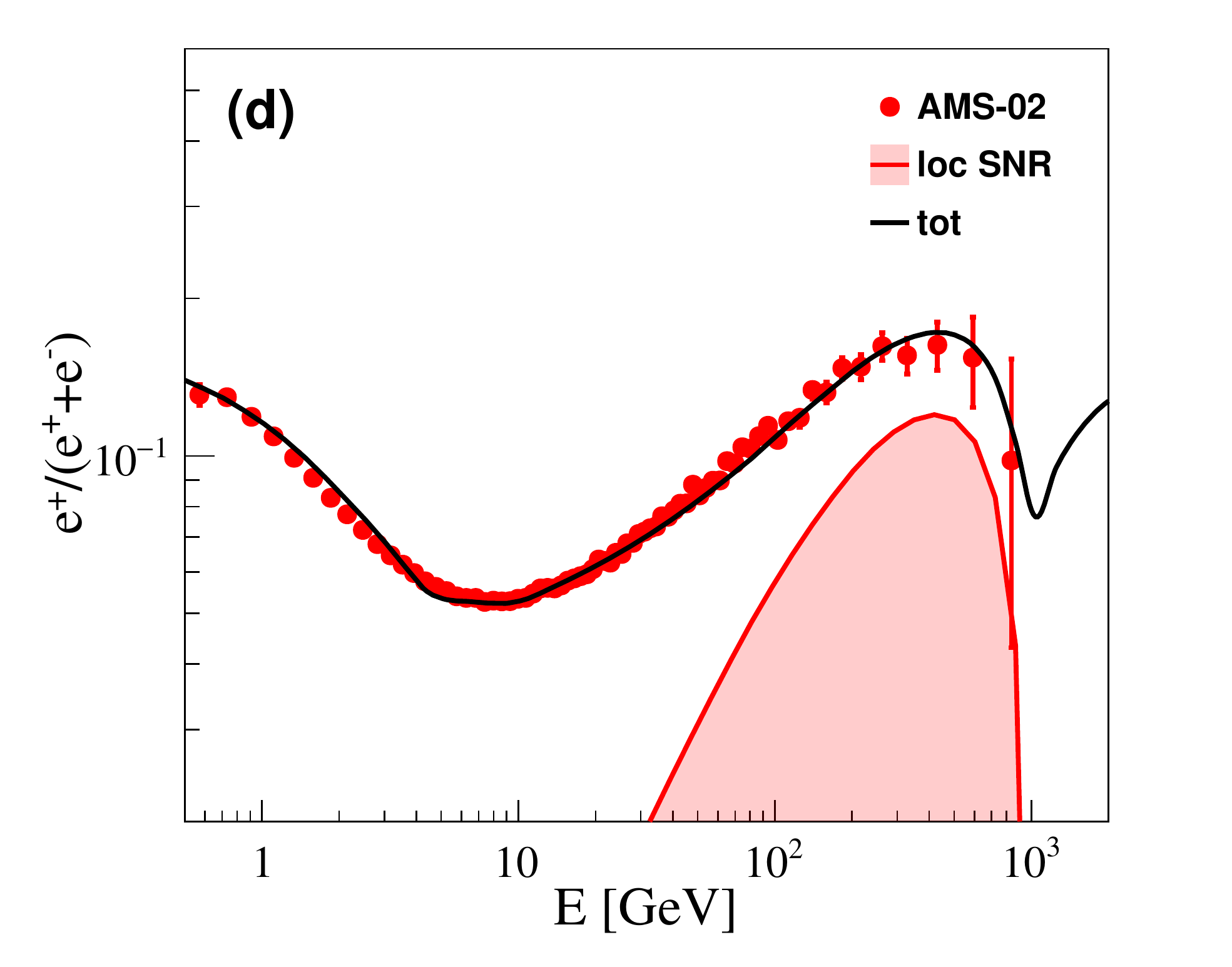}
	\caption{Similar to the Figure \ref{fig:pHespec}, the spectra of positron, electron and their sum  from panel (a) to (c) and the ratio of positron to the sum of positron and electron in panel (d). The data points are adopted from AMS-02 experiments \citep{2019PhRvL.122d1102A,2019PhRvL.122j1101A}.}
	\label{fig:posielec}
\end{figure}

It is a hot topic for the positron excess and electron spectral hardening at high energy. Similar to nuclei secondary particles, the positron is also composed of two parts as around Geminga SNR and global background component from pp-collision. Figure \ref{fig:posielec} shows the spectra of positron, electron, their sum and the ratio of positron to the sum of positron and electron. The model calculations are good reproduction of experimental data. Particularly, the positron will take $5\%$ energy from its mother species proton in pp-collision. Owing to the spectral break-off around 5 TeV of proton, the positron spectrum has cut-off around 250 GeV and successfully reproduce the AMS-02 measurements as shown in panel (a). As the discussion in section 2, the age of Geminga SNR is $3.3\times10^5$ yrs, which leads to the energy break of electron spectrum about TeV as shown in panel (b) \citep{1995PhRvD..52.3265A}. Then the difference of energy break-off between positron and electron can be naturally understood. For the ratio of positron to the sum of positron and electron, it increases with increasing energy above about TeV, which results from the cut-off of the total electron spectrum at $\sim$TeV.

\subsection{Ratios}
\label{subsec:figures}
The ratios are important to understand the acceleration, propagation and interaction properties of CRs. Thanks to the unprecedented precise measurements from AMS-02, the ratios of primary to primary, secondary to secondary and secondary to secondary species can be
well measured and clearly shown the difference \citep{2017PhRvL.119y1101A,2018PhRvL.121e1103A,2021PhRvL.127b1101A,2020PhRvL.124u1102A,2021PhRvL.126d1104A}. In this section, the corresponding model calculations are obtained to reproduce those observations.


\subsubsection{Primary to primary species}
\label{subsec:pritopri}

\begin{figure*}[htp]
	\centering
	\includegraphics[width=5.5cm]{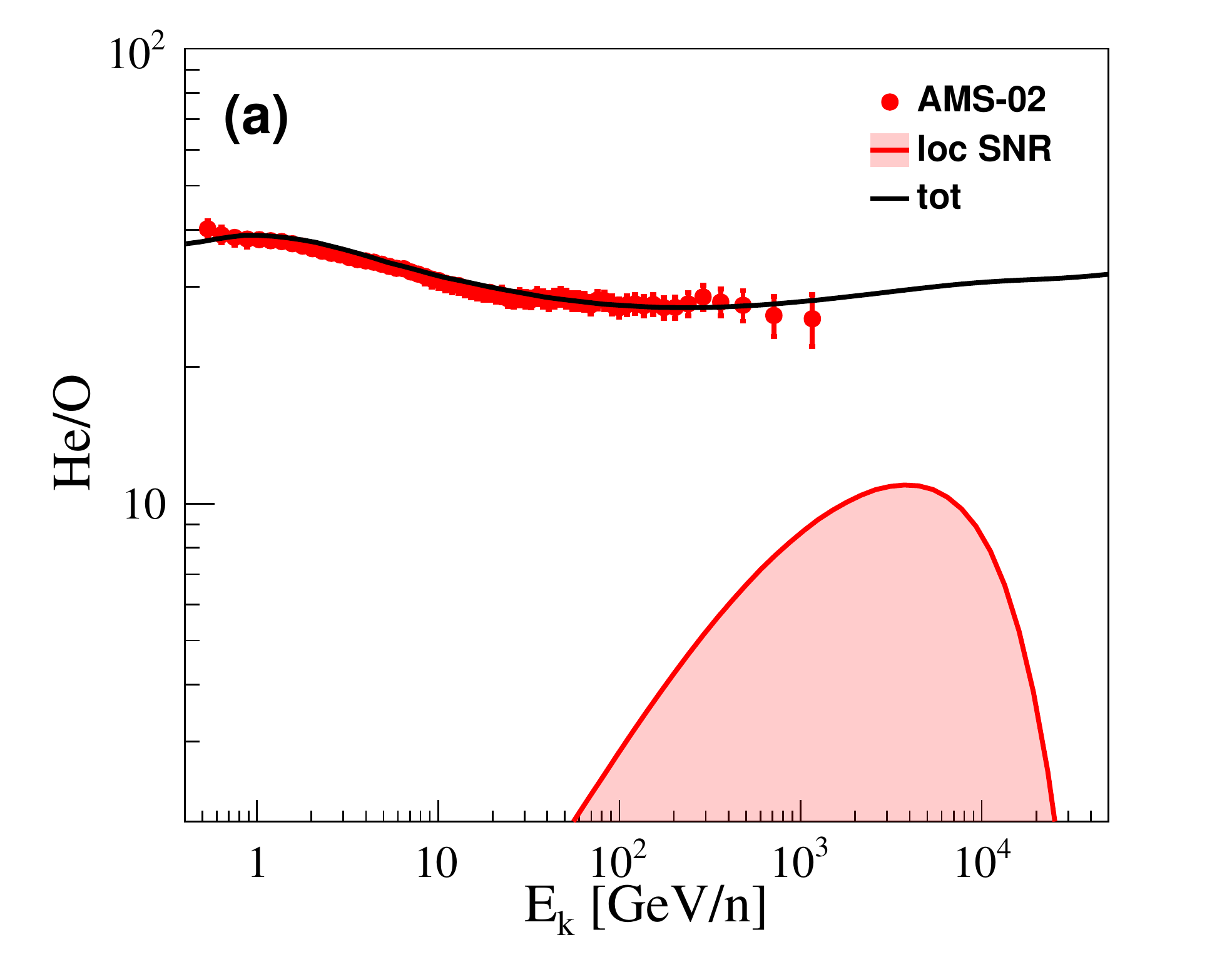}
	\includegraphics[width=5.5cm]{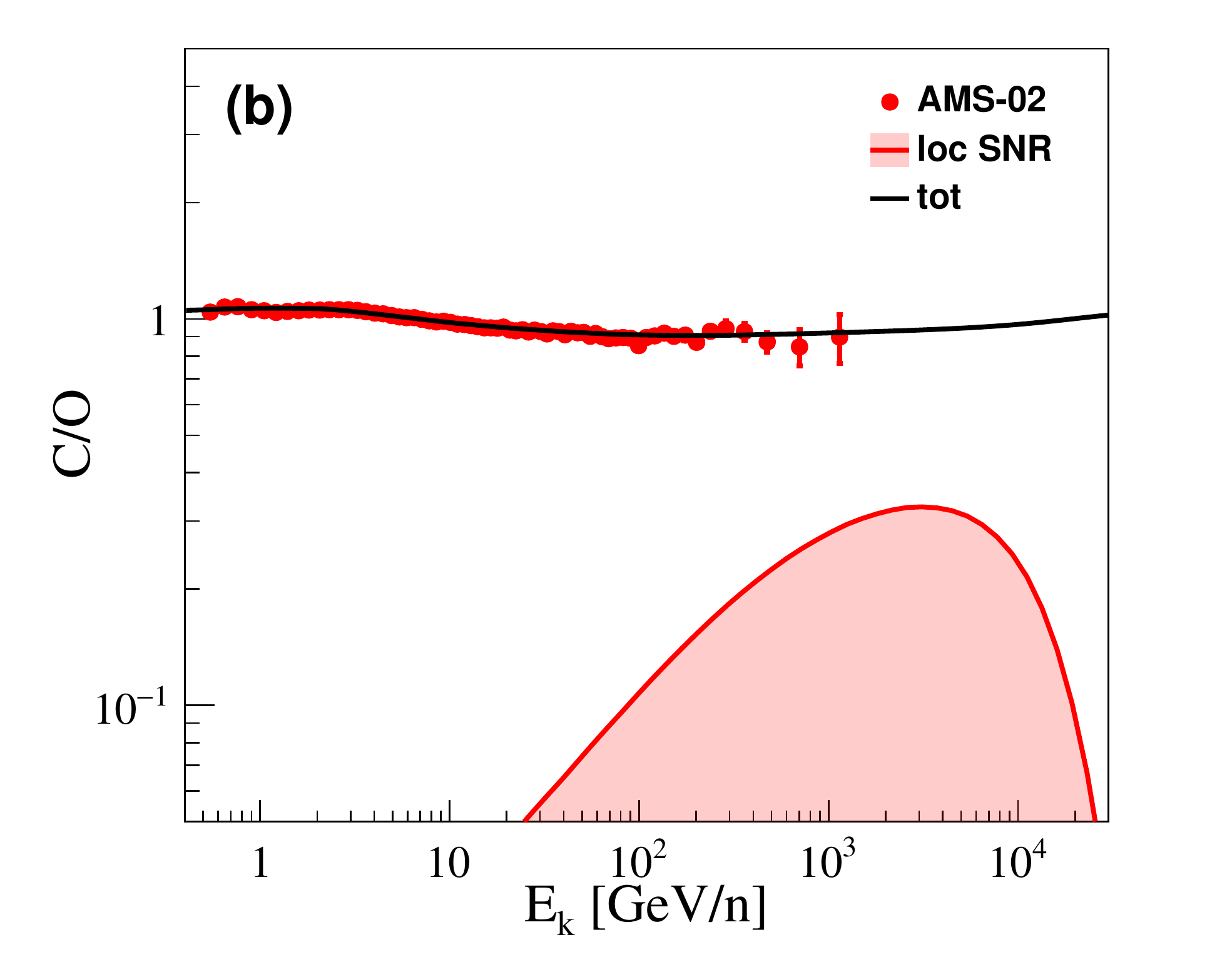}
	\includegraphics[width=5.5cm]{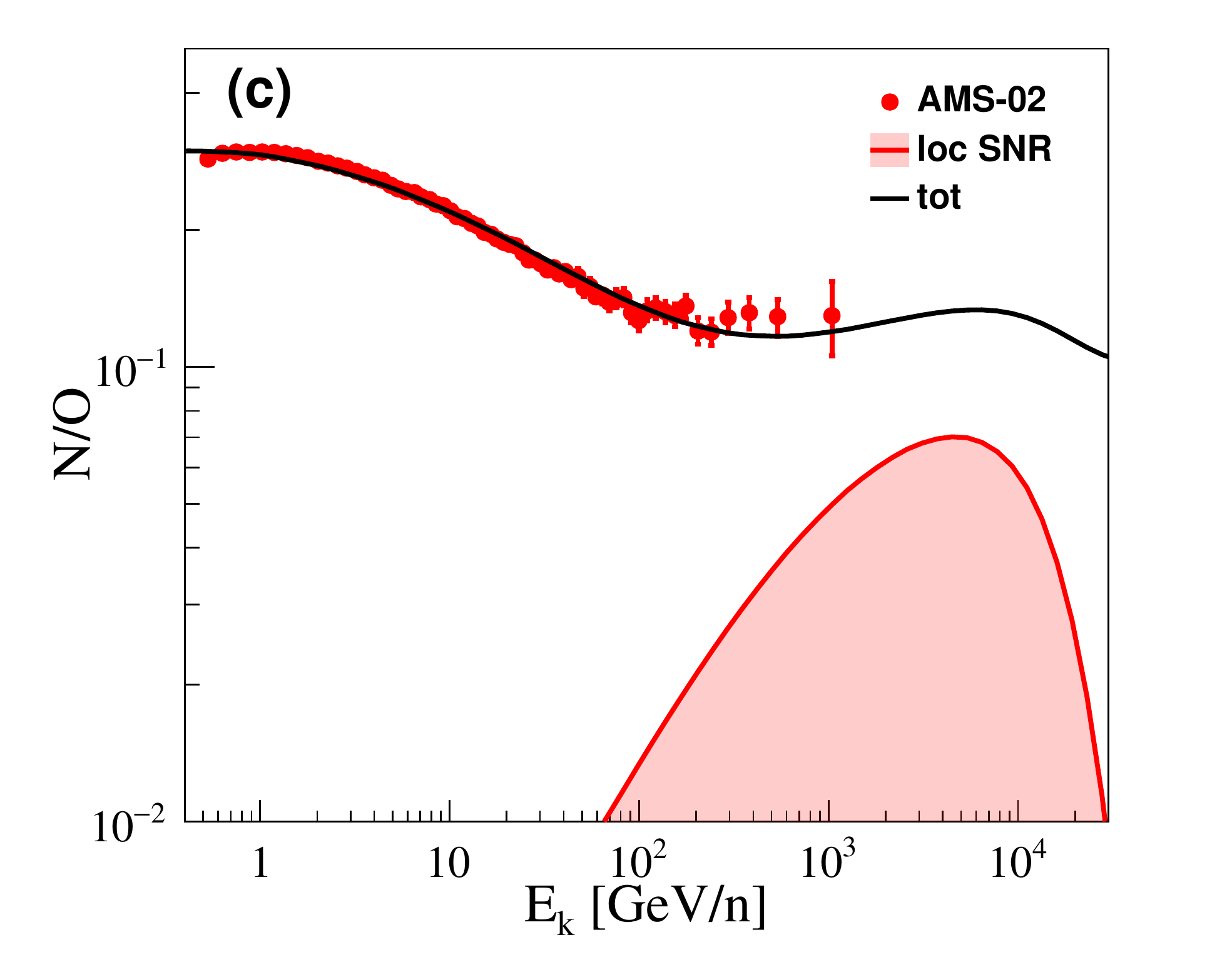}
	\includegraphics[width=5.5cm]{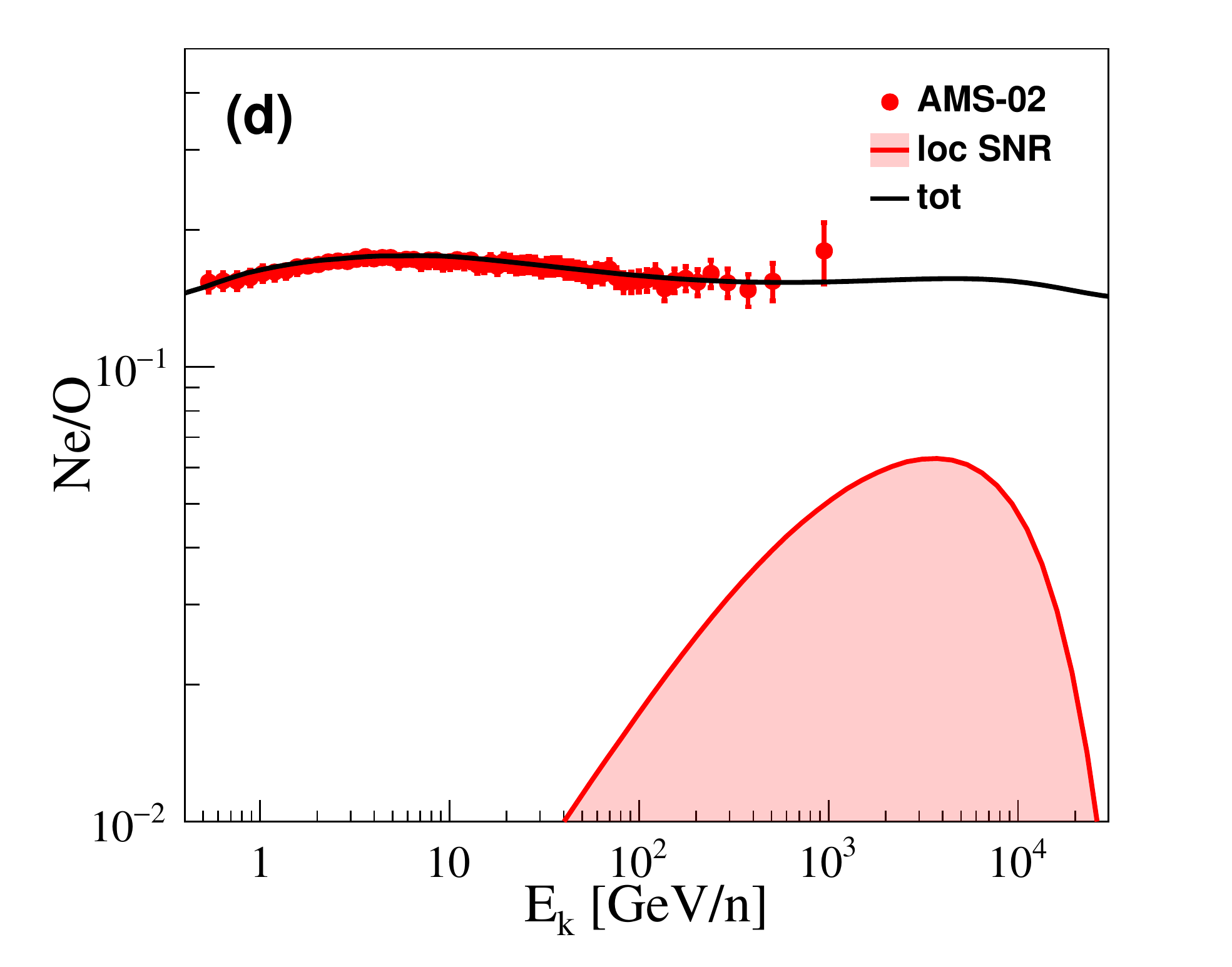}
	\includegraphics[width=5.5cm]{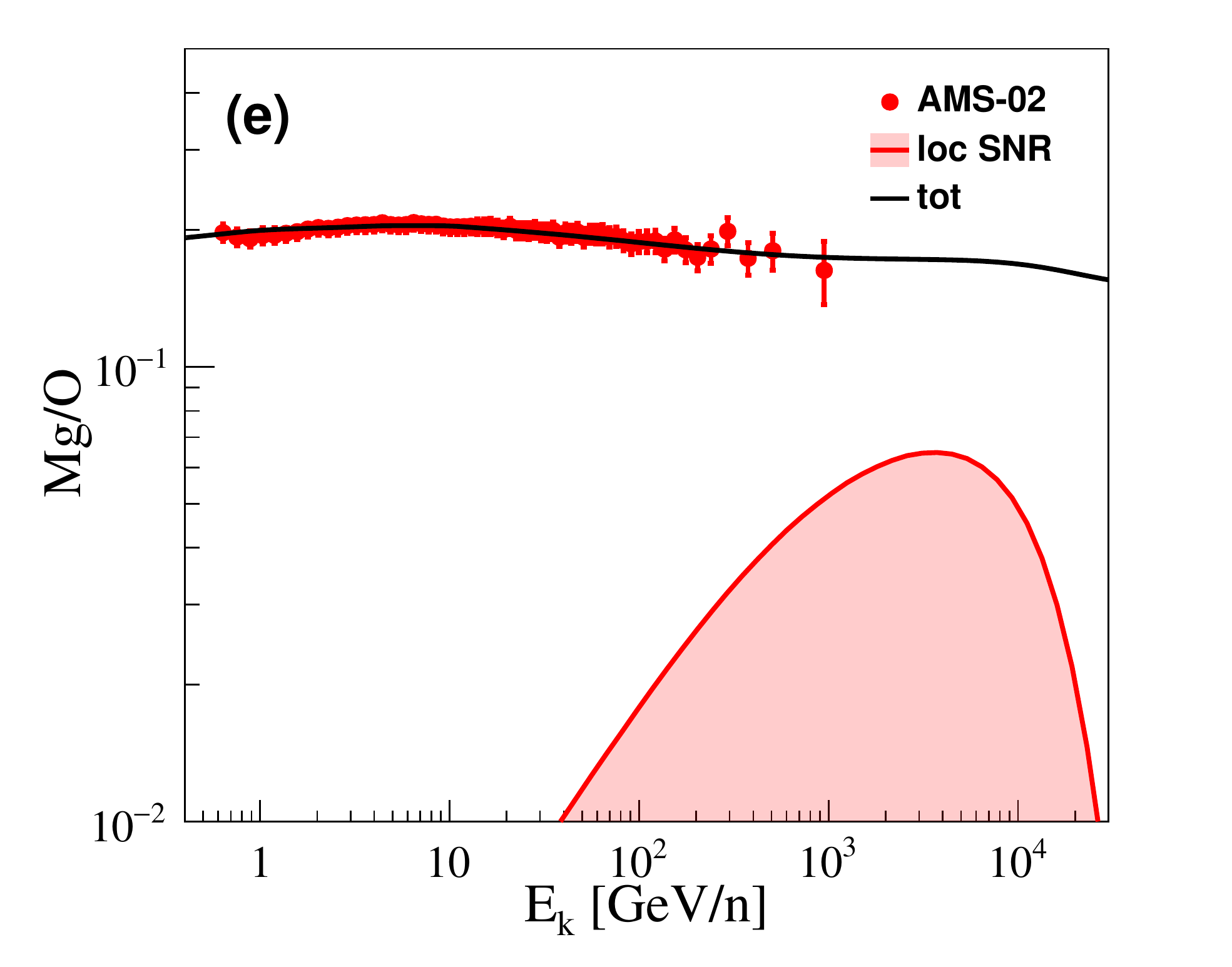}
	\includegraphics[width=5.5cm]{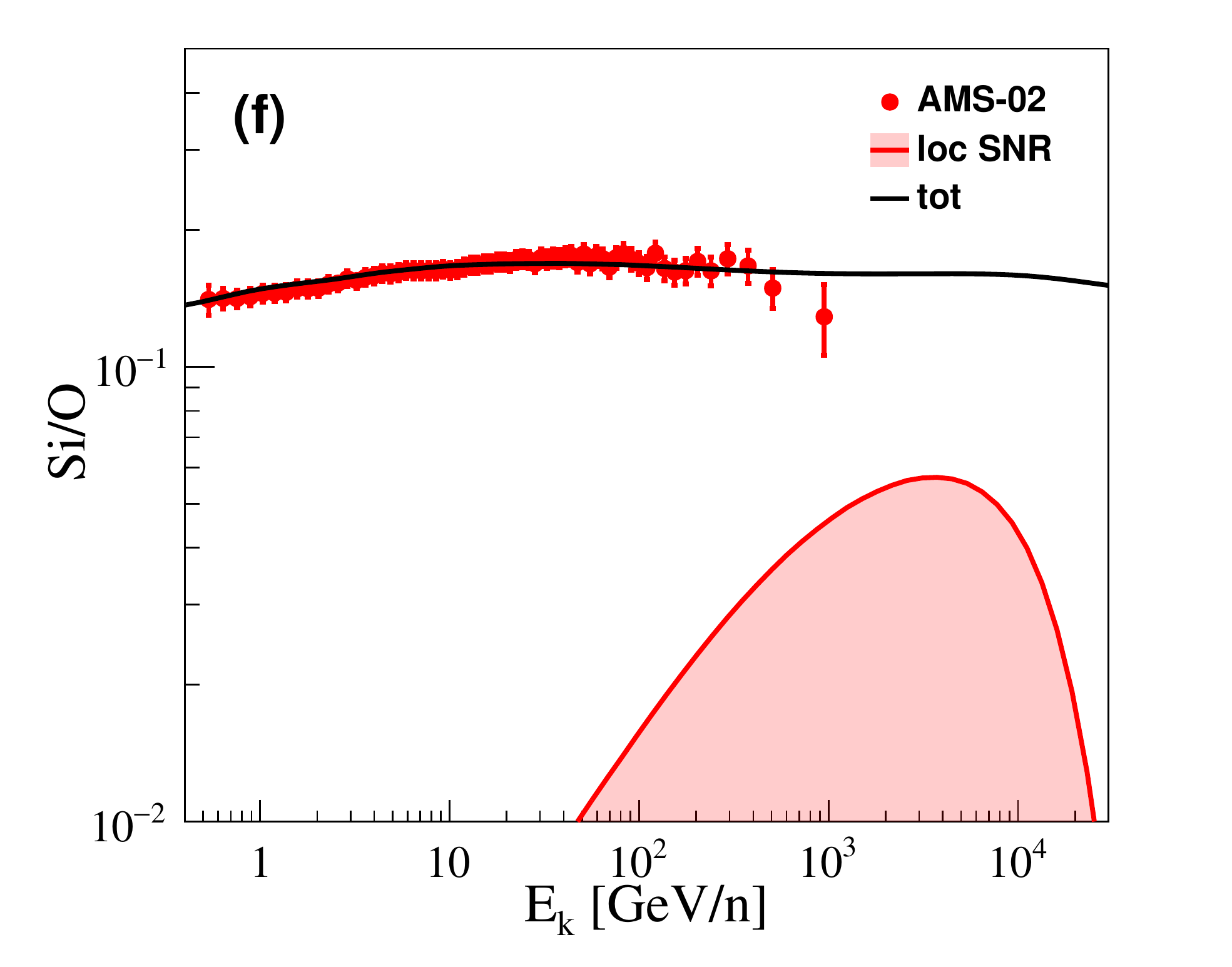}
	\includegraphics[width=5.5cm]{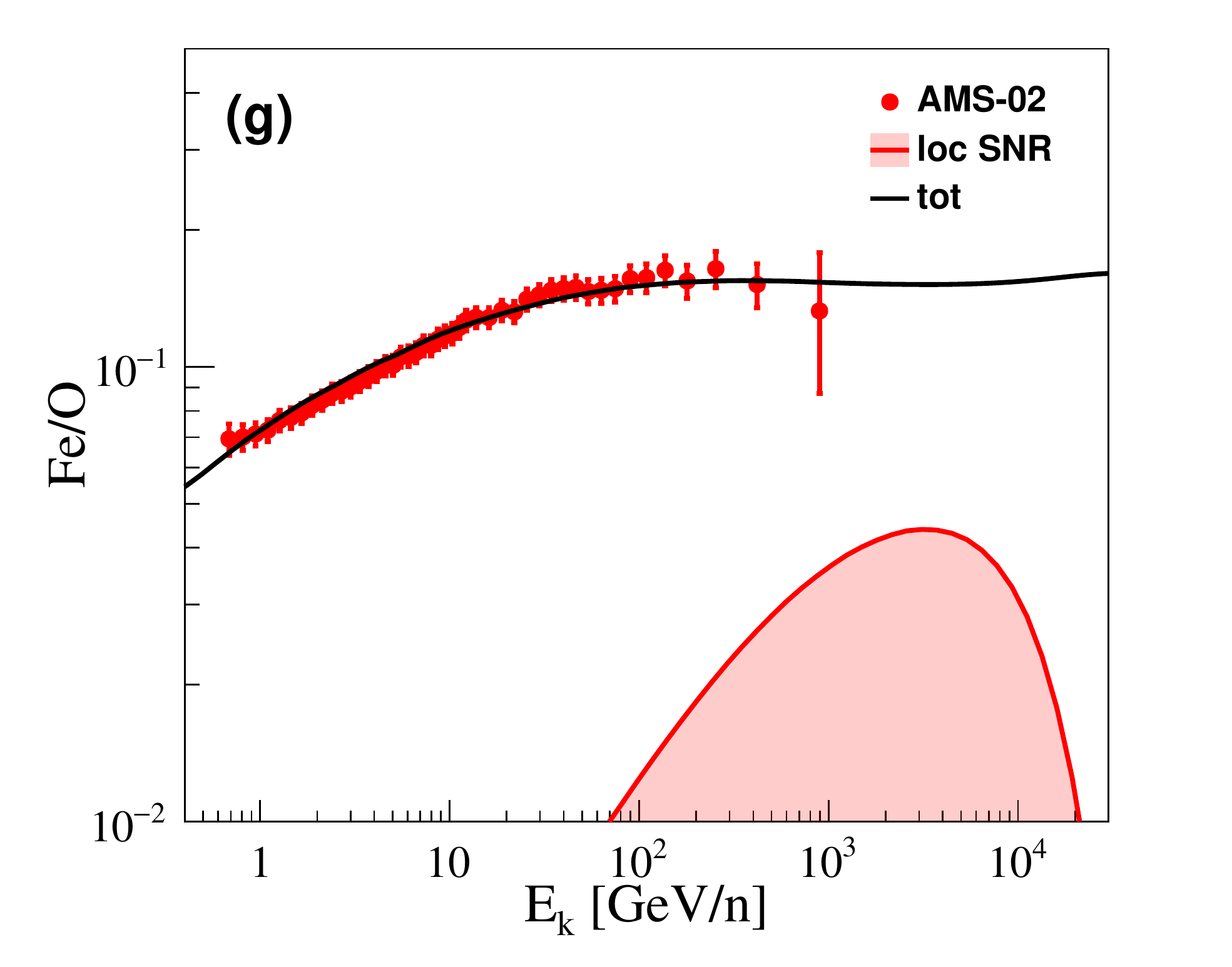}
	\includegraphics[width=5.5cm]{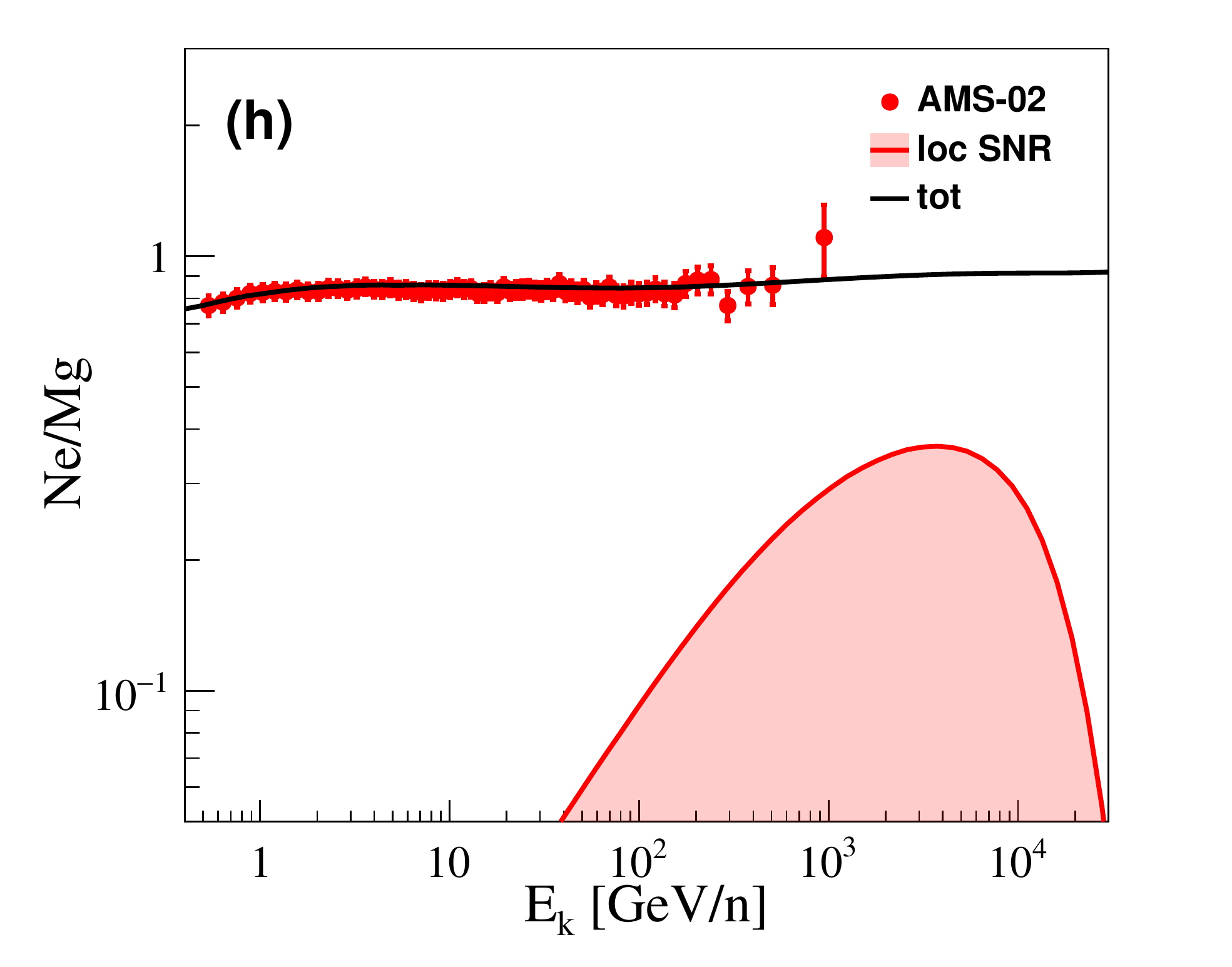}
	\includegraphics[width=5.5cm]{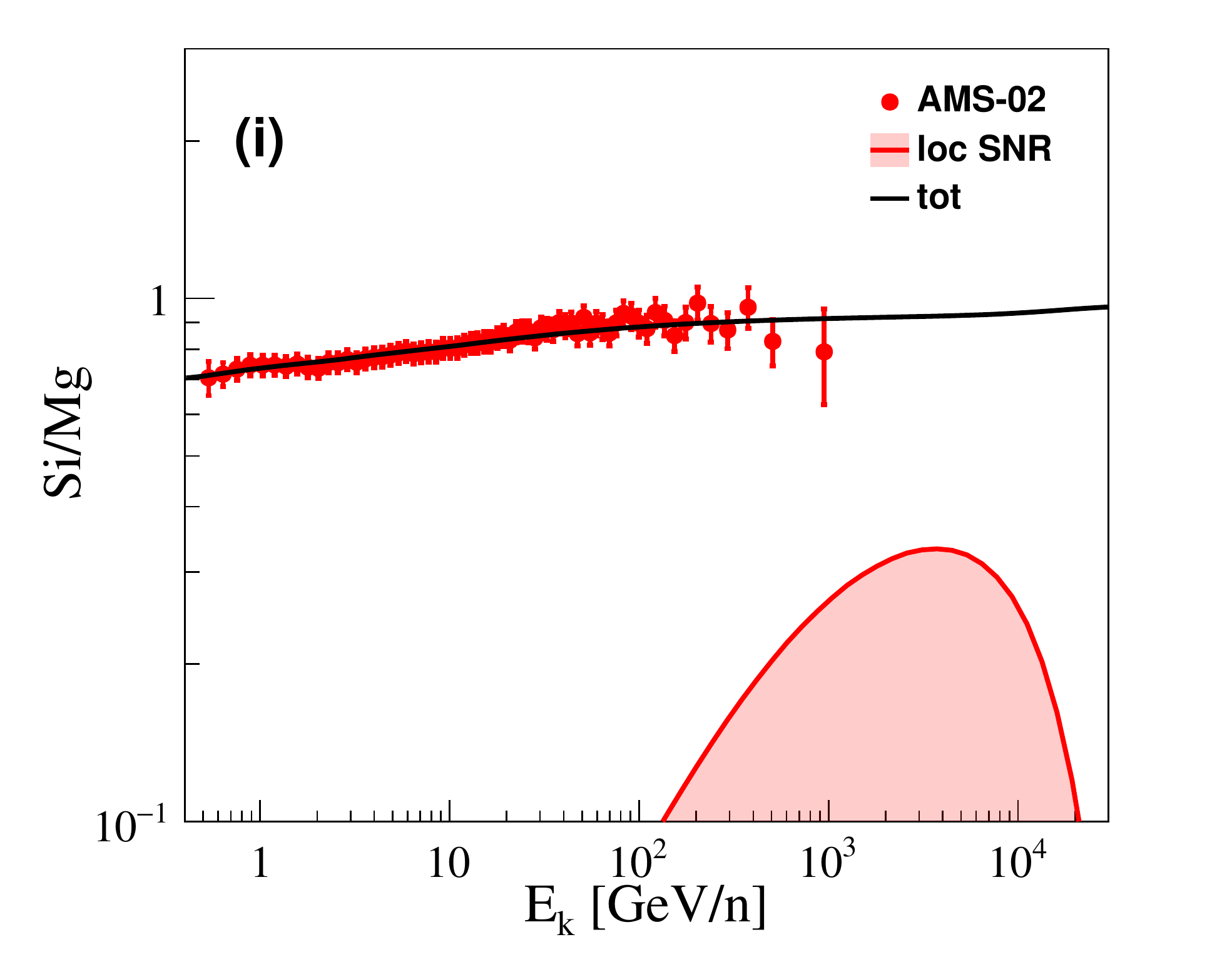}
	\includegraphics[width=5.5cm]{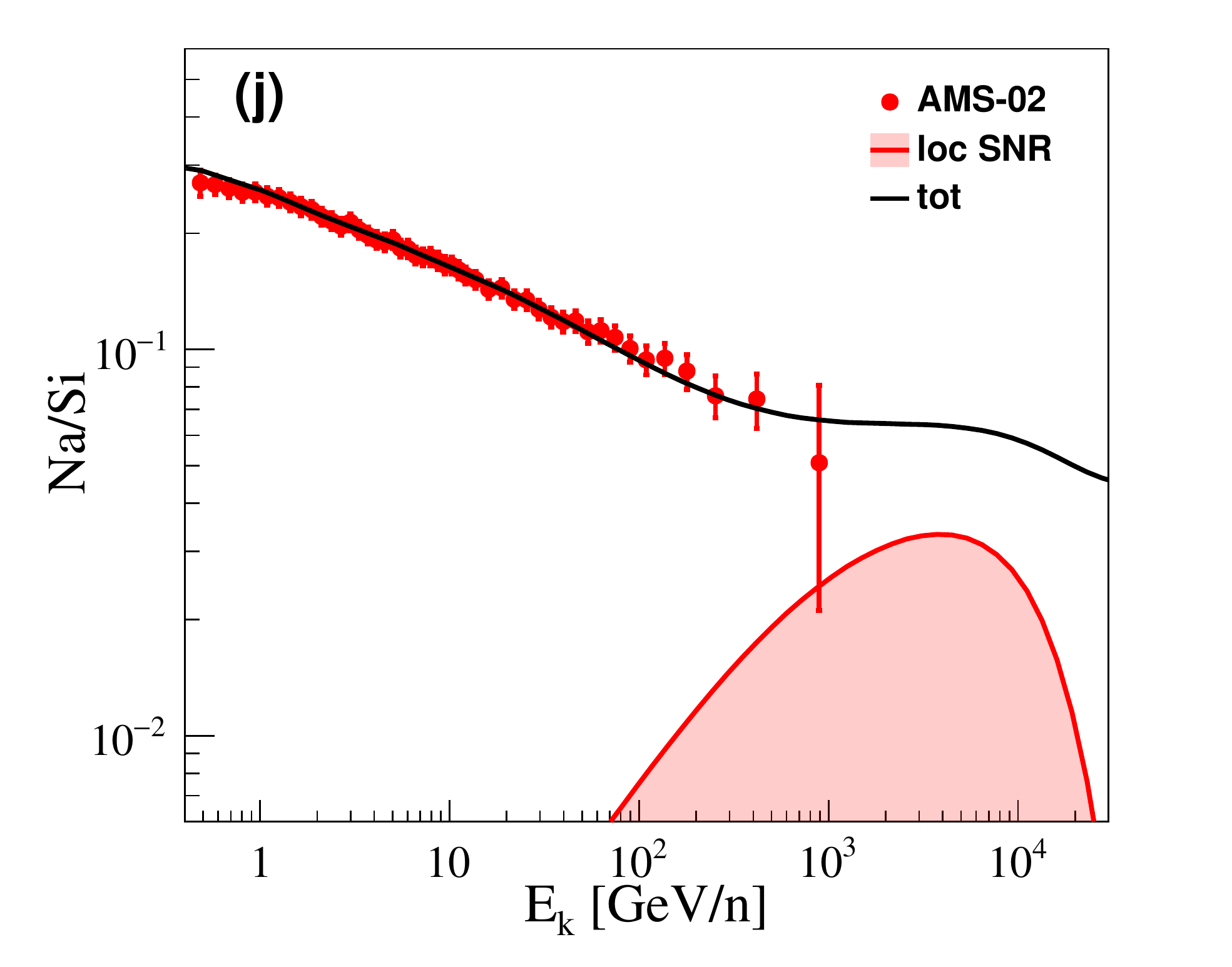}
	\includegraphics[width=5.5cm]{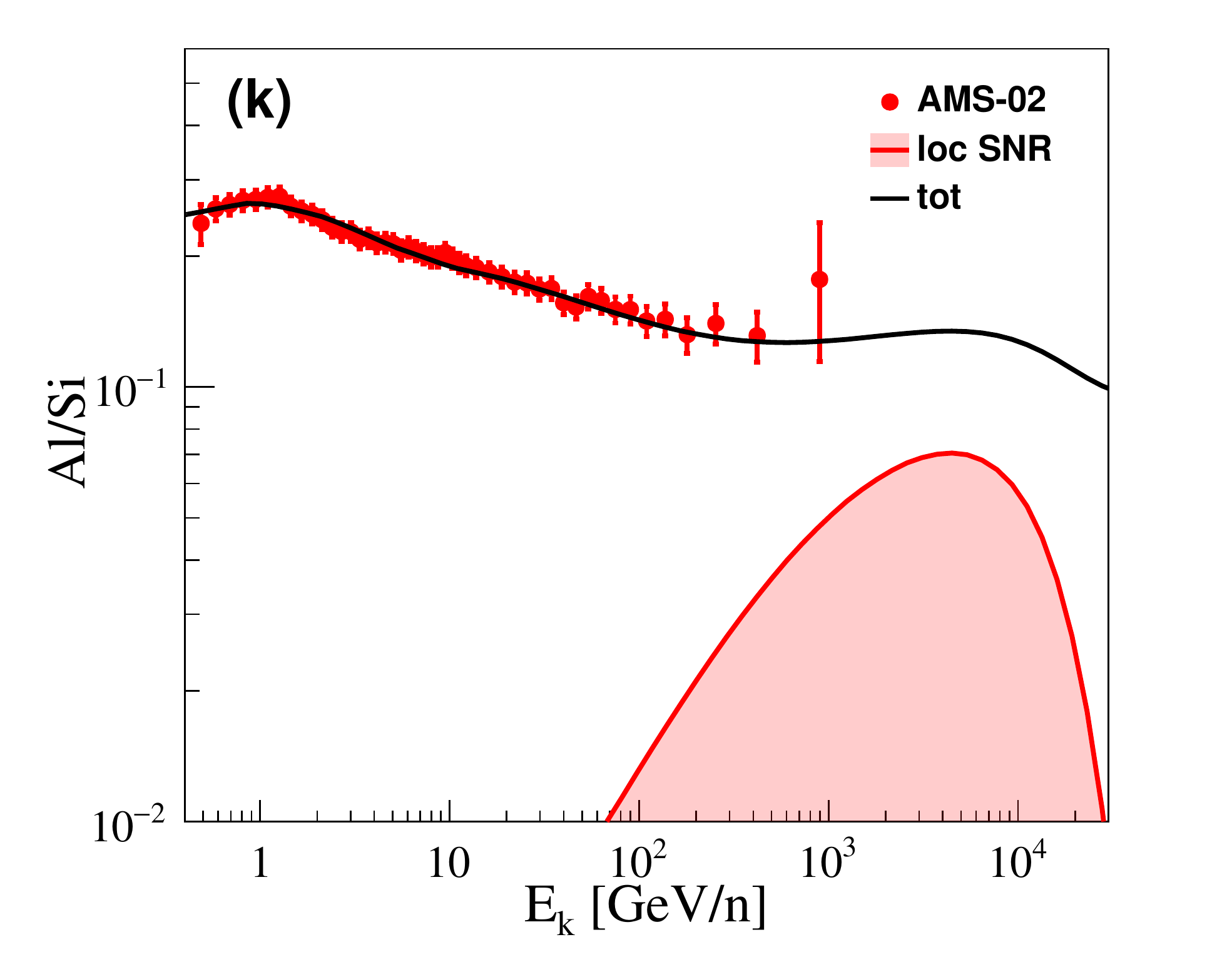}
	\includegraphics[width=5.5cm]{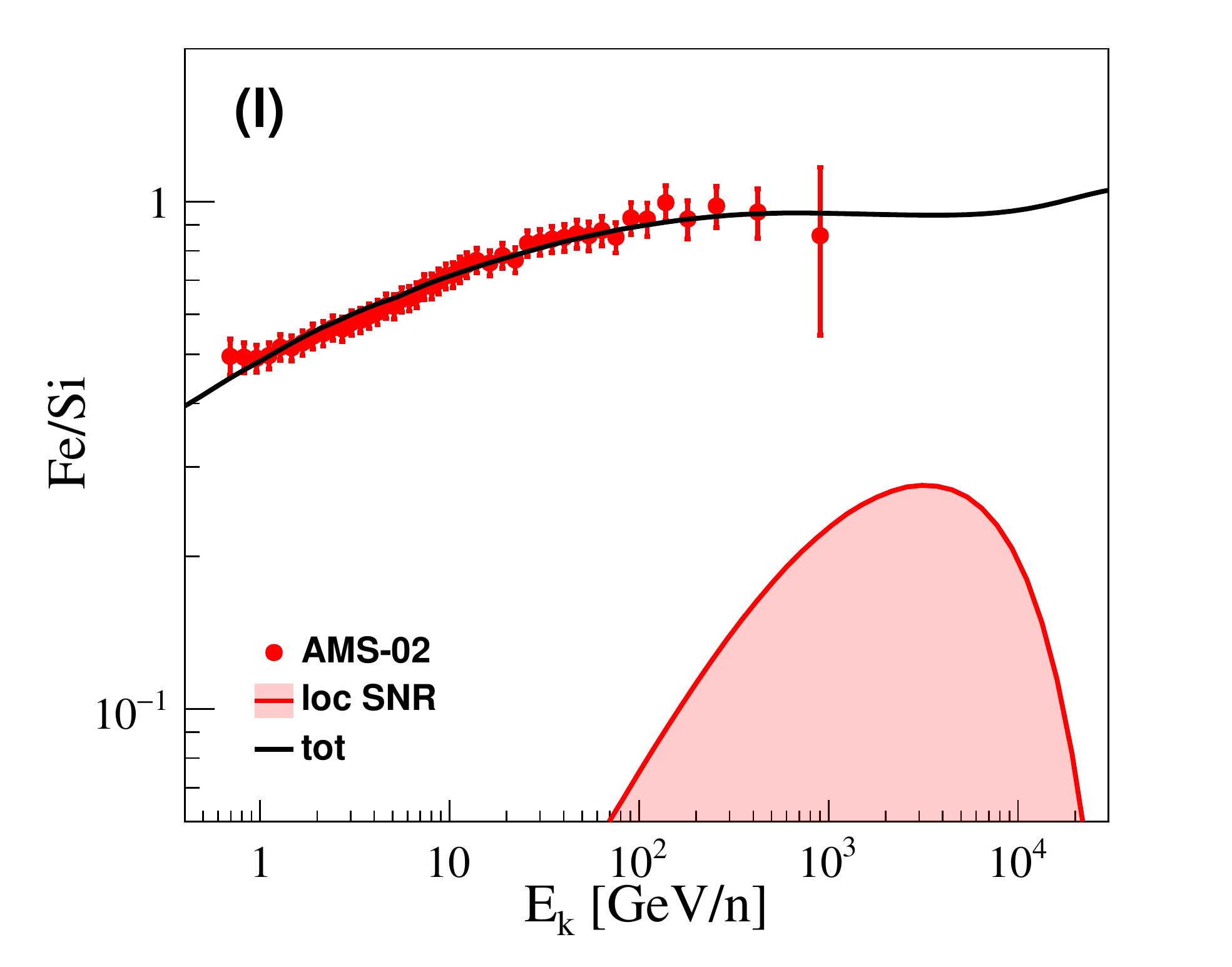}
	\includegraphics[width=5.5cm]{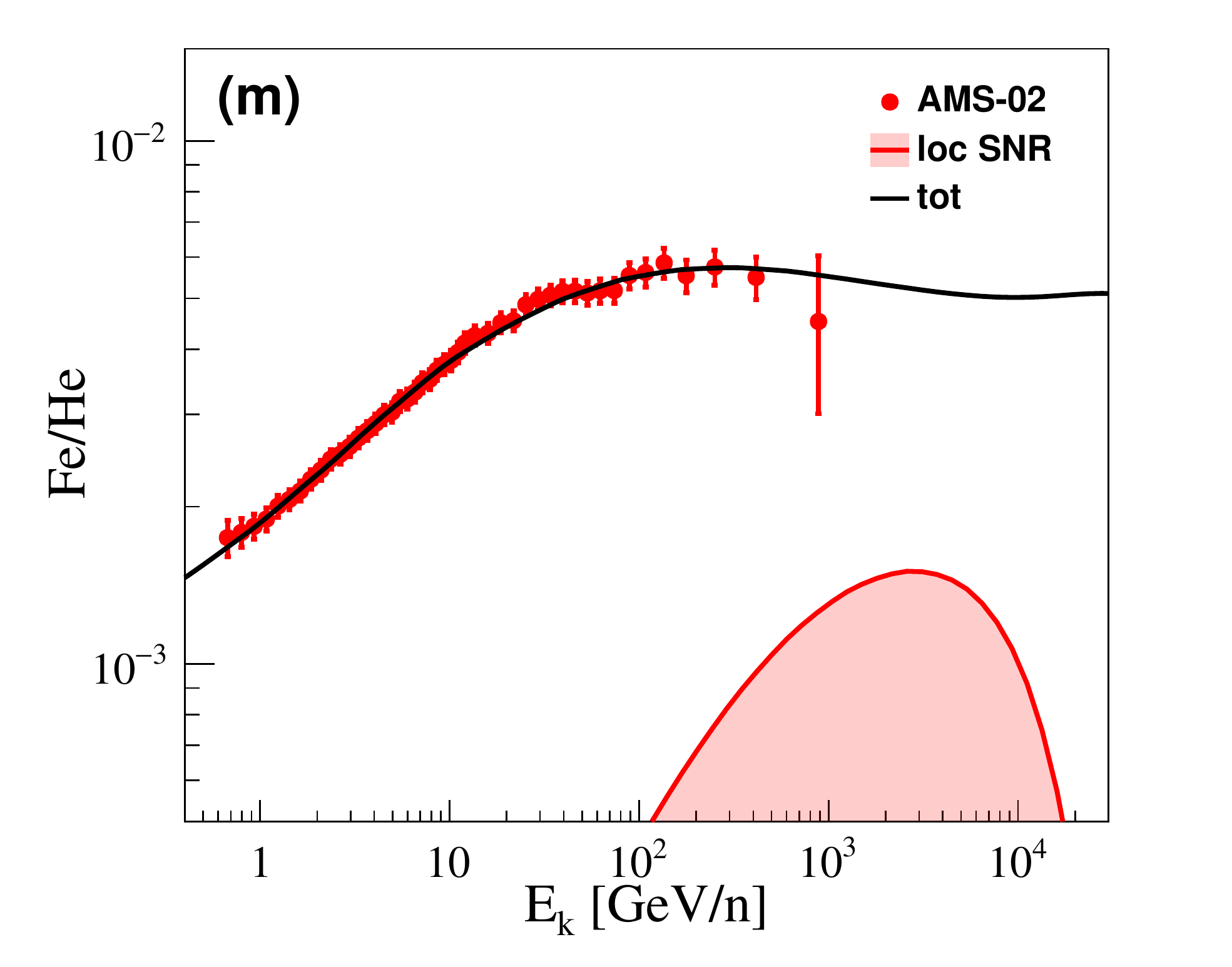}
	\caption{Comparison between model calculations and observations for the ratios of primary to primary species. 
		The red solid lines with shadow indicate the contributions from the nearby SNR. The black solid lines represent the ratio results of model expectation with taking into account the nearby source component.
		The data points are adopted from AMS-02 measurements \citep{2017PhRvL.119y1101A,2018PhRvL.121e1103A,2021PhRvL.127b1101A,2020PhRvL.124u1102A,2021PhRvL.126d1104A}.}
	\label{fig:pprat}
\end{figure*}

The ratios of primary to primary species carry the acceleration information in the source region. The model of diffusive shock acceleration predicts that the individual species should have identical power law spectrum \citep{2015ICRC...34....9S,2021PhR...894....1A}. Figure \ref{fig:pprat}
shows the comparison between model calculations and observations for the ratios of primary to primary species for $He/O$, $C/O$,$N/O$, $Ne/O$, $Mg/O$, $Si/O$, $Ne/Mg$, $Si/Mg$, $Na/Si$, $Al/Si$, $Fe/He$, $Fe/O$ and $Fe/Si$. The red solid lines with shadow indicate the contributions from the nearby SNR. The black solid lines represent the ratio results of model expectation, taking into account the contributions from the nearby source.  
In fact, the individual spectrum has well reproduced the observations as shown in Figure \ref{fig:pHespec}, \ref{fig:prispec} and \ref{fig:prisecspec}, so the ratios should be also consistent between data and model calculations. Figure \ref{fig:pprat} lists the ratios of primary to primary species, which has good consistency with observations. 

\subsubsection{secondary to primary species}
\begin{figure}[htp]
	\centering
	
	\includegraphics[width=4.cm]{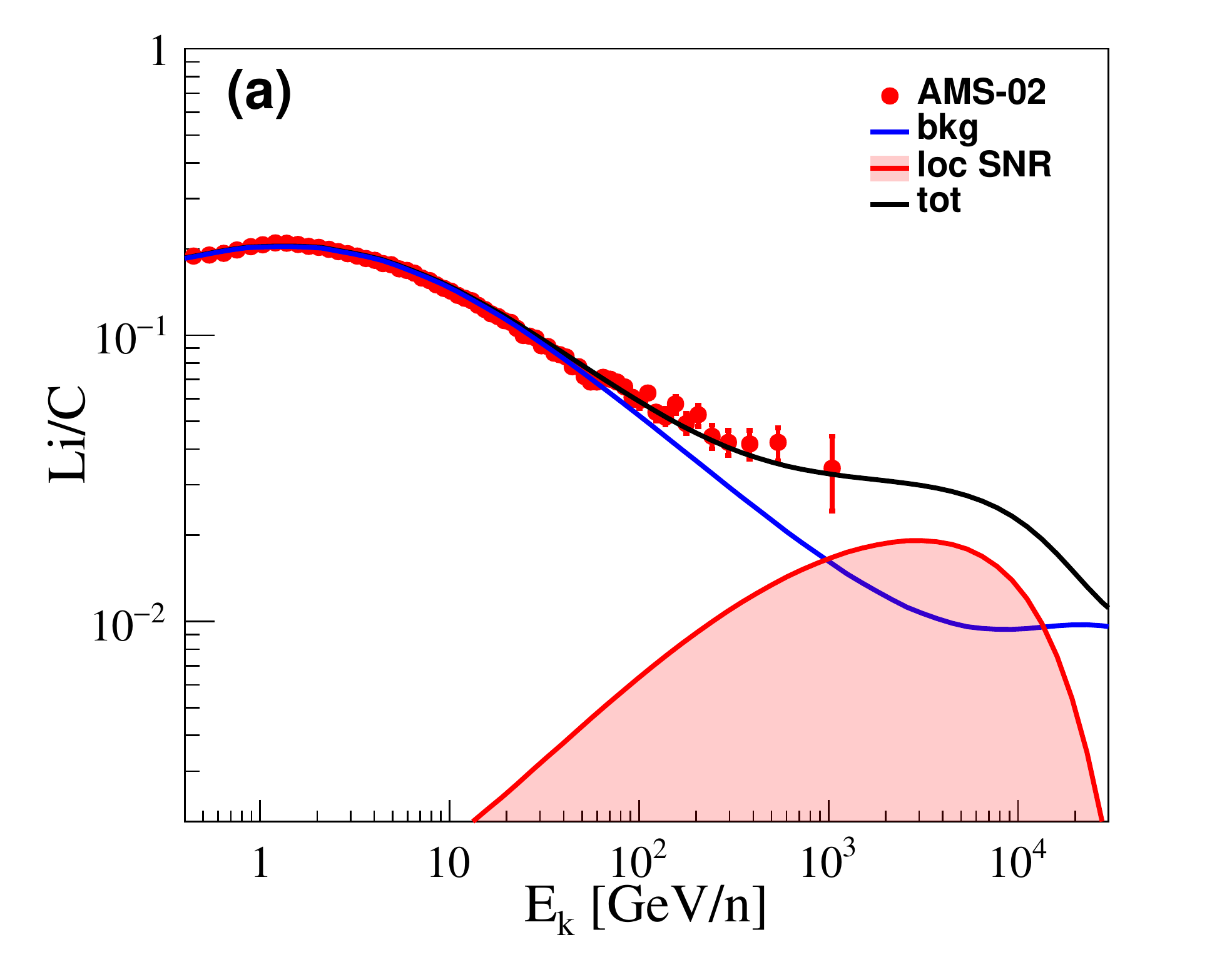}
	\includegraphics[width=4.cm]{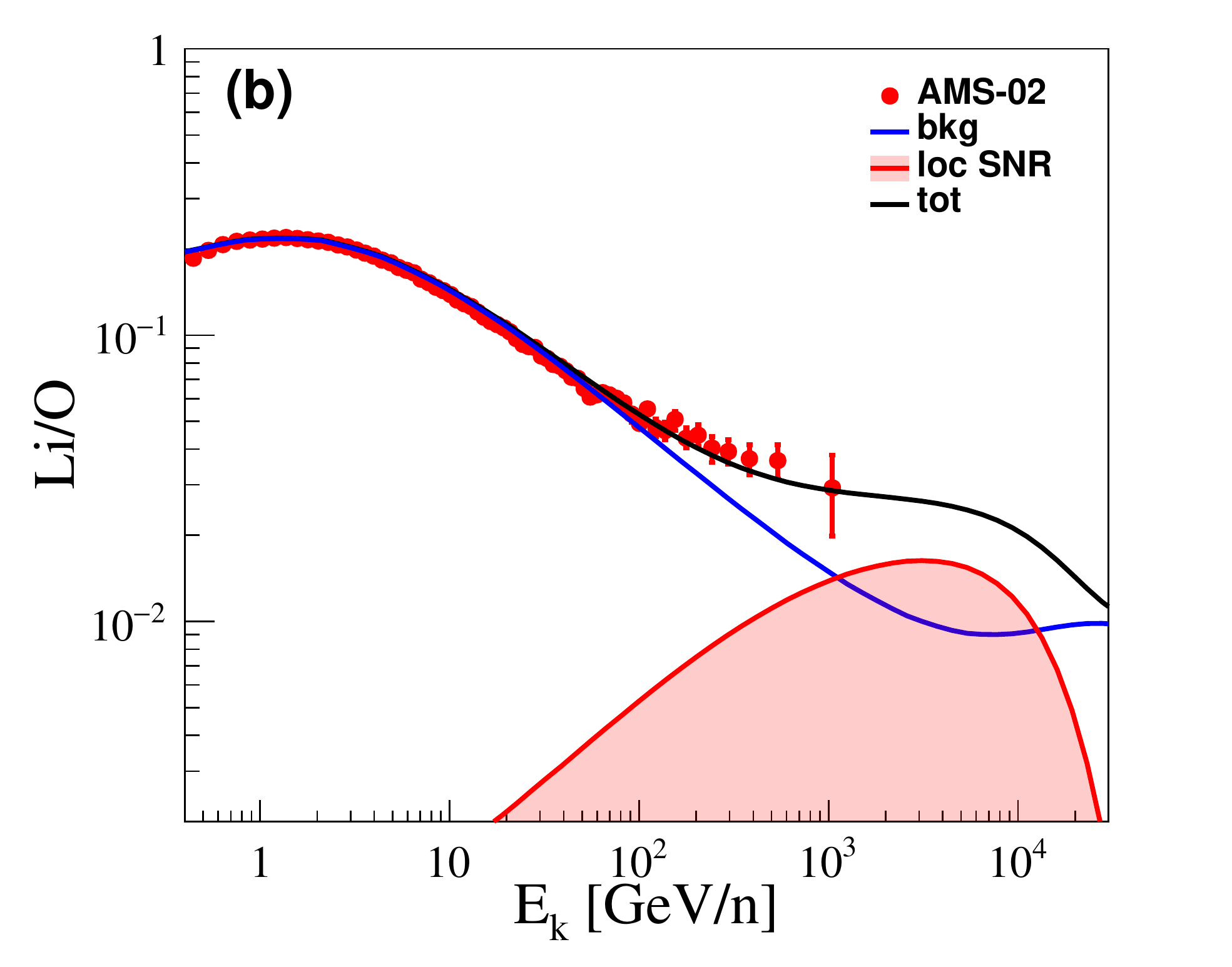}
	\includegraphics[width=4.cm]{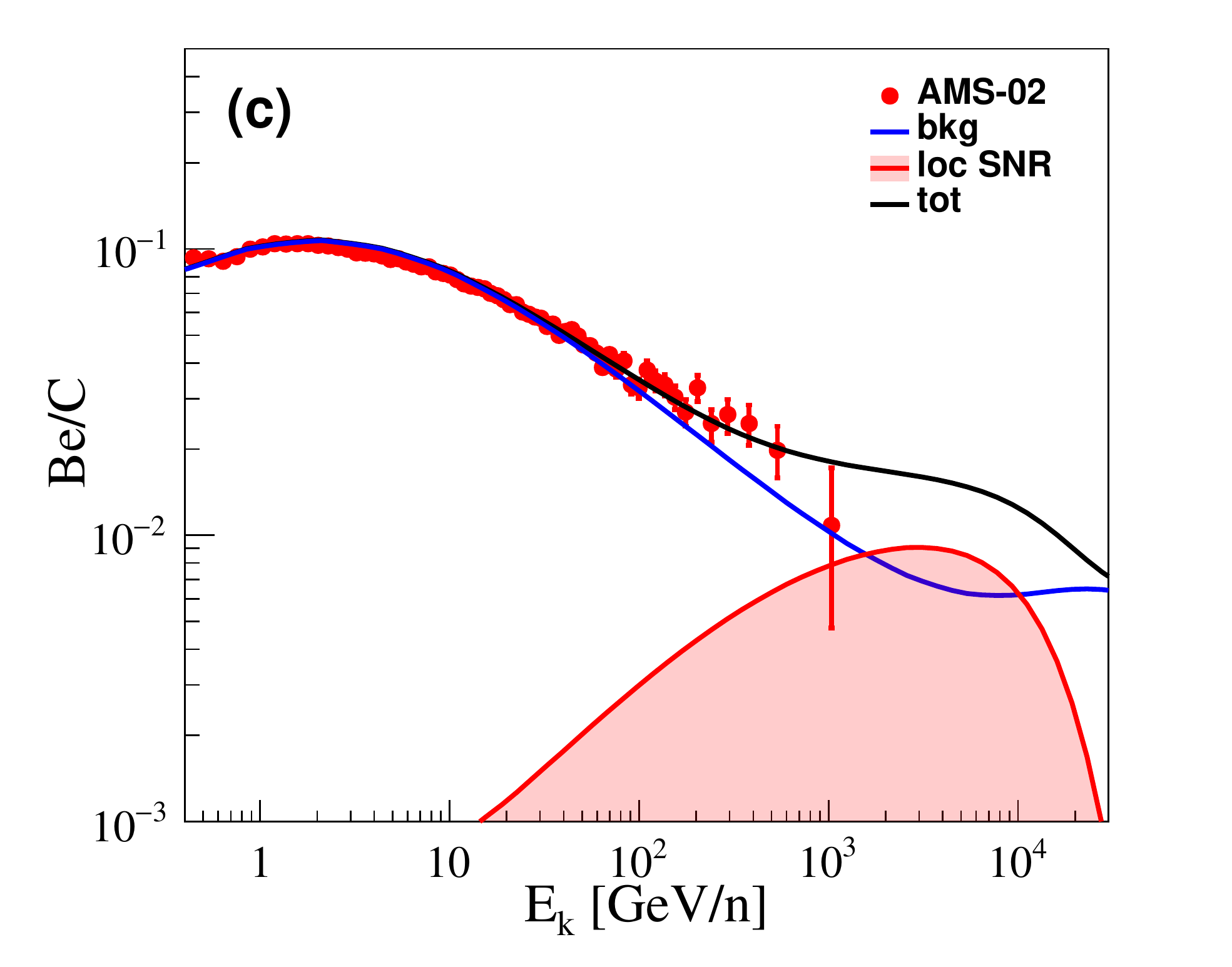}
	\includegraphics[width=4.cm]{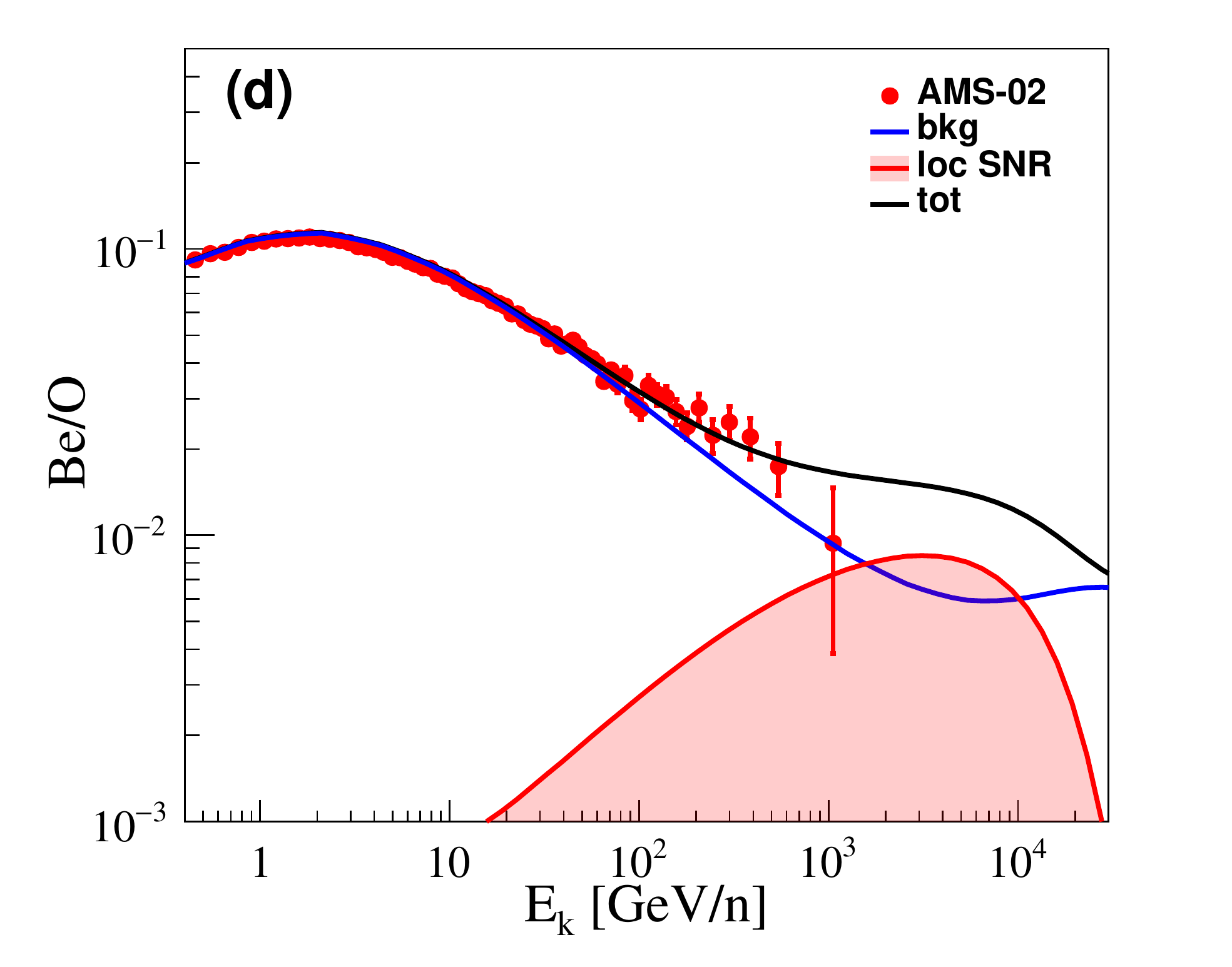}
	\includegraphics[width=4.cm]{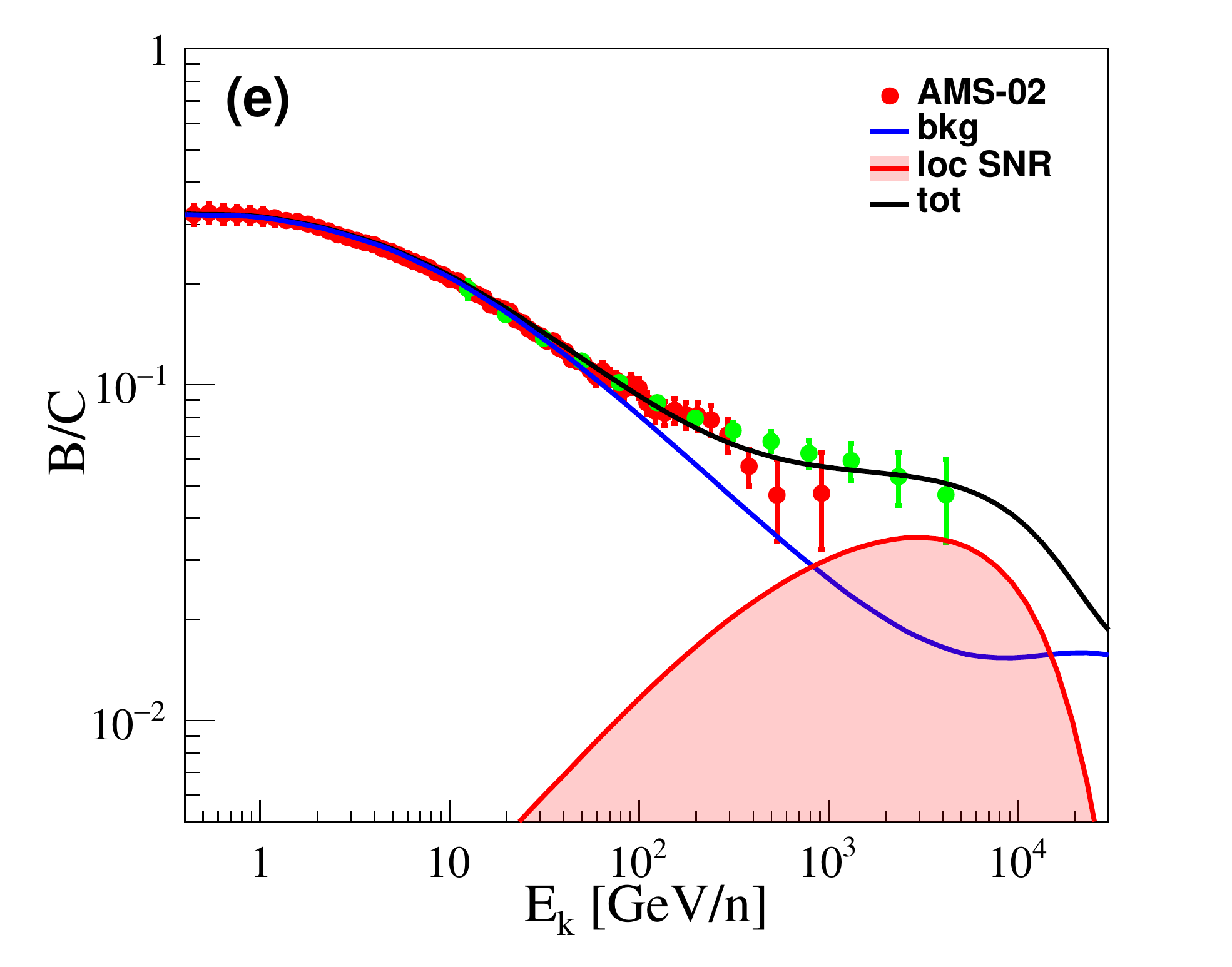}
	\includegraphics[width=4.cm]{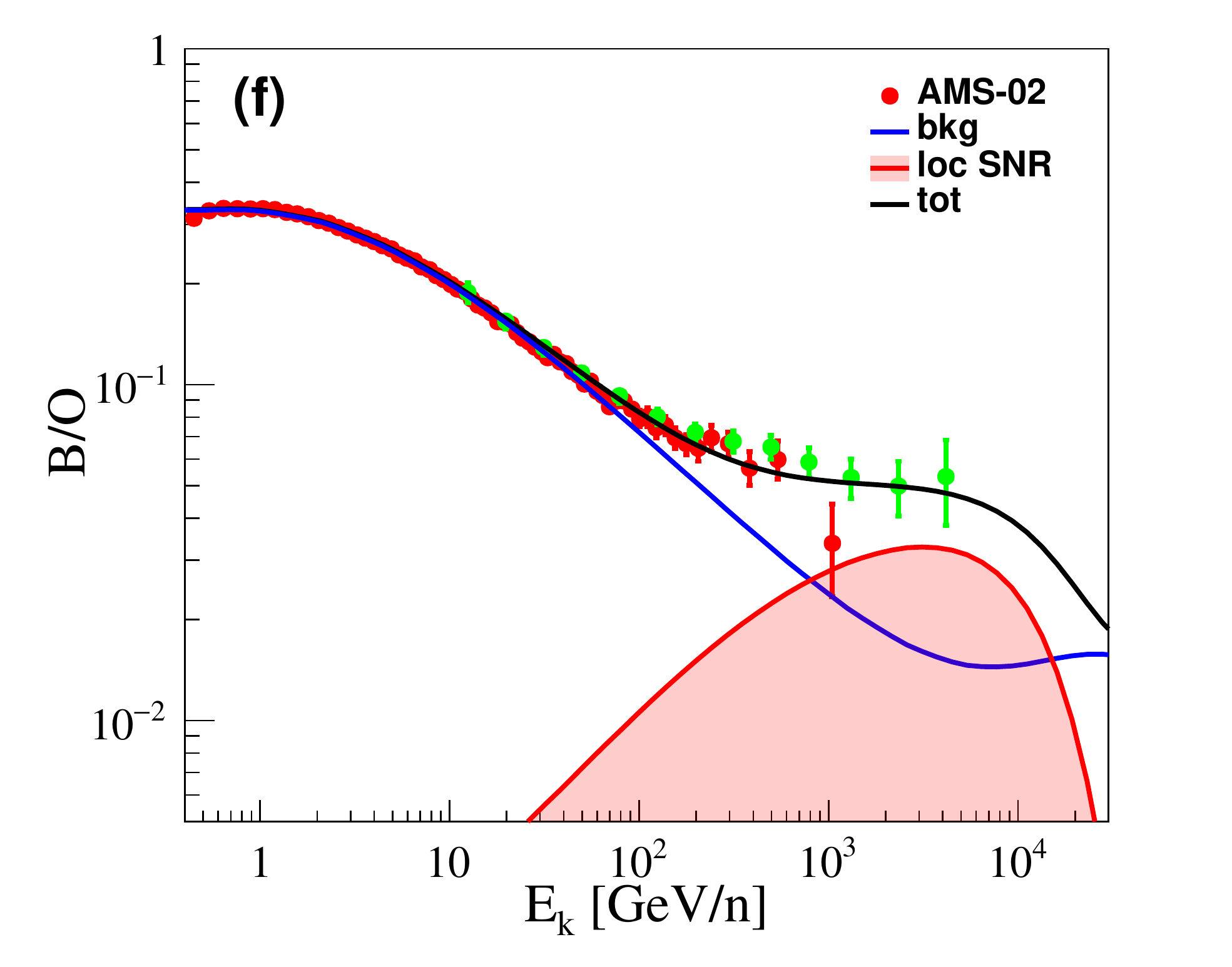}
	\includegraphics[width=4.cm]{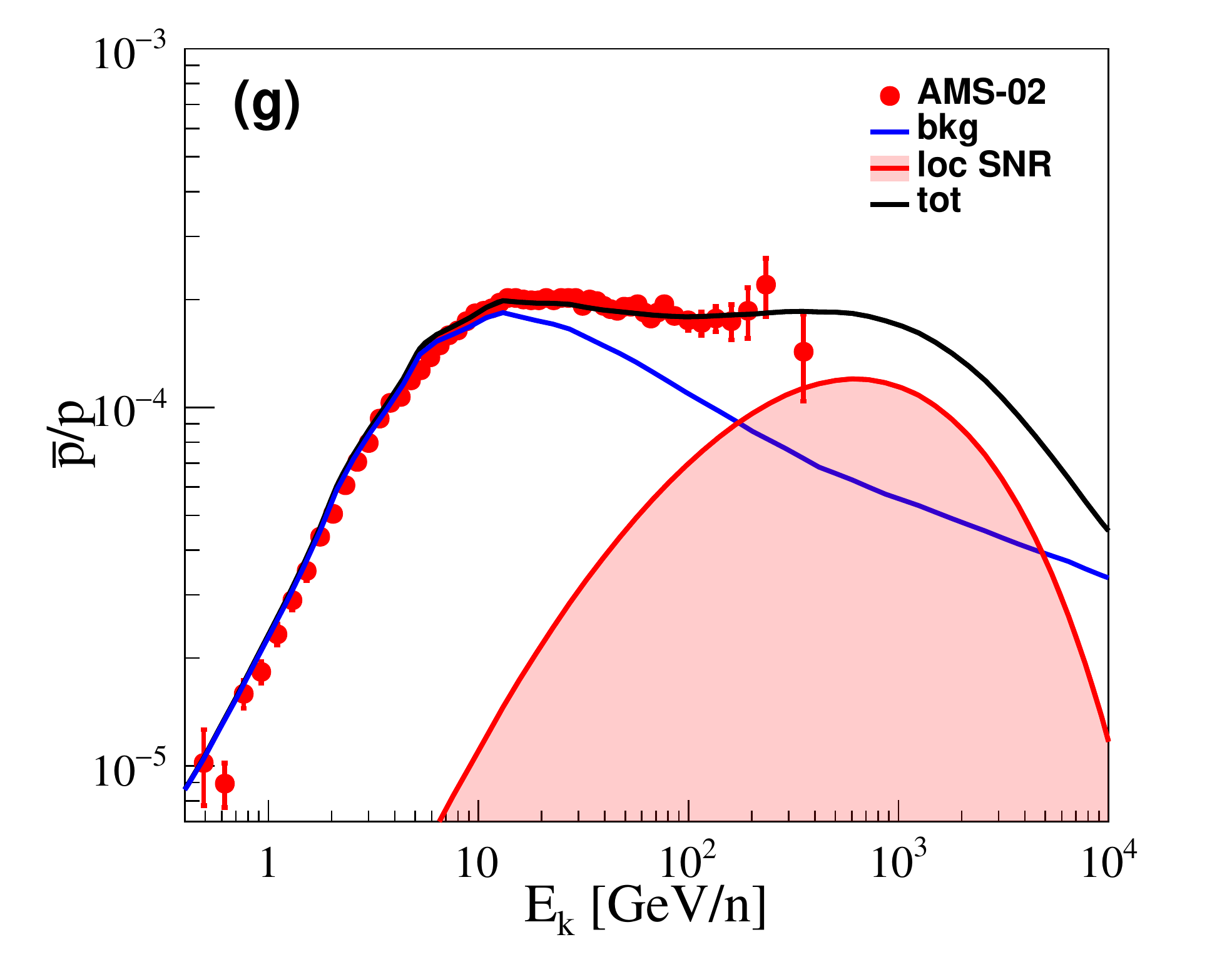}
	\caption{
		Comparison between model calculations and observations for the ratios of secondary to primary species.  The data points are adopted from AMS-02 measurements \citep{2016PhRvL.117i1103A,2018PhRvL.120b1101A,2016PhRvL.117w1102A}.}
	\label{fig:sprat}
\end{figure}

It is believed that most of the secondary nuclei originate from collisions of
CRs with ISM in propagation. Therefore the information about CR propagation can be extracted from comparison between the spectra of secondary particles and those of primary CRs \citep{2017PhRvD..95h3007Y,2019SCPMA..6249511Y,2019FrPhy..1524601Y}.
Figure \ref{fig:sprat} shows the ratios of secondary to primary species for $\bar p/p$, $Li/C$, $Be/C$, $B/C$, $Li/O$, $Be/O$ and $B/O$. The blue solid lines represent the ratios of background component of secondary particles to the total amount of primary particles, the red solid lines show the ratios of nearby SNR component of secondary particles to the total amount of primary particles and the black solid line is the ratio of total secondary to total primary.
The model calculations work well to reproduce the observations.
Here except $\bar p/p$ with energy independent (constant) distribution above 10 GeV, all other heavier nuclei have hardening above 200 GeV for model calculations. Just recently, the similar hardening has also been discovered in $B/C$ and $B/O$ above 100 GeV/n by DAMPE experiment \citep{DAMPECOLLABORATION2022}. It is obvious that our model calculations work well to reproduce this hardening. Similar to the spectra of secondary species, the ratios have a break-off around TeV for $\bar p/p$ and 
5 TeV for heavier nuclei, which can offer a crucial and definitive identification with other kinds of models. 

\subsubsection{Secondary to secondary species}
\begin{figure*}[htp]
	\centering
	\includegraphics[width=5.5cm]{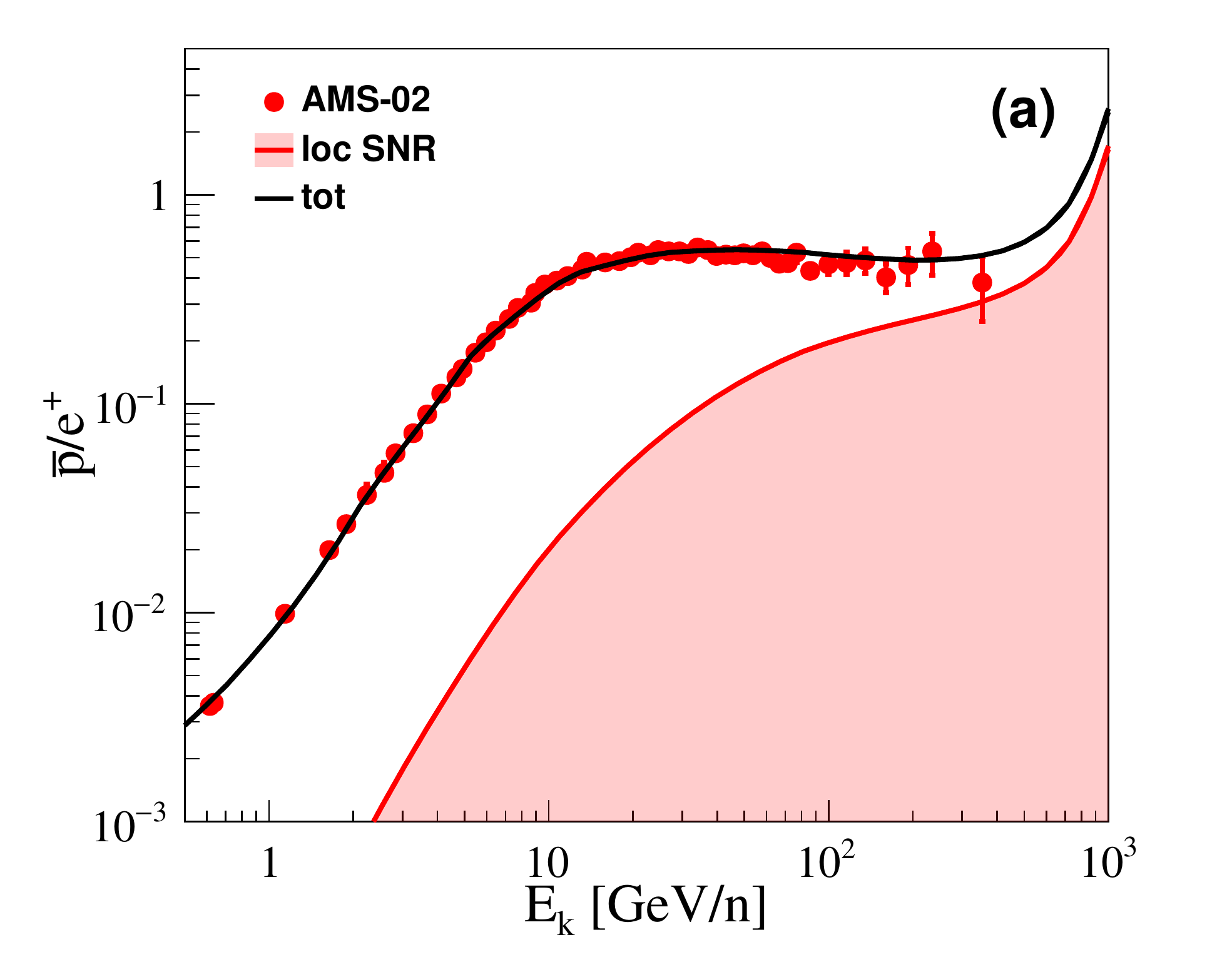}
	\includegraphics[width=5.5cm]{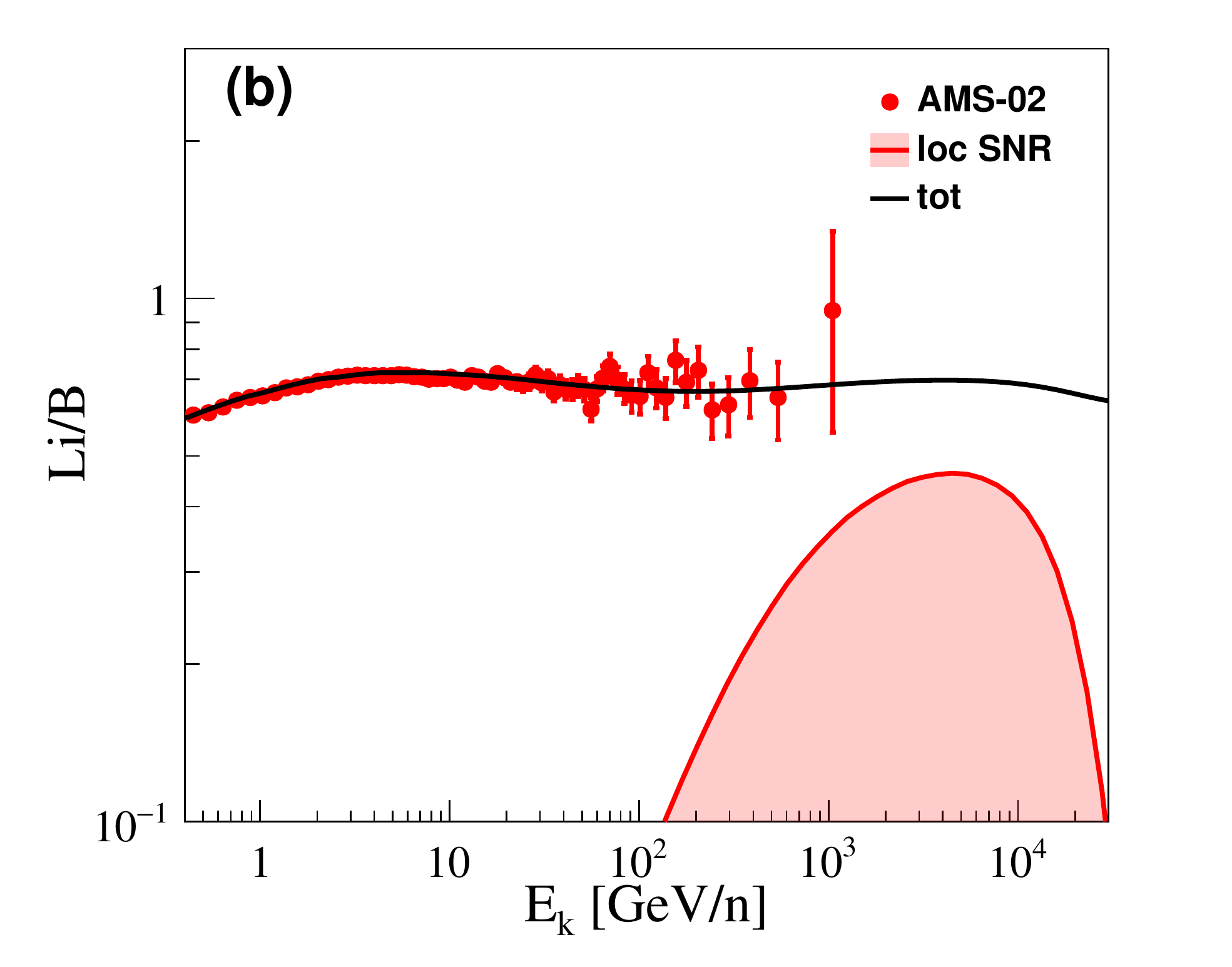}
	\includegraphics[width=5.5cm]{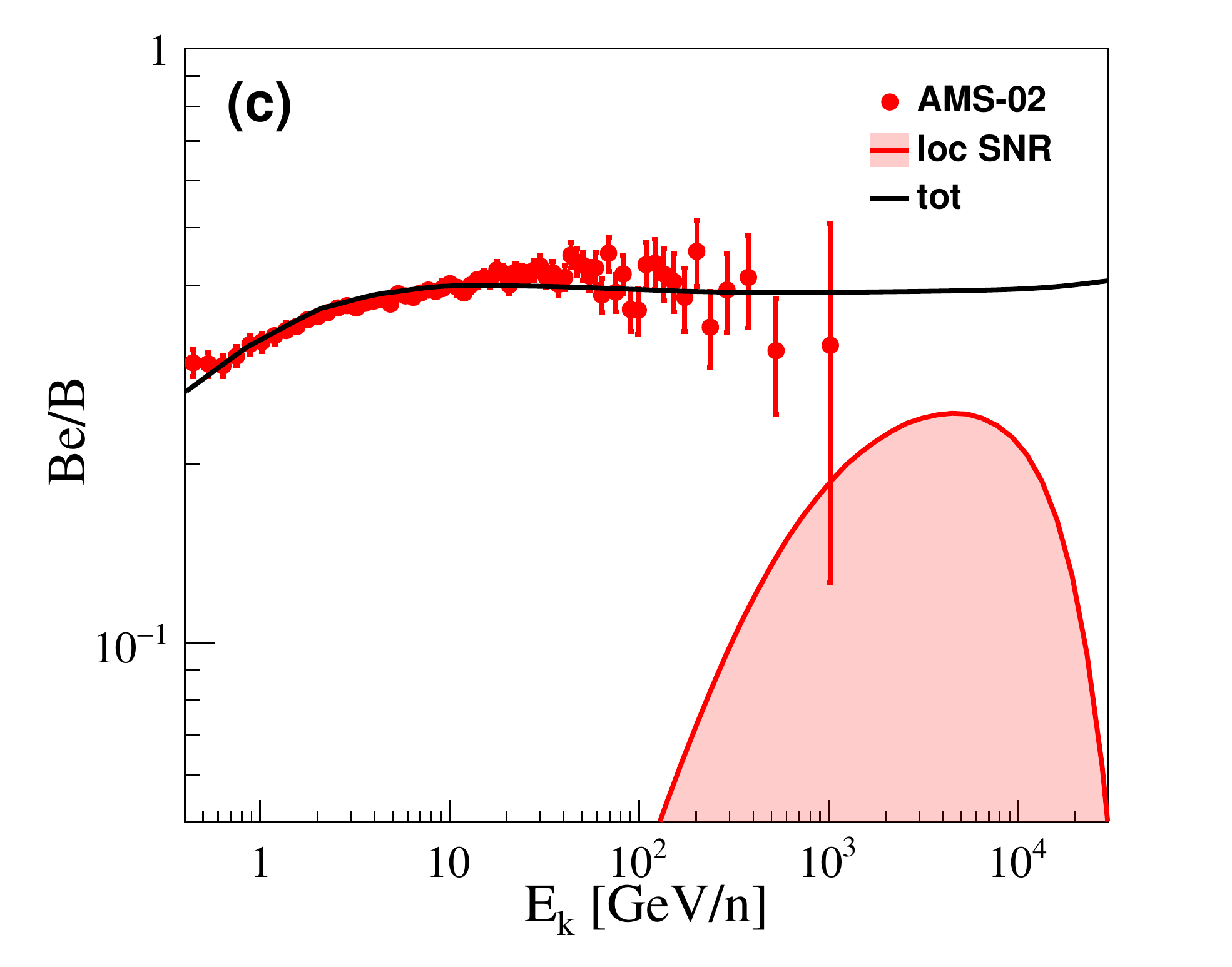}
	\caption{Similar to Figure \ref{fig:pprat}. Comparison between model calculations and observations for the ratios of secondary to secondary species. The data points are adopted from AMS-02 measurements \citep{2016PhRvL.117i1103A,2018PhRvL.120b1101A}.}
	\label{fig:ssrat}
\end{figure*}

The ratios of secondary to secondary species take important information of interaction in CR propagation. If they originate from the same mother particle, the spectral behavior reflects the common pecularity considering the known interaction cross-section, the same ISM and the same interaction time. This property can be served to understand the CR origin puzzle, such as positron \citep{2009Natur.458..607A}. Figure \ref{fig:ssrat} shows the ratio comparison between model calculations and measurements for $\bar p/e^+$, $Li/B$ and $Be/B$. Our model calculations are consistent with the measurements. More interesting is that the energy independent distribution is clear shown from them above 10 GeV. The model calculation of $\bar p/e^+$ rises up sharply above 300 GeV. This is because the energy cut-off of positron around $300$ GeV. 

\subsection{Anisotropy}
\begin{figure*}[htp]
	\centering
	\includegraphics[width=18.cm]{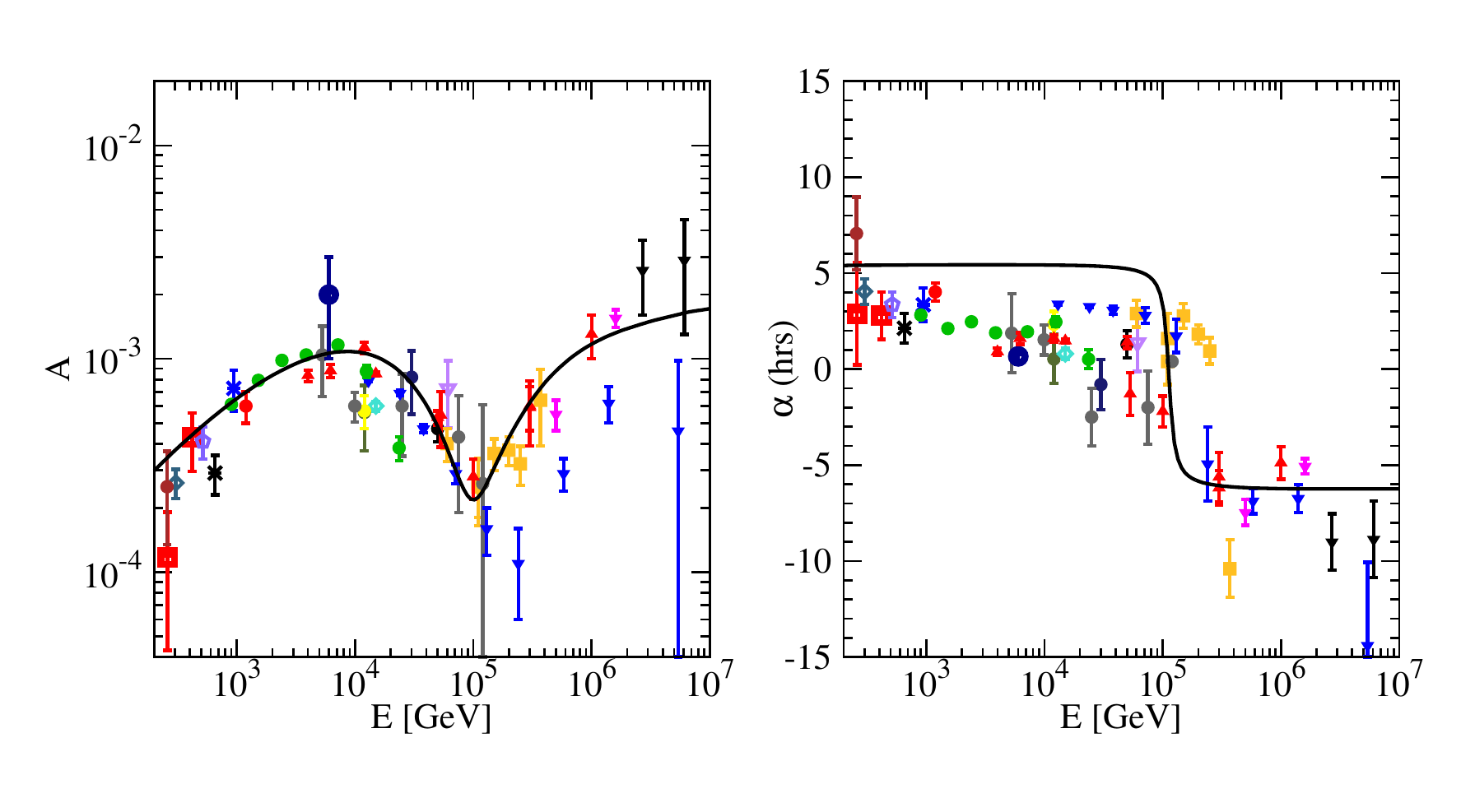}
	\caption{Comparison between model calculations and observations of the amplitude and phase of anisotropy for CRs. The data points are taken from underground muon detectors:
		Norikura (\cite{1973ICRC....2.1058S}),
		Ottawa (\cite{1981ICRC...10..246B}),
		London (\cite{1983ICRC....3..383T}),
		Bolivia (\cite{1985P&SS...33.1069S}),
		Budapest (\cite{1985P&SS...33.1069S}),
		Hobart (\cite{1985P&SS...33.1069S}),
		London (\cite{1985P&SS...33.1069S}),
		Misato (\cite{1985P&SS...33.1069S}),
		Socorro (\cite{1985P&SS...33.1069S}),
		Yakutsk (\cite{1985P&SS...33.1069S}),
		Banksan (\cite{1987ICRC....2...22A}),
		Hong Kong (\cite{1987ICRC....2...18L}),
		Sakashita (\cite{1990ICRC....6..361U}),
		Utah (\cite{1991ApJ...376..322C}),
		Liapootah (\cite{1995ICRC....4..639M}),
		Matsushiro (\cite{1995ICRC....4..648M}),
		Poatina (\cite{1995ICRC....4..635F}),
		Kamiokande (\cite{1997PhRvD..56...23M}),
		Marco (\cite{2003PhRvD..67d2002A}),
		SuperKamiokande (\cite{2007PhRvD..75f2003G});
		and air shower array experiments:
		PeakMusala (\cite{1975ICRC....2..586G}),
		Baksan (\cite{1981ICRC....2..146A}),
		Norikura (\cite{1989NCimC..12..695N}),
		EAS-TOP (\cite{1995ICRC....2..800A,
			1996ApJ...470..501A, 2009ApJ...692L.130A}),
		Baksan (\cite{2009NuPhS.196..179A}),
		Milagro (\cite{2009ApJ...698.2121A}),
		IceCube (\cite{2010ApJ...718L.194A, 2012ApJ...746...33A}),
		Ice-Top (\cite{2013ApJ...765...55A}),
		ARGO-YBJ (\cite{2015ApJ...809...90B}),
		Tibet (\cite{2005ApJ...626L..29A,
			2015ICRC...34..355A, 2017ApJ...836..153A})
		K-Grande(\citep{2015ICRC...34..281C}).}
	\label{fig:aniso}
\end{figure*}

\begin{figure*}[htp]
	\centering
	\includegraphics[width=8.5cm]{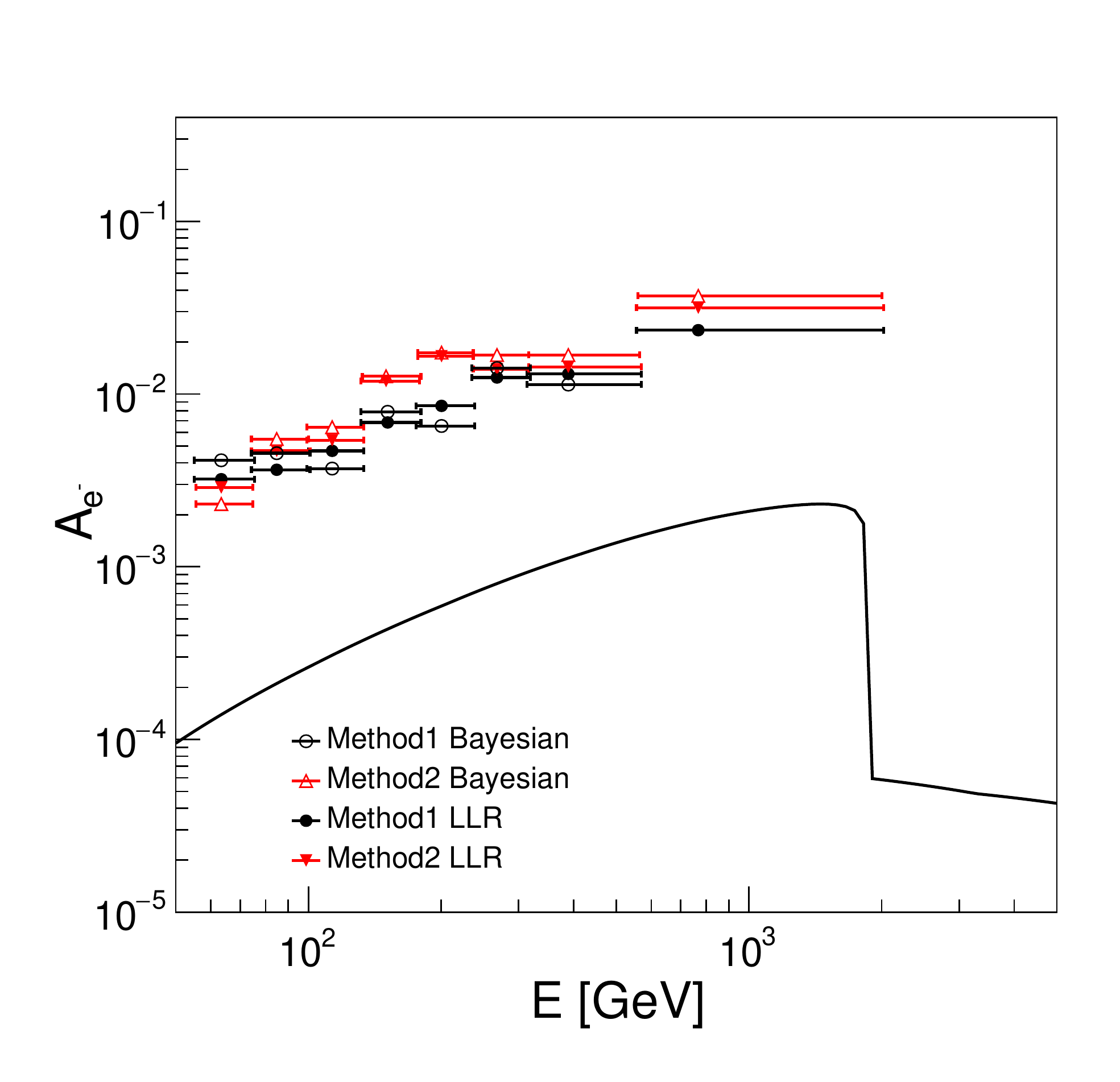}
	\includegraphics[width=8.5cm]{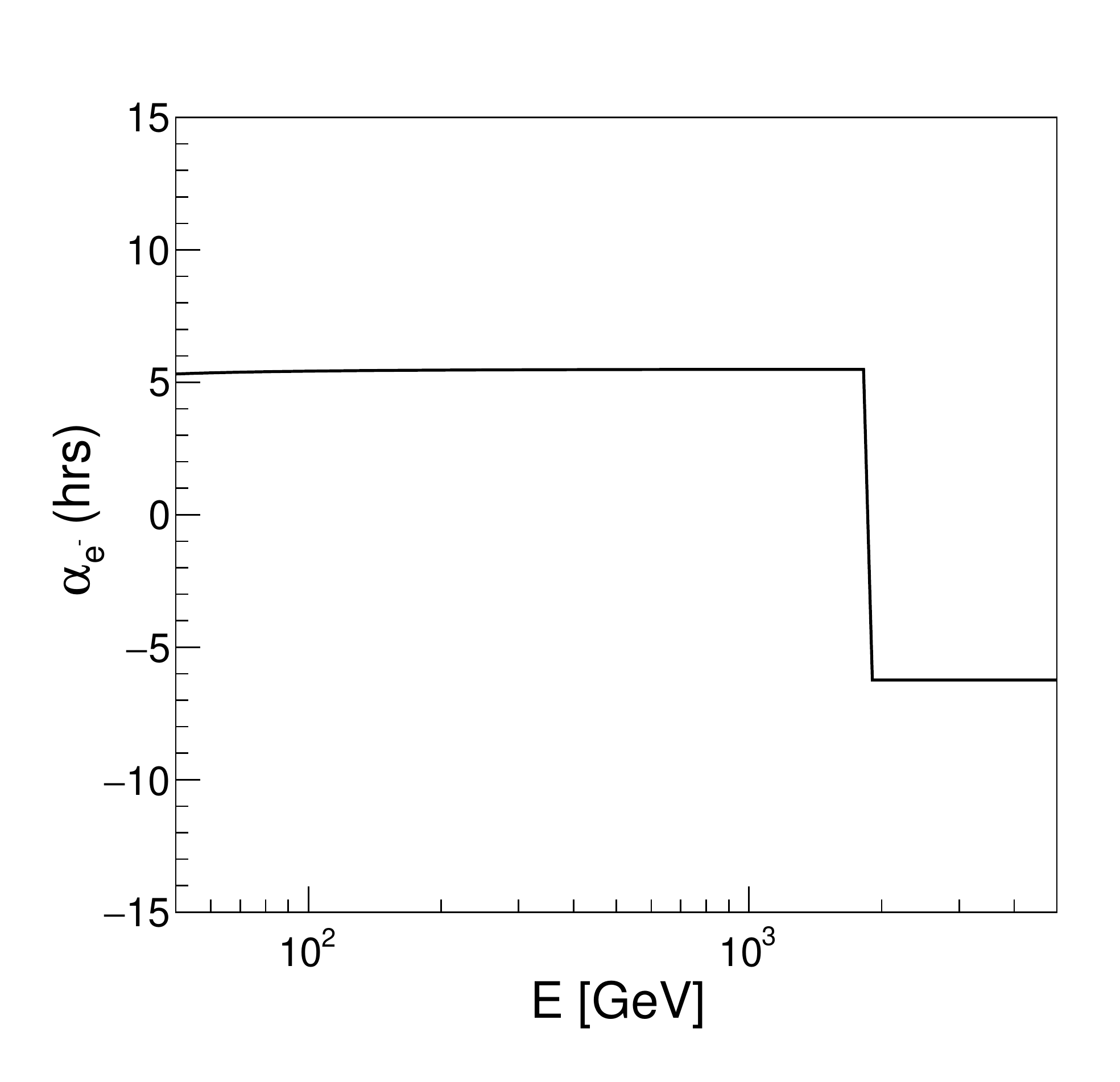}
	\caption{Comparison between model calculations and up-limit of the amplitude and phase of anisotropy for CREs. The data points are adopted from \citep{2017PhRvL.118i1103A}}
	\label{fig:anisoE}
\end{figure*}

Similar to the results of spectra and ratios, the anisotropy is also the joint contributions from the background and local sources of CRs. Due to more abundant sources in the inner disk, the phase of anisotropy directs to the galactic center. However, the observations of phase roughly point to the direction of anti-galactic center from 100 GeV to 100 TeV. The local source, located at the outer of galactic disk, plays the dominant roles in this energy region. So the anisotropy demonstrates mutually repressive competition between local source and background. The dip structure at 100 TeV is the transition energy point for the two kinds of sources.  
Figure \ref{fig:aniso} shows the amplitude and phase of anisotropy for CRs. The CR anistropy can be fitted well under this physical scenario. Figure \ref{fig:anisoE} presents amplitude and phase for CREs. The expection of amplitude for CREs is lower than the up-limit of Fermi-LAT experiment \citep{2017PhRvL.118i1103A}. 

\section{Summary} \label{sec:summary}
The new generation of space-borne and ground-based experiments took unprecedented precise measurements for CR spectra and anisotropy and revealed multi-messenger anomalies for them. 
In this work, we propose that the local source, Geminga SNR, is the common origin for all those anomalies. 
The Geminga SNR has three critical advantages: perfect age of 330 kyrs, suitable position with a distance of 330 pc and assumed DMC around it.

The physical figure can be summarized as following. Firstly, the diffusive shock around Geminga SNR can accelerate the CR nuclei and electrons to very high energy together. Here we adopt the break-off rigidity to be $\sim$5 TV for the sake of 14 TV energy cut-off observation by DAMPE experiment \citep{2019SciA....5.3793A}. The electron will suffer the energy loss for 330 kyrs, which lead to the spectral break around TeV observed by DAMPE, HESS \citep{2008PhRvL.101z1104A,2009A&A...508..561A,2017Natur.552...63D}. Due to the contributions of the local source, the spectral hardening and drop-off for all nuclei and electron respective can be well reproduced.  
Secondarily, the interaction of pp-collision and fragmentation with DMC happen to produce secondary particles, such as $\bar p, e^+, \gamma, Li,Be,B$ and so on. Inheriting the properties from their mother particles,
the $\bar p, e^+, \gamma$ as products of pp-collision will take $10\%, 5\%, 10\%$ of proton energy, which result in the corresponding energy break-off at 250, 500, 500 GeV respectively for them. So the hardening and drop-off for secondary particles can also be well consistent with the measurements. 
Simultaneously, the heavier secondary nuclei from fragmentation process, such as $Li,Be,B$, will have the same energy break-off with their mothers around $\sim$5 TeV. This property is typical for this kind model, which can be served to test the model in future. Lastly, thanks to the appropriate location of Geminga SNR, the special evolution with energy of anisotropy in amplitude and phase can roughly reproduce the observations. 

Summarily, the multi-messenger anomaly of spectra, ratios and anisotropy can be well reproduced in one unified physical mechanism. Particularly, the difference of spectral break-off between positron and electron can also be explained in the same scenario.
More interesting is that we expect the typical energy break-off feature for secondary nuclei $Li,Be,B$ and their ratios with primary nuclei. This feature will also play important role to resolve the origin puzzle of positron and $\bar p$ excesses. We hope that this feature can be observed by HERD experiment in future.

\begin{acknowledgments}
	This work is supported by the National Key Research and Development Program 
	(No. 2018YFA0404202), the National Natural Science Foundation of China 
	(Nos. 12275279, U2031110, 12175248).
\end{acknowledgments}

%






\appendix
\section{The propagation of CRs}
It has been recognized in recent years that the propagation of CRs in the Milky Way should depend on the spatial locations, as inferred by the HAWC and LHAASO observations of extended $\gamma$-ray halos around pulsars \citep{2017Sci...358..911A,2021PhRvL.126x1103A} and the spatial variations of the CR intensities and spectral indices from Fermi-LAT observations \citep{2016PhRvD..93l3007Y,2016ApJS..223...26A}. The spatially-dependent propagation (SDP) model was also proposed to explain the observed hardening of CRs \citep{2012ApJ...752L..13T,2015PhRvD..92h1301T,2016PhRvD..94l3007F,2016ApJ...819...54G,2018ApJ...869..176L,2018PhRvD..97f3008G}, and also the large-scale anisotropies by means of
a nearby source\citep{2019JCAP...10..010L,2019JCAP...12..007Q}.

In the SDP model, the diffusive halo is divided into two parts, the inner halo (disk) and the outer halo. In the inner halo, 
the diffusion coefficient is much smaller than that in the outer halo, as indicated by the HAWC observations. The propagation equation of CRs in the magnetic halo and the descriptions of specific iterms in it can be refered to \citep{2016ApJ...819...54G}. The spatial diffusion coefficient $D_{xx}$ can be parameterized as

\begin{equation}
	D_{xx}(r,z, {\cal R} )= D_{0}F(r,z)\beta^{\eta} \left(\frac{\cal R}
	{{\cal R}_{0}} \right)^{\delta_{0}F(r,z)},
	\label{eq:diffusion}
\end{equation}
where $r$ and $z$ are cylindrical coordinate, ${\cal R}$ is the particle's rigidity, $\beta$ is the particle's velocity in unit of light speed, $D_0$ and $\delta_0$ are constants representing the diffusion coefficient and its high-energy rigidity dependence in the outer halo, $\eta$ is a phenomenological constant in order to fit the low-energy data. The spatial dependent function $F(r, z)$ is given as 
\begin{equation}
	F(r,z) = \left\{
	\begin{array}{ll}
		{\frac{N_m}{1+f(r,z)}+\left[1-\frac{N_{m}}{1+f(r,z)}\right]}\left(\frac{{z}}{\xi{z}_{\rm h}} \right)^{n},  &  {{|z|} \leq \xi{z}_{\rm h}} \\
		\\
		1,  &  { {|z|} > \xi{z}_{\rm h}} \\
	\end{array},
	\right.
\end{equation}
where the total half-thickness of the propagation halo is $z_h$, and the half-thickness of the inner halo is $\xi z_{h}$. The constant $n$ describes the smoothness of the parameters at the transition between the two halos. The expression $f(r, z)$ is the source density distribution. In this work, we adopt the diffusion re-acceleration model, with the diffusive re-acceleration coefficient $D_{pp}$, which correlated with $D_{xx}$ via $D_{pp}D_{xx} = \frac{4p^{2}v_{A}^{2}}{3\delta(4-\delta^{2})(4-\delta)}$, where $v_A$ is the Alfv\'en velocity, $p$ is the momentum, and $\delta$ is the rigidity dependence slope of the diffusion coefficient \citep{1994ApJ...431..705S}. The parameters of SDP model used here are listed in Table \ref{table_1}.

\section{Background sources}
All of the SNRs other than the nearby one are labeled as
background sources. The source
density distribution is approximated as an
axisymmetric form parametrized as
\begin{equation}
	f(r, z) = \left(\frac{r}{r_\odot} \right)^\alpha \exp \left[-\frac{\beta(r-r_\odot)}{r_\odot} \right] \exp \left(-\frac{|z|}{z_s} \right) ~,
	\label{eq:radial_dis}
\end{equation}
where $r_\odot \equiv 8.5$ kpc represents the distance from the Galactic center to the solar system. Parameters $\alpha$ and $\beta$ are taken to be 1.69 and 3.33 \citep{1996A&AS..120C.437C}. The density of the source distribution decreases exponentially along the vertical height from the Galactic plane, with $z_s$ = 200 pc. The parameters of SDP model used here are listed in Table \ref{table_1}.

The injection spectrum of nuclei and primary electrons are assumed to be an exponentially cutoff broken power-law function of particle rigidity ${\cal R}$, i.e.
\begin{equation}
	\label{eq:spectrum_CRE}
	q({\cal R}) = Q_{0} \left\{
	\begin{array}{lll}
		\left(\frac{{\cal R}}{{\cal R}_{\rm br}} \right)^{\nu_{1}},  &  {{\cal R} \leq {\cal R}_{\rm br}} \\
		\\
		\left(\frac{{\cal R}}{{\cal R}_{\rm br}} \right)^{\nu_{2}} \exp\left[-\frac{\cal R}{{\cal R}_{\rm c}} \right],  &  { {\cal R} > {\cal R}_{\rm br}} \\
	\end{array}
	\right.,
\end{equation}
where $Q_{0}$ is the normalization factor, ${\cal R}_{\rm br}$ is break rigidity, $\nu_{1,2}$ are the spectral incides before and after the break rigidity, ${\cal R}_{\rm c}$ is the cutoff rigidity. Table \ref{para_inj} shows the injection spectrum parameters of different CR nuclei species. The numerical package DRAGON is used to solve the propagation equation of CRs \citep{2017JCAP...02..015E}.
For energies smaller than tens of GeV, the fluxes of CRs are suppressed
by the solar modulation effect. We use the force-field approximation
\citep{1968ApJ...154.1011G} to account for the solar modulation.

The secondary CR nuclei, such as Li, Be, B, can be brought forth from the fragmentation of heavier parent nuclei throughout the transport. The production rate is expressed as follows
\begin{equation}
	Q_j = \sum_{i = \rm C, N, O} (n_{\rm H} \sigma_{i+{\rm H}\rightarrow j} +n_{\rm He} \sigma_{i+{\rm He} \rightarrow j} ) v \psi_i ~,
\end{equation}
where $n_{{\rm H}/{\rm He}}$ is the number density of hydrogen/helium in the ISM and $\sigma_{i+{\rm H/He}\rightarrow j}$ is the total cross section of the corresponding hadronic interaction. Unlike above secondary CR nuclei, secondary $\rm e^+$ and $\rm \bar{p}$ are produced through the pp collisions between the primary CR nuclei from background sources and ISM. Therefore the source term of both $\rm e^+$ and $\rm \bar{p}$ is the convolution of the energy spectra of primary nuclei $\Phi_i(E)$ and the relevant differential cross section $d \sigma_{i + {\rm H/He} \to j}/d E_j$, i.e.
\begin{eqnarray}
	\nonumber Q_j &=& \sum_{i = \rm p, He} \int dp_i v \left\lbrace n_{\rm H} \frac{ \sigma_{i+{\rm H}\rightarrow j}(p_i, p_j)}{ dp_j} \right. \\
	& & \left. +n_{\rm He} \frac{\sigma_{i+{\rm He}\rightarrow j}(p_i, p_j)}{dp_j} \right\rbrace  \psi_i(p_i) ~,
\end{eqnarray}
Furthermore, antiprotons may still undergo non-annihilated inelastic scattering with ISM protons during propagation, which can generate the tertiary production.

\section{Nearby supernova remnant}
We assume that one supernova explosion in the vicinity of the solar neighborhood occurred in a giant MC about $10^5-10^6$ years ago. The CR charged particles were continually accelerated by passing back and forth across the shock front with the expansion of supernova ejecta. Under the SDP scenario of GCR propagation, \citet{2022ApJ...930...82L} demonstrated that, 
among the observed local SNRs, only Geminga SNR is able to explain both the spectra and anisotropies observations of CRs simultaneously. Since the location of Monogem is similar to that of Geminga, their impacts on CR flux may degenerate with each other. Here, for simplicity, we take Geminga SNR as the major contributor of the local source in this work.

The Geminga SNR locates at its birth place with the distance and age of r = 330 pc and $t_{inj} = 3.3\times10^5$ yrs. The direction in the galactic coordinate is $l=194^{\circ}$(galactic longitude) and $b=-13^{\circ}$(galactic latitude) \citep{1994A&A...281L..41S}. Its distance, age and direction jointly decide the important role as an optimal candidate of nearby source to CRs \citep{2019JCAP...10..010L,2019JCAP...12..007Q, 2022ApJ...926...41Z}. The injection process of SNR is approximated as burst-like. So the primary and secondary particles have experienced 300 kyrs and a tiny part of them enter into solar system in the end. In this work, the propagated spectrum 
from Geminga SNR is thus a convolution of the Green's function and the time-dependent injection rate $Q_0(t)$\citep{1995PhRvD..52.3265A}, i.e.

\begin{equation}
	\varphi(\vec{r}, {\cal R}, t) = \int_{t_i}^{t} G(\vec{r}-\vec{r}^\prime, t-t^\prime, {\cal R}) Q_0(t^\prime) d t^\prime .
\end{equation}
The normalization is determined through fitting Galactic cosmic rays energy spectra and the detailed parameters is listed in Table \ref{para_inj}.

Besides, the CR nuclei generated by the local SNR also collide with the molecular gas around them and give birth to prolific daughter particles, like B, $\rm e^{\pm}$, $\rm \bar{p}$, and so forth. The yields of B and $\rm e^{\pm}$, $\rm \bar{p}$ inside the MC are respectively
\begin{eqnarray}
	Q_j = \sum_{i = \rm C, N, O} (n_{\rm H} \sigma_{i+{\rm H}\rightarrow j} +n_{\rm He} \sigma_{i+{\rm He}\rightarrow j} )v Q_i(E) t_{\rm col}
\end{eqnarray}
and 
\begin{equation}
		Q_j  \; = \;
		\sum_{i = \rm p, He} {\displaystyle \int\limits_{E_{\rm th}}^{+ \infty}} \; d E_i \;
		v \; \left\lbrace n_{\, \rm H}
		{\displaystyle \frac{d \sigma_{i + {\rm H} \to j}}{d E_j }} \right. \\
		\left. +n_{\, \rm He} {\displaystyle \frac{d \sigma_{i + {\rm He} \to j}}{d E_j }} \right\rbrace
		Q_i(E_i) t_{\rm col} ~,
		\label{sec_source}
\end{equation}
where $n_{\rm H/He}$ is the number density of hydrogen/helium in MC. In this work, we assume that it is $1000$ times greater than the mean value of ISM. $t_{\rm col}$ is the duration of collision of $420$ yrs. $Q_i(E)$ is the accelerated spectrum of primary nuclei inside local SNR.

Summarily, the local source, Geminga SNR, is responsible for the spectral hardening at 200 GV and cut-off at 14 TV for all the nuclei species. The primary electron can be accelerated to $\sim$5 TeV, similar to nuclei, but will undergo the energy loss in ISRF and have a spectral cut-off around TeV due to its age of $3.3\times10^5$ yrs. For the secondary particles, the spectral cut-off depend on their mother ones and the interaction modes. For positron, the energy break-off will located at $\sim$300 GeV. The different energy break-off between positron and electron can be natural understood in this scenario. The secondary heavier nuclei will have the same spectral cut-off as their mothers at $\sim$5 TV, which can be tested in future experiments like HERD
\citep{2022PhyS...97e4010K}.

\bibliography{refs}{}

\begin{thebibliography}{}
\expandafter\ifx\csname natexlab\endcsname\relax\def\natexlab#1{#1}\fi
\providecommand{\url}[1]{\href{#1}{#1}}
\providecommand{\dodoi}[1]{doi:~\href{http://doi.org/#1}{\nolinkurl{#1}}}
\providecommand{\doeprint}[1]{\href{http://ascl.net/#1}{\nolinkurl{http://ascl.net/#1}}}
\providecommand{\doarXiv}[1]{\href{https://arxiv.org/abs/#1}{\nolinkurl{https://arxiv.org/abs/#1}}}

\bibitem[{{Aartsen} {et~al.}(2013){Aartsen}, {Abbasi}, {Abdou}, {Ackermann},
  {Adams}, {Aguilar}, {Ahlers}, {Altmann}, {Andeen}, {Auffenberg}, \&
  et~al.}]{2013ApJ...765...55A}
{Aartsen}, M.~G., {Abbasi}, R., {Abdou}, Y., {et~al.} 2013, \apj, 765, 55,
  \dodoi{10.1088/0004-637X/765/1/55}

\bibitem[{{Aartsen} {et~al.}(2016){Aartsen}, {Abraham}, {Ackermann}, {Adams},
  {Aguilar}, {Ahlers}, {Ahrens}, {Altmann}, {Anderson}, {Ansseau}, {Anton},
  {Archinger}, {Arguelles}, {Arlen}, {Auffenberg}, {Bai}, {Barwick}, {Baum},
  {Bay}, {Beatty}, {Becker Tjus}, {Becker}, {Beiser}, {BenZvi}, {Berghaus},
  {Berley}, {Bernardini}, {Bernhard}, {Besson}, {Binder}, {Bindig}, {Bissok},
  {Blaufuss}, {Blumenthal}, {Boersma}, {Bohm}, {B{\"o}rner}, {Bos}, {Bose},
  {B{\"o}ser}, {Botner}, {Braun}, {Brayeur}, {Bretz}, {Buzinsky}, {Casey},
  {Casier}, {Cheung}, {Chirkin}, {Christov}, {Clark}, {Classen}, {Coenders},
  {Collin}, {Conrad}, {Cowen}, {Cruz Silva}, {Daughhetee}, {Davis}, {Day}, {de
  Andr{\'e}}, {De Clercq}, {del Pino Rosendo}, {Dembinski}, {De Ridder},
  {Desiati}, {de Vries}, {de Wasseige}, {de With}, {DeYoung},
  {D{\'\i}az-V{\'e}lez}, {di Lorenzo}, {Dujmovic}, {Dumm}, {Dunkman},
  {Eberhardt}, {Ehrhardt}, {Eichmann}, {Euler}, {Evenson}, {Fahey}, {Fazely},
  {Feintzeig}, {Felde}, {Filimonov}, {Finley}, {Flis}, {F{\"o}sig}, {Fuchs},
  {Gaisser}, {Gaior}, {Gallagher}, {Gerhardt}, {Ghorbani}, {Gier}, {Gladstone},
  {Glagla}, {Gl{\"u}senkamp}, {Goldschmidt}, {Golup}, {Gonzalez}, {G{\'o}ra},
  {Grant}, {Griffith}, {Ha}, {Haack}, {Haj Ismail}, {Hallgren}, {Halzen},
  {Hansen}, {Hansmann}, {Hansmann}, {Hanson}, {Hebecker}, {Heereman},
  {Helbing}, {Hellauer}, {Hickford}, {Hignight}, {Hill}, {Hoffman}, {Hoffmann},
  {Holzapfel}, {Homeier}, {Hoshina}, {Huang}, {Huber}, {Huelsnitz}, {Hulth},
  {Hultqvist}, {In}, {Ishihara}, {Jacobi}, {Japaridze}, {Jeong}, {Jero},
  {Jones}, {Jurkovic}, {Kappes}, {Karg}, {Karle}, {Katz}, {Kauer}, {Keivani},
  {Kelley}, {Kemp}, {Kheirandish}, {Kim}, {Kintscher}, {Kiryluk}, {Klein},
  {Kohnen}, {Koirala}, {Kolanoski}, {Konietz}, {K{\"o}pke}, {Kopper}, {Kopper},
  {Koskinen}, {Kowalski}, {Krings}, {Kroll}, {Kroll}, {Kr{\"u}ckl}, {Kunnen},
  {Kunwar}, {Kurahashi}, {Kuwabara}, {Labare}, {Lanfranchi}, {Larson},
  {Lennarz}, {Lesiak-Bzdak}, {Leuermann}, {Leuner}, {Lu}, {L{\"u}nemann},
  {Madsen}, {Maggi}, {Mahn}, {Mandelartz}, {Maruyama}, {Mase}, {Matis},
  {Maunu}, {McNally}, {Meagher}, {Medici}, {Meier}, {Meli}, {Menne}, {Merino},
  {Meures}, {Miarecki}, {Middell}, {Mohrmann}, {Montaruli}, {Morse},
  {Nahnhauer}, {Naumann}, {Neer}, {Niederhausen}, {Nowicki}, {Nygren},
  {Obertacke Pollmann}, {Olivas}, {Omairat}, {O'Murchadha}, {Palczewski},
  {Pandya}, {Pankova}, {Paul}, {Pepper}, {P{\'e}rez de los Heros}, {Pfendner},
  {Pieloth}, {Pinat}, {Posselt}, {Price}, {Przybylski}, {Quinnan}, {Raab},
  {R{\"a}del}, {Rameez}, {Rawlins}, {Reimann}, {Relich}, {Resconi}, {Rhode},
  {Richman}, {Richter}, {Riedel}, {Robertson}, {Rongen}, {Rott}, {Ruhe},
  {Ryckbosch}, {Sabbatini}, {Sander}, {Sandrock}, {Sandroos}, {Sarkar},
  {Schatto}, {Schimp}, {Schlunder}, {Schmidt}, {Schoenen}, {Sch{\"o}neberg},
  {Sch{\"o}nwald}, {Schumacher}, {Seckel}, {Seunarine}, {Soldin}, {Song},
  {Spiczak}, {Spiering}, {Stahlberg}, {Stamatikos}, {Stanev}, {Stasik},
  {Steuer}, {Stezelberger}, {Stokstad}, {St{\"o}ssl}, {Str{\"o}m},
  {Strotjohann}, {Sullivan}, {Sutherland}, {Taavola}, {Taboada}, {Tatar},
  {Ter-Antonyan}, {Terliuk}, {Te{\v{s}}i{\'c}}, {Tilav}, {Toale}, {Tobin},
  {Toscano}, {Tosi}, {Tselengidou}, {Turcati}, {Unger}, {Usner}, {Vallecorsa},
  {Vandenbroucke}, {van Eijndhoven}, {Vanheule}, {van Santen}, {Veenkamp},
  {Vehring}, {Voge}, {Vraeghe}, {Walck}, {Wallace}, {Wallraff}, {Wandkowsky},
  {Weaver}, {Wendt}, {Westerhoff}, {Whelan}, {Wiebe}, {Wiebusch}, {Wille},
  {Williams}, {Wills}, {Wissing}, {Wolf}, {Wood}, {Woschnagg}, {Xu}, {Xu},
  {Xu}, {Yanez}, {Yodh}, {Yoshida}, {Zoll}, \& {IceCube
  Collaboration}}]{2016ApJ...826..220A}
{Aartsen}, M.~G., {Abraham}, K., {Ackermann}, M., {et~al.} 2016, \apj, 826,
  220, \dodoi{10.3847/0004-637X/826/2/220}

\bibitem[{{Aartsen} {et~al.}(2019){Aartsen}, {Ackermann}, {Adams}, {Aguilar},
  {Ahlers}, {Ahrens}, {Alispach}, {Andeen}, {Anderson}, {Ansseau}, {Anton},
  {Arg{\"u}elles}, {Auffenberg}, {Axani}, {Backes}, {Bagherpour}, {Bai},
  {Barbano}, {Barwick}, {Baum}, {Baur}, {Bay}, {Beatty}, {Becker}, {Becker
  Tjus}, {BenZvi}, {Berley}, {Bernardini}, {Besson}, {Binder}, {Bindig},
  {Blaufuss}, {Blot}, {Bohm}, {B{\"o}rner}, {B{\"o}ser}, {Botner},
  {B{\"o}ttcher}, {Bourbeau}, {Bourbeau}, {Bradascio}, {Braun}, {Bretz},
  {Bron}, {Brostean-Kaiser}, {Burgman}, {Buscher}, {Busse}, {Carver}, {Chen},
  {Cheung}, {Chirkin}, {Clark}, {Classen}, {Collin}, {Conrad}, {Coppin},
  {Correa}, {Cowen}, {Cross}, {Dave}, {de Andr{\'e}}, {De Clercq}, {DeLaunay},
  {Dembinski}, {Deoskar}, {De Ridder}, {Desiati}, {de Vries}, {de Wasseige},
  {de With}, {DeYoung}, {Diaz}, {D{\'\i}az-V{\'e}lez}, {Dujmovic}, {Dunkman},
  {Dvorak}, {Eberhardt}, {Ehrhardt}, {Eller}, {Evenson}, {Fahey}, {Fazely},
  {Felde}, {Feusels}, {Filimonov}, {Finley}, {Franckowiak}, {Friedman},
  {Fritz}, {Gaisser}, {Gallagher}, {Ganster}, {Garrappa}, {Gerhardt},
  {Ghorbani}, {Glauch}, {Gl{\"u}senkamp}, {Goldschmidt}, {Gonzalez}, {Grant},
  {Griffith}, {G{\"u}nder}, {G{\"u}nd{\"u}z}, {Haack}, {Hallgren}, {Halve},
  {Halzen}, {Hanson}, {Hebecker}, {Heereman}, {Heix}, {Helbing}, {Hellauer},
  {Henningsen}, {Hickford}, {Hignight}, {Hill}, {Hoffman}, {Hoffmann},
  {Hoinka}, {Hokanson-Fasig}, {Hoshina}, {Huang}, {Huber}, {Hultqvist},
  {H{\"u}nnefeld}, {Hussain}, {In}, {Iovine}, {Ishihara}, {Jacobi},
  {Japaridze}, {Jeong}, {Jero}, {Jones}, {Jonske}, {Joppe}, {Kang}, {Kappes},
  {Kappesser}, {Karg}, {Karl}, {Karle}, {Katz}, {Kauer}, {Kelley},
  {Kheirandish}, {Kim}, {Kintscher}, {Kiryluk}, {Kittler}, {Klein}, {Koirala},
  {Kolanoski}, {K{\"o}pke}, {Kopper}, {Kopper}, {Koskinen}, {Kowalski},
  {Krings}, {Kr{\"u}ckl}, {Kulacz}, {Kunwar}, {Kurahashi}, {Kyriacou},
  {Labare}, {Lanfranchi}, {Larson}, {Lauber}, {Lazar}, {Leonard}, {Leuermann},
  {Liu}, {Lohfink}, {Lozano Mariscal}, {Lu}, {Lucarelli}, {L{\"u}nemann},
  {Luszczak}, {Madsen}, {Maggi}, {Mahn}, {Makino}, {Mallik}, {Mallot},
  {Mancina}, {Mari{\c{s}}}, {Maruyama}, {Mase}, {Maunu}, {Meagher}, {Medici},
  {Medina}, {Meier}, {Meighen-Berger}, {Menne}, {Merino}, {Meures}, {Miarecki},
  {Micallef}, {Moment{\'e}}, {Montaruli}, {Moore}, {Morse}, {Moulai}, {Muth},
  {Nagai}, {Nahnhauer}, {Nakarmi}, {Naumann}, {Neer}, {Niederhausen},
  {Nowicki}, {Nygren}, {Obertacke Pollmann}, {Olivas}, {O'Murchadha},
  {O'Sullivan}, {Palczewski}, {Pandya}, {Pankova}, {Park}, {Peiffer},
  {P{\'e}rez de los Heros}, {Philippen}, {Pieloth}, {Pinat}, {Pizzuto}, {Plum},
  {Porcelli}, {Price}, {Przybylski}, {Raab}, {Raissi}, {Rameez}, {Rauch},
  {Rawlins}, {Rea}, {Reimann}, {Relethford}, {Renzi}, {Resconi}, {Rhode},
  {Richman}, {Robertson}, {Rongen}, {Rott}, {Ruhe}, {Ryckbosch}, {Rysewyk},
  {Safa}, {Sanchez Herrera}, {Sandrock}, {Sandroos}, {Santander}, {Sarkar},
  {Sarkar}, {Satalecka}, {Schaufel}, {Schlunder}, {Schmidt}, {Schneider},
  {Schneider}, {Schumacher}, {Sclafani}, {Seckel}, {Seunarine}, {Shefali},
  {Silva}, {Snihur}, {Soedingrekso}, {Soldin}, {Song}, {Spiczak}, {Spiering},
  {Stachurska}, {Stamatikos}, {Stanev}, {Stasik}, {Stein}, {Stettner},
  {Steuer}, {Stezelberger}, {Stokstad}, {St{\"o}{\ss}l}, {Strotjohann},
  {St{\"u}rwald}, {Stuttard}, {Sullivan}, {Sutherland}, {Taboada}, {Tenholt},
  {Ter-Antonyan}, {Terliuk}, {Tilav}, {Tomankova}, {T{\"o}nnis}, {Toscano},
  {Tosi}, {Tselengidou}, {Tung}, {Turcati}, {Turcotte}, {Turley}, {Ty},
  {Unger}, {Unland Elorrieta}, {Usner}, {Vandenbroucke}, {Van Driessche}, {van
  Eijk}, {van Eijndhoven}, {Vanheule}, {van Santen}, {Vraeghe}, {Walck},
  {Wallace}, {Wallraff}, {Wandkowsky}, {Watson}, {Weaver}, {Weiss}, {Weldert},
  {Wendt}, {Werthebach}, {Westerhoff}, {Whelan}, {Whitehorn}, {Wiebe},
  {Wiebusch}, {Wille}, {Williams}, {Wills}, {Wolf}, {Wood}, {Wood},
  {Woschnagg}, {Wrede}, {Xu}, {Xu}, {Xu}, {Yanez}, {Yodh}, {Yoshida}, {Yuan},
  {Z{\"o}cklein}, \& {IceCube Collaboration}}]{2019PhRvD.100h2002A}
{Aartsen}, M.~G., {Ackermann}, M., {Adams}, J., {et~al.} 2019, \prd, 100,
  082002, \dodoi{10.1103/PhysRevD.100.082002}

\bibitem[{{Aartsen} {et~al.}(2020){Aartsen}, {Abbasi}, {Ackermann}, {Adams},
  {Aguilar}, {Ahlers}, {Ahrens}, {Alispach}, {Amin}, {Andeen}, {Anderson},
  {Ansseau}, {Anton}, {Arg{\"u}elles}, {Auffenberg}, {Axani}, {Bagherpour},
  {Bai}, {Balagopal V.}, {Barbano}, {Barwick}, {Bastian}, {Baum}, {Baur},
  {Bay}, {Beatty}, {Becker}, {Becker Tjus}, {BenZvi}, {Berley}, {Bernardini},
  {Besson}, {Binder}, {Bindig}, {Blaufuss}, {Blot}, {Bohm}, {B{\"o}ser},
  {Botner}, {B{\"o}ttcher}, {Bourbeau}, {Bourbeau}, {Bradascio}, {Braun},
  {Bron}, {Brostean-Kaiser}, {Burgman}, {Buscher}, {Busse}, {Carver}, {Chen},
  {Cheung}, {Chirkin}, {Choi}, {Clark}, {Clark}, {Classen}, {Coleman},
  {Collin}, {Conrad}, {Coppin}, {Correa}, {Cowen}, {Cross}, {Dave}, {De
  Clercq}, {DeLaunay}, {Dembinski}, {Deoskar}, {De Ridder}, {Desiati}, {de
  Vries}, {de Wasseige}, {de With}, {DeYoung}, {Dharani}, {Diaz},
  {D{\'\i}az-V{\'e}lez}, {Dujmovic}, {Dvorak}, {Eberhardt}, {Ehrhardt},
  {Eller}, {Engel}, {Evenson}, {Fahey}, {Fazely}, {Felde}, {Fienberg},
  {Filimonov}, {Finley}, {Fox}, {Franckowiak}, {Friedman}, {Fritz}, {Gaisser},
  {Gallagher}, {Ganster}, {Garrappa}, {Gerhardt}, {Ghorbani}, {Glauch},
  {Gl{\"u}senkamp}, {Goldschmidt}, {Gonzalez}, {Grant}, {Gr{\'e}goire},
  {Griffith}, {Griswold}, {G{\"u}nder}, {G{\"u}nd{\"u}z}, {Haack}, {Hallgren},
  {Halliday}, {Halve}, {Halzen}, {Hanson}, {Haungs}, {Hauser}, {Hebecker},
  {Heereman}, {Heix}, {Helbing}, {Hellauer}, {Henningsen}, {Hickford},
  {Hignight}, {Hill}, {Hill}, {Hoffman}, {Hoffmann}, {Hoinka},
  {Hokanson-Fasig}, {Hoshina}, {Huber}, {Huber}, {Hultqvist}, {H{\"u}nnefeld},
  {Hussain}, {In}, {Iovine}, {Ishihara}, {Jansson}, {Japaridze}, {Jeong},
  {Jero}, {Jones}, {Jonske}, {Joppe}, {Kang}, {Kang}, {Kappes}, {Kappesser},
  {Karg}, {Karl}, {Karle}, {Katz}, {Kauer}, {Kellermann}, {Kelley},
  {Kheirandish}, {Kim}, {Kintscher}, {Kiryluk}, {Kittler}, {Klein}, {Koirala},
  {Kolanoski}, {K{\"o}pke}, {Kopper}, {Kopper}, {Koskinen}, {Koundal},
  {Kowalski}, {Krings}, {Kr{\"u}ckl}, {Kulacz}, {Kurahashi}, {Kyriacou},
  {Lanfranchi}, {Larson}, {Lauber}, {Lazar}, {Leonard}, {Leszczy{\'n}ska},
  {Li}, {Liu}, {Lohfink}, {Lozano Mariscal}, {Lu}, {Lucarelli}, {Ludwig},
  {L{\"u}nemann}, {Luszczak}, {Lyu}, {Ma}, {Madsen}, {Maggi}, {Mahn}, {Mallik},
  {Mallot}, {Mancina}, {Mari{\c{s}}}, {Maruyama}, {Mase}, {Maunu}, {McNally},
  {Meagher}, {Medici}, {Medina}, {Meier}, {Meighen-Berger}, {Merino}, {Merz},
  {Meures}, {Micallef}, {Mockler}, {Moment{\'e}}, {Montaruli}, {Moore},
  {Morse}, {Moulai}, {Muth}, {Nagai}, {Naumann}, {Neer}, {Nguy{\"e}n},
  {Niederhausen}, {Nisa}, {Nowicki}, {Nygren}, {Obertacke Pollmann}, {Oehler},
  {Olivas}, {O'Murchadha}, {O'Sullivan}, {Pandya}, {Pankova}, {Park}, {Parker},
  {Paudel}, {Peiffer}, {P{\'e}rez de los Heros}, {Philippen}, {Pieloth},
  {Pieper}, {Pinat}, {Pizzuto}, {Plum}, {Popovych}, {Porcelli}, {Price},
  {Przybylski}, {Raab}, {Raissi}, {Rameez}, {Rauch}, {Rawlins}, {Rea},
  {Rehman}, {Reimann}, {Relethford}, {Renschler}, {Renzi}, {Resconi}, {Rhode},
  {Richman}, {Robertson}, {Rongen}, {Rott}, {Ruhe}, {Ryckbosch}, {Rysewyk
  Cantu}, {Safa}, {Sanchez Herrera}, {Sandrock}, {Sandroos}, {Santander},
  {Sarkar}, {Sarkar}, {Satalecka}, {Scharf}, {Schaufel}, {Schieler},
  {Schlunder}, {Schmidt}, {Schneider}, {Schneider}, {Schr{\"o}der},
  {Schumacher}, {Sclafani}, {Seckel}, {Seunarine}, {Shefali}, {Silva},
  {Smithers}, {Snihur}, {Soedingrekso}, {Soldin}, {Song}, {Spiczak},
  {Spiering}, {Stachurska}, {Stamatikos}, {Stanev}, {Stein}, {Stettner},
  {Steuer}, {Stezelberger}, {Stokstad}, {Strotjohann}, {St{\"u}rwald},
  {Stuttard}, {Sullivan}, {Taboada}, {Tenholt}, {Ter-Antonyan}, {Terliuk},
  {Tilav}, {Tollefson}, {Tomankova}, {T{\"o}nnis}, {Toscano}, {Tosi},
  {Trettin}, {Tselengidou}, {Tung}, {Turcati}, {Turcotte}, {Turley}, {Ty},
  {Unger}, {Unland Elorrieta}, {Usner}, {Vandenbroucke}, {Van Driessche}, {van
  Eijk}, {van Eijndhoven}, {Vannerom}, {van Santen}, {Verpoest}, {Vraeghe},
  {Walck}, {Wallace}, {Wallraff}, {Wandkowsky}, {Watson}, {Weaver}, {Weindl},
  {Weldert}, {Wendt}, {Werthebach}, {Whelan}, {Whitehorn}, {Wiebe}, {Wiebusch},
  {Wille}, {Williams}, {Wills}, {Wolf}, {Wood}, {Wood}, {Woschnagg}, {Wrede},
  {Wulff}, {Xu}, {Xu}, {Xu}, {Yanez}, {Yodh}, {Yoshida}, {Yuan}, {Zhang},
  {Z{\"o}cklein}, \& {IceCube Collaboration}}]{2020PhRvD.102l2001A}
{Aartsen}, M.~G., {Abbasi}, R., {Ackermann}, M., {et~al.} 2020, \prd, 102,
  122001, \dodoi{10.1103/PhysRevD.102.122001}

\bibitem[{{Abbasi} {et~al.}(2010){Abbasi}, {Abdou}, {Abu-Zayyad}, {Adams},
  {Aguilar}, {Ahlers}, {Andeen}, {Auffenberg}, {Bai}, {Baker}, \&
  et~al.}]{2010ApJ...718L.194A}
{Abbasi}, R., {Abdou}, Y., {Abu-Zayyad}, T., {et~al.} 2010, \apjl, 718, L194,
  \dodoi{10.1088/2041-8205/718/2/L194}

\bibitem[{{Abbasi} {et~al.}(2012){Abbasi}, {Abdou}, {Abu-Zayyad}, {Ackermann},
  {Adams}, {Aguilar}, {Ahlers}, {Allen}, {Altmann}, {Andeen}, \&
  et~al.}]{2012ApJ...746...33A}
---. 2012, \apj, 746, 33, \dodoi{10.1088/0004-637X/746/1/33}

\bibitem[{{Abbasi} {et~al.}(2018){Abbasi}, {Abe}, {Abu-Zayyad}, {Allen},
  {Azuma}, {Barcikowski}, {Belz}, {Bergman}, {Blake}, {Cady}, {Cheon}, {Chiba},
  {Chikawa}, {Di Matteo}, {Fujii}, {Fujita}, {Fukushima}, {Furlich}, {Goto},
  {Hanlon}, {Hayashi}, {Hayashi}, {Hayashida}, {Hibino}, {Honda}, {Ikeda},
  {Inoue}, {Ishii}, {Ishimori}, {Ito}, {Ivanov}, {Jeong}, {Jeong}, {Jui},
  {Kadota}, {Kakimoto}, {Kalashev}, {Kasahara}, {Kawai}, {Kawakami}, {Kawana},
  {Kawata}, {Kido}, {Kim}, {Kim}, {Kim}, {Kishigami}, {Kitamura}, {Kitamura},
  {Kuzmin}, {Kuznetsov}, {Kwon}, {Lee}, {Lubsandorzhiev}, {Lundquist},
  {Machida}, {Martens}, {Matsuyama}, {Matthews}, {Mayta}, {Minamino}, {Mukai},
  {Myers}, {Nagasawa}, {Nagataki}, {Nakamura}, {Nakamura}, {Nonaka}, {Nozato},
  {Oda}, {Ogio}, {Ogura}, {Ohnishi}, {Ohoka}, {Okuda}, {Omura}, {Ono}, {Onogi},
  {Oshima}, {Ozawa}, {Park}, {Pshirkov}, {Rodriguez}, {Rubtsov}, {Ryu},
  {Sagawa}, {Sahara}, {Saito}, {Saito}, {Sakaki}, {Sakurai}, {Scott}, {Seki},
  {Sekino}, {Shah}, {Shibata}, {Shibata}, {Shimodaira}, {Shin}, {Shin},
  {Smith}, {Sokolsky}, {Stokes}, {Stratton}, {Stroman}, {Suzawa}, {Takagi},
  {Takahashi}, {Takamura}, {Takeda}, {Takeishi}, {Taketa}, {Takita}, {Tameda},
  {Tanaka}, {Tanaka}, {Tanaka}, {Thomas}, {Thomson}, {Tinyakov}, {Tkachev},
  {Tokuno}, {Tomida}, {Troitsky}, {Tsunesada}, {Tsutsumi}, {Uchihori}, {Udo},
  {Urban}, {Wong}, {Yamamoto}, {Yamane}, {Yamaoka}, {Yamazaki}, {Yang},
  {Yashiro}, {Yoneda}, {Yoshida}, {Yoshii}, {Zhezher}, {Zundel}, \& {Telescope
  Array Collaboration}}]{2018ApJ...865...74A}
{Abbasi}, R.~U., {Abe}, M., {Abu-Zayyad}, T., {et~al.} 2018, \apj, 865, 74,
  \dodoi{10.3847/1538-4357/aada05}

\bibitem[{{Abdo} {et~al.}(2009){Abdo}, {Allen}, {Aune}, {Berley}, {Casanova},
  {Chen}, {Dingus}, {Ellsworth}, {Fleysher}, {Fleysher}, {Gonzalez}, {Goodman},
  {Hoffman}, {Hopper}, {H{\"u}ntemeyer}, {Kolterman}, {Lansdell}, {Linnemann},
  {McEnery}, {Mincer}, {Nemethy}, {Noyes}, {Pretz}, {Ryan}, {Parkinson},
  {Shoup}, {Sinnis}, {Smith}, {Sullivan}, {Vasileiou}, {Walker}, {Williams}, \&
  {Yodh}}]{2009ApJ...698.2121A}
{Abdo}, A.~A., {Allen}, B.~T., {Aune}, T., {et~al.} 2009, \apj, 698, 2121,
  \dodoi{10.1088/0004-637X/698/2/2121}

\bibitem[{{Abdollahi} {et~al.}(2017){Abdollahi}, {Ackermann}, {Ajello},
  {Albert}, {Atwood}, {Baldini}, {Barbiellini}, {Bellazzini}, {Bissaldi},
  {Bloom}, {Bonino}, {Bottacini}, {Brandt}, {Bruel}, {Buson}, {Caragiulo},
  {Cavazzuti}, {Chekhtman}, {Ciprini}, {Costanza}, {Cuoco}, {Cutini},
  {D'Ammando}, {de Palma}, {Desiante}, {Digel}, {Di Lalla}, {Di Mauro}, {Di
  Venere}, {Donaggio}, {Drell}, {Favuzzi}, {Focke}, {Fukazawa}, {Funk},
  {Fusco}, {Gargano}, {Gasparrini}, {Giglietto}, {Giordano}, {Giroletti},
  {Green}, {Guiriec}, {Harding}, {Jogler}, {J{\'o}hannesson}, {Kamae}, {Kuss},
  {Larsson}, {Latronico}, {Li}, {Longo}, {Loparco}, {Lubrano}, {Magill},
  {Malyshev}, {Manfreda}, {Mazziotta}, {Meehan}, {Michelson}, {Mitthumsiri},
  {Mizuno}, {Moiseev}, {Monzani}, {Morselli}, {Negro}, {Nuss}, {Ohsugi},
  {Omodei}, {Paneque}, {Perkins}, {Pesce-Rollins}, {Piron}, {Pivato},
  {Principe}, {Rain{\`o}}, {Rando}, {Razzano}, {Reimer}, {Reimer}, {Sgr{\`o}},
  {Simone}, {Siskind}, {Spada}, {Spandre}, {Spinelli}, {Strong}, {Tajima},
  {Thayer}, {Torres}, {Troja}, {Vandenbroucke}, {Zaharijas}, {Zimmer}, \&
  {Fermi-LAT Collaboration}}]{2017PhRvL.118i1103A}
{Abdollahi}, S., {Ackermann}, M., {Ajello}, M., {et~al.} 2017, \prl, 118,
  091103, \dodoi{10.1103/PhysRevLett.118.091103}

\bibitem[{{Abeysekara} {et~al.}(2017){Abeysekara}, {Albert}, {Alfaro},
  {Alvarez}, {{\'A}lvarez}, {Arceo}, {Arteaga-Vel{\'a}zquez}, {Avila Rojas},
  {Ayala Solares}, {Barber}, {Bautista-Elivar}, {Becerril}, {Belmont-Moreno},
  {BenZvi}, {Berley}, {Bernal}, {Braun}, {Brisbois}, {Caballero-Mora},
  {Capistr{\'a}n}, {Carrami{\~n}ana}, {Casanova}, {Castillo}, {Cotti},
  {Cotzomi}, {Couti{\~n}o de Le{\'o}n}, {De Le{\'o}n}, {De la Fuente},
  {Dingus}, {DuVernois}, {D{\'\i}az-V{\'e}lez}, {Ellsworth}, {Engel},
  {Enr{\'\i}quez-Rivera}, {Fiorino}, {Fraija}, {Garc{\'\i}a-Gonz{\'a}lez},
  {Garfias}, {Gerhardt}, {Gonz{\'a}lez Mu{\~n}oz}, {Gonz{\'a}lez}, {Goodman},
  {Hampel-Arias}, {Harding}, {Hern{\'a}ndez}, {Hern{\'a}ndez-Almada}, {Hinton},
  {Hona}, {Hui}, {H{\"u}ntemeyer}, {Iriarte}, {Jardin-Blicq}, {Joshi},
  {Kaufmann}, {Kieda}, {Lara}, {Lauer}, {Lee}, {Lennarz}, {Vargas},
  {Linnemann}, {Longinotti}, {Luis Raya}, {Luna-Garc{\'\i}a}, {L{\'o}pez-Coto},
  {Malone}, {Marinelli}, {Martinez}, {Martinez-Castellanos},
  {Mart{\'\i}nez-Castro}, {Mart{\'\i}nez-Huerta}, {Matthews}, {Mirand
  a-Romagnoli}, {Moreno}, {Mostaf{\'a}}, {Nellen}, {Newbold}, {Nisa},
  {Noriega-Papaqui}, {Pelayo}, {Pretz}, {P{\'e}rez-P{\'e}rez}, {Ren}, {Rho},
  {Rivi{\`e}re}, {Rosa-Gonz{\'a}lez}, {Rosenberg}, {Ruiz-Velasco}, {Salazar},
  {Salesa Greus}, {Sand oval}, {Schneider}, {Schoorlemmer}, {Sinnis}, {Smith},
  {Springer}, {Surajbali}, {Taboada}, {Tibolla}, {Tollefson}, {Torres},
  {Ukwatta}, {Vianello}, {Weisgarber}, {Westerhoff}, {Wisher}, {Wood},
  {Yapici}, {Yodh}, {Younk}, {Zepeda}, {Zhou}, {Guo}, {Hahn}, {Li}, \&
  {Zhang}}]{2017Sci...358..911A}
{Abeysekara}, A.~U., {Albert}, A., {Alfaro}, R., {et~al.} 2017, Science, 358,
  911, \dodoi{10.1126/science.aan4880}

\bibitem[{{Accardo} {et~al.}(2014){Accardo}, {Aguilar}, {Aisa}, {Alvino},
  {Ambrosi}, {Andeen}, {Arruda}, {Attig}, {Azzarello}, {Bachlechner}, \&
  et~al.}]{2014PhRvL.113l1101A}
{Accardo}, L., {Aguilar}, M., {Aisa}, D., {et~al.} 2014, Physical Review
  Letters, 113, 121101, \dodoi{10.1103/PhysRevLett.113.121101}

\bibitem[{{Acero} {et~al.}(2016){Acero}, {Ackermann}, {Ajello}, {Albert},
  {Baldini}, {Ballet}, {Barbiellini}, {Bastieri}, {Bellazzini}, {Bissaldi},
  {Bloom}, {Bonino}, {Bottacini}, {Brandt}, {Bregeon}, {Bruel}, {Buehler},
  {Buson}, {Caliandro}, {Cameron}, {Caragiulo}, {Caraveo}, {Casandjian},
  {Cavazzuti}, {Cecchi}, {Charles}, {Chekhtman}, {Chiang}, {Chiaro}, {Ciprini},
  {Claus}, {Cohen-Tanugi}, {Conrad}, {Cuoco}, {Cutini}, {D'Ammando}, {de
  Angelis}, {de Palma}, {Desiante}, {Digel}, {Di Venere}, {Drell}, {Favuzzi},
  {Fegan}, {Ferrara}, {Focke}, {Franckowiak}, {Funk}, {Fusco}, {Gargano},
  {Gasparrini}, {Giglietto}, {Giordano}, {Giroletti}, {Glanzman}, {Godfrey},
  {Grenier}, {Guiriec}, {Hadasch}, {Harding}, {Hayashi}, {Hays}, {Hewitt},
  {Hill}, {Horan}, {Hou}, {Jogler}, {J{\'o}hannesson}, {Kamae}, {Kuss},
  {Landriu}, {Larsson}, {Latronico}, {Li}, {Li}, {Longo}, {Loparco},
  {Lovellette}, {Lubrano}, {Maldera}, {Malyshev}, {Manfreda}, {Martin},
  {Mayer}, {Mazziotta}, {McEnery}, {Michelson}, {Mirabal}, {Mizuno}, {Monzani},
  {Morselli}, {Nuss}, {Ohsugi}, {Omodei}, {Orienti}, {Orlando}, {Ormes},
  {Paneque}, {Pesce-Rollins}, {Piron}, {Pivato}, {Rain{\`o}}, {Rando},
  {Razzano}, {Razzaque}, {Reimer}, {Reimer}, {Remy}, {Renault},
  {S{\'a}nchez-Conde}, {Schaal}, {Schulz}, {Sgr{\`o}}, {Siskind}, {Spada},
  {Spandre}, {Spinelli}, {Strong}, {Suson}, {Tajima}, {Takahashi}, {Thayer},
  {Thompson}, {Tibaldo}, {Tinivella}, {Torres}, {Tosti}, {Troja}, {Vianello},
  {Werner}, {Wood}, {Wood}, {Zaharijas}, \& {Zimmer}}]{2016ApJS..223...26A}
{Acero}, F., {Ackermann}, M., {Ajello}, M., {et~al.} 2016, \apjs, 223, 26,
  \dodoi{10.3847/0067-0049/223/2/26}

\bibitem[{{Adriani} {et~al.}(2009){Adriani}, {Barbarino}, {Bazilevskaya},
  {Bellotti}, {Boezio}, {Bogomolov}, {Bonechi}, {Bongi}, {Bonvicini}, {Bottai},
  {Bruno}, {Cafagna}, {Campana}, {Carlson}, {Casolino}, {Castellini}, {de
  Pascale}, {de Rosa}, {de Simone}, {di Felice}, {Galper}, {Grishantseva},
  {Hofverberg}, {Koldashov}, {Krutkov}, {Kvashnin}, {Leonov}, {Malvezzi},
  {Marcelli}, {Menn}, {Mikhailov}, {Mocchiutti}, {Orsi}, {Osteria}, {Papini},
  {Pearce}, {Picozza}, {Ricci}, {Ricciarini}, {Simon}, {Sparvoli},
  {Spillantini}, {Stozhkov}, {Vacchi}, {Vannuccini}, {Vasilyev}, {Voronov},
  {Yurkin}, {Zampa}, {Zampa}, \& {Zverev}}]{2009Natur.458..607A}
{Adriani}, O., {Barbarino}, G.~C., {Bazilevskaya}, G.~A., {et~al.} 2009, \nat,
  458, 607, \dodoi{10.1038/nature07942}

\bibitem[{{Adriani} {et~al.}(2011){Adriani}, {Barbarino}, {Bazilevskaya},
  {Bellotti}, {Boezio}, {Bogomolov}, {Bonechi}, {Bongi}, {Bonvicini},
  {Borisov}, {Bottai}, {Bruno}, {Cafagna}, {Campana}, {Carbone}, {Carlson},
  {Casolino}, {Castellini}, {Consiglio}, {De Pascale}, {De Santis}, {De
  Simone}, {Di Felice}, {Galper}, {Gillard}, {Grishantseva}, {Jerse},
  {Karelin}, {Koldashov}, {Krutkov}, {Kvashnin}, {Leonov}, {Malakhov},
  {Malvezzi}, {Marcelli}, {Mayorov}, {Menn}, {Mikhailov}, {Mocchiutti},
  {Monaco}, {Mori}, {Nikonov}, {Osteria}, {Palma}, {Papini}, {Pearce},
  {Picozza}, {Pizzolotto}, {Ricci}, {Ricciarini}, {Rossetto}, {Sarkar},
  {Simon}, {Sparvoli}, {Spillantini}, {Stozhkov}, {Vacchi}, {Vannuccini},
  {Vasilyev}, {Voronov}, {Yurkin}, {Wu}, {Zampa}, {Zampa}, \&
  {Zverev}}]{2011Sci...332...69A}
---. 2011, Science, 332, 69, \dodoi{10.1126/science.1199172}

\bibitem[{{Aglietta} {et~al.}(1995){Aglietta}, {Alessandro}, {Antonioli},
  {Arneodo}, {Bergamasco}, {Bertaina}, {Bosio}, {Castellina}, {Castagnoli},
  {Chaivasa}, {Cini}, {D' Ettorre Piazzoli}, {Di Sciascio}, {Fulgione},
  {Galeotti}, {Ghia}, {Iacovacci}, {Mannocchi}, {Melagrana}, {Mengotti Silva},
  {Morello}, {Navarra}, {Riccati}, {Saavedra}, {Trinchero}, {Vallania}, \&
  {Vernetto}}]{1995ICRC....2..800A}
{Aglietta}, M., {Alessandro}, B., {Antonioli}, P., {et~al.} 1995, International
  Cosmic Ray Conference, 2, 800

\bibitem[{{Aglietta} {et~al.}(1996){Aglietta}, {Alessandro}, {Antonioli},
  {Arneodo}, {Bergamasco}, {Bertaina}, {Bosio}, {Castellina}, {Castagnoli},
  {Chiavassa}, {Cini Castagnoli}, {D'Ettorre Piazzoli}, {di Sciascio},
  {Fulgione}, {Galeotti}, {Ghia}, {Iacovacci}, {Mannocchi}, {Melagrana},
  {Mengotti Silva}, {Morello}, {Navarra}, {Riccati}, {Saavedra}, {Trinchero},
  {Vallania}, {Vernetto}, \& {EAS-Top Collaboration}}]{1996ApJ...470..501A}
---. 1996, \apj, 470, 501, \dodoi{10.1086/177881}

\bibitem[{{Aglietta} {et~al.}(2009){Aglietta}, {Alekseenko}, {Alessandro},
  {Antonioli}, {Arneodo}, {Bergamasco}, {Bertaina}, {Bonino}, {Castellina},
  {Chiavassa}, {D'Ettorre Piazzoli}, {Di Sciascio}, {Fulgione}, {Galeotti},
  {Ghia}, {Iacovacci}, {Mannocchi}, {Morello}, {Navarra}, {Saavedra},
  {Stamerra}, {Trinchero}, {Valchierotti}, {Vallania}, {Vernetto}, \&
  {Vigorito}}]{2009ApJ...692L.130A}
{Aglietta}, M., {Alekseenko}, V.~V., {Alessandro}, B., {et~al.} 2009, \apjl,
  692, L130, \dodoi{10.1088/0004-637X/692/2/L130}

\bibitem[{{Aguilar} {et~al.}(2015{\natexlab{a}}){Aguilar}, {Aisa}, {Alpat},
  {Alvino}, {Ambrosi}, {Andeen}, {Arruda}, {Attig}, {Azzarello}, {Bachlechner},
  \& et~al.}]{2015PhRvL.114q1103A}
{Aguilar}, M., {Aisa}, D., {Alpat}, B., {et~al.} 2015{\natexlab{a}}, Physical
  Review Letters, 114, 171103, \dodoi{10.1103/PhysRevLett.114.171103}

\bibitem[{{Aguilar} {et~al.}(2015{\natexlab{b}}){Aguilar}, {Aisa}, {Alpat},
  {Alvino}, {Ambrosi}, {Andeen}, {Arruda}, {Attig}, {Azzarello}, {Bachlechner},
  \& et~al.}]{2015PhRvL.115u1101A}
---. 2015{\natexlab{b}}, Physical Review Letters, 115, 211101,
  \dodoi{10.1103/PhysRevLett.115.211101}

\bibitem[{{Aguilar} {et~al.}(2016{\natexlab{a}}){Aguilar}, {Ali Cavasonza},
  {Alpat}, {Ambrosi}, {Arruda}, {Attig}, {Aupetit}, {Azzarello}, {Bachlechner},
  {Barao}, {Barrau}, {Barrin}, {Bartoloni}, {Basara},
  {Ba{\c{s}}e{\c{C}}{\textsection}mez-du Pree}, {Battarbee}, {Battiston},
  {Bazo}, {Becker}, {Behlmann}, {Beischer}, {Berdugo}, {Bertucci}, {Bindi},
  {Boella}, {de Boer}, {Bollweg}, {Bonnivard}, {Borgia}, {Boschini},
  {Bourquin}, {Bueno}, {Burger}, {Cadoux}, {Cai}, {Capell}, {Caroff}, {Casaus},
  {Castellini}, {Cernuda}, {Cervelli}, {Chae}, {Chang}, {Chen}, {Chen}, {Chen},
  {Cheng}, {Chou}, {Choumilov}, {Choutko}, {Chung}, {Clark}, {Clavero},
  {Coignet}, {Consolandi}, {Contin}, {Corti}, {Coste}, {Creus}, {Crispoltoni},
  {Cui}, {Dai}, {Delgado}, {Della Torre}, {Demirk{\"o}z}, {Derome}, {Di Falco},
  {Dimiccoli}, {D{\'\i}az}, {von Doetinchem}, {Dong}, {Donnini}, {Duranti},
  {D'Urso}, {Egorov}, {Eline}, {Eronen}, {Feng}, {Fiandrini}, {Finch},
  {Fisher}, {Formato}, {Galaktionov}, {Gallucci}, {Garc{\'\i}a},
  {Garc{\'\i}a-L{\'o}pez}, {Gargiulo}, {Gast}, {Gebauer}, {Gervasi}, {Ghelfi},
  {Giovacchini}, {Goglov}, {G{\'o}mez-Coral}, {Gong}, {Goy}, {Grabski},
  {Grandi}, {Graziani}, {Guerri}, {Guo}, {Habiby}, {Haino}, {Han}, {He},
  {Heil}, {Hoffman}, {Hsieh}, {Huang}, {Huang}, {Huh}, {Incagli}, {Ionica},
  {Jang}, {Jinchi}, {Kang}, {Kanishev}, {Kim}, {Kim}, {Kirn}, {Konak},
  {Kounina}, {Kounine}, {Koutsenko}, {Krafczyk}, {La Vacca}, {Laudi},
  {Laurenti}, {Lazzizzera}, {Lebedev}, {Lee}, {Lee}, {Leluc}, {Li}, {Li}, {Li},
  {Li}, {Li}, {Li}, {Li}, {Li}, {Lim}, {Lin}, {Lipari}, {Lippert}, {Liu},
  {Liu}, {Lu}, {Lu}, {Luebelsmeyer}, {Luo}, {Luo}, {Lv}, {Majka},
  {Ma{\~n}{\'a}}, {Mar{\'\i}n}, {Martin}, {Mart{\'\i}nez}, {Masi}, {Maurin},
  {Menchaca-Rocha}, {Meng}, {Mo}, {Morescalchi}, {Mott}, {Nelson}, {Ni},
  {Nikonov}, {Nozzoli}, {Nunes}, {Oliva}, {Orcinha}, {Palmonari}, {Palomares},
  {Paniccia}, {Pauluzzi}, {Pensotti}, {Pereira}, {Picot-Clemente}, {Pilo},
  {Pizzolotto}, {Plyaskin}, {Pohl}, {Poireau}, {Putze}, {Quadrani}, {Qi},
  {Qin}, {Qu}, {R{\"a}ih{\"a}}, {Rancoita}, {Rapin}, {Ricol}, {Rodr{\'\i}guez},
  {Rosier-Lees}, {Rozhkov}, {Rozza}, {Sagdeev}, {Sandweiss}, {Saouter},
  {Schael}, {Schmidt}, {Schulz von Dratzig}, {Schwering}, {Seo}, {Shan}, {Shi},
  {Siedenburg}, {Son}, {Song}, {Sun}, {Tacconi}, {Tang}, {Tang}, {Tao},
  {Tescaro}, {Ting}, {Ting}, {Tomassetti}, {Torsti}, {T{\"u}rko{\v{g}}lu},
  {Urban}, {Vagelli}, {Valente}, {Vannini}, {Valtonen}, {V{\'a}zquez Acosta},
  {Vecchi}, {Velasco}, {Vialle}, {Vitale}, {Vitillo}, {Wang}, {Wang}, {Wang},
  {Wang}, {Wang}, {Wang}, {Wei}, {Weng}, {Whitman}, {Wienkenh{\"o}ver},
  {Willenbrock}, {Wu}, {Wu}, {Xia}, {Xiong}, {Xu}, {Yan}, {Yang}, {Yang},
  {Yang}, {Yi}, {Yu}, {Yu}, {Zeissler}, {Zhang}, {Zhang}, {Zhang}, {Zhang},
  {Zhang}, {Zhang}, {Zheng}, {Zhu}, {Zhuang}, {Zhukov}, {Zichichi},
  {Zimmermann}, {Zuccon}, \& {AMS Collaboration}}]{2016PhRvL.117i1103A}
{Aguilar}, M., {Ali Cavasonza}, L., {Alpat}, B., {et~al.} 2016{\natexlab{a}},
  \prl, 117, 091103, \dodoi{10.1103/PhysRevLett.117.091103}

\bibitem[{{Aguilar} {et~al.}(2016{\natexlab{b}}){Aguilar}, {Ali Cavasonza},
  {Ambrosi}, {Arruda}, {Attig}, {Aupetit}, {Azzarello}, {Bachlechner}, {Barao},
  {Barrau}, {Barrin}, {Bartoloni}, {Basara}, {Ba{\c{s}}e{\v{g}}mez-du Pree},
  {Battarbee}, {Battiston}, {Becker}, {Behlmann}, {Beischer}, {Berdugo},
  {Bertucci}, {Bindel}, {Bindi}, {Boella}, {de Boer}, {Bollweg}, {Bonnivard},
  {Borgia}, {Boschini}, {Bourquin}, {Bueno}, {Burger}, {Cadoux}, {Cai},
  {Capell}, {Caroff}, {Casaus}, {Castellini}, {Cervelli}, {Chae}, {Chang},
  {Chen}, {Chen}, {Chen}, {Cheng}, {Chou}, {Choumilov}, {Choutko}, {Chung},
  {Clark}, {Clavero}, {Coignet}, {Consolandi}, {Contin}, {Corti}, {Creus},
  {Crispoltoni}, {Cui}, {Dai}, {Delgado}, {Della Torre}, {Demakov},
  {Demirk{\"o}z}, {Derome}, {Di Falco}, {Dimiccoli}, {D{\'\i}az}, {von
  Doetinchem}, {Dong}, {Donnini}, {Duranti}, {D'Urso}, {Egorov}, {Eline},
  {Eronen}, {Feng}, {Fiandrini}, {Finch}, {Fisher}, {Formato}, {Galaktionov},
  {Gallucci}, {Garc{\'\i}a}, {Garc{\'\i}a-L{\'o}pez}, {Gargiulo}, {Gast},
  {Gebauer}, {Gervasi}, {Ghelfi}, {Giovacchini}, {Goglov}, {G{\'o}mez-Coral},
  {Gong}, {Goy}, {Grabski}, {Grandi}, {Graziani}, {Guo}, {Haino}, {Han}, {He},
  {Heil}, {Hoffman}, {Hsieh}, {Huang}, {Huang}, {Huh}, {Incagli}, {Ionica},
  {Jang}, {Jinchi}, {Kang}, {Kanishev}, {Kim}, {Kim}, {Kirn}, {Konak},
  {Kounina}, {Kounine}, {Koutsenko}, {Krafczyk}, {La Vacca}, {Laudi},
  {Laurenti}, {Lazzizzera}, {Lebedev}, {Lee}, {Lee}, {Leluc}, {Li}, {Li}, {Li},
  {Li}, {Li}, {Li}, {Li}, {Li}, {Li}, {Lim}, {Lin}, {Lipari}, {Lippert}, {Liu},
  {Liu}, {Lordello}, {Lu}, {Lu}, {Luebelsmeyer}, {Luo}, {Luo}, {Lv}, {Machate},
  {Majka}, {Ma{\~n}{\'a}}, {Mar{\'\i}n}, {Martin}, {Mart{\'\i}nez}, {Masi},
  {Maurin}, {Menchaca-Rocha}, {Meng}, {Mikuni}, {Mo}, {Morescalchi}, {Mott},
  {Nelson}, {Ni}, {Nikonov}, {Nozzoli}, {Oliva}, {Orcinha}, {Palmonari},
  {Palomares}, {Paniccia}, {Pauluzzi}, {Pensotti}, {Pereira}, {Picot-Clemente},
  {Pilo}, {Pizzolotto}, {Plyaskin}, {Pohl}, {Poireau}, {Putze}, {Quadrani},
  {Qi}, {Qin}, {Qu}, {R{\"a}ih{\"a}}, {Rancoita}, {Rapin}, {Ricol},
  {Rosier-Lees}, {Rozhkov}, {Rozza}, {Sagdeev}, {Sandweiss}, {Saouter},
  {Schael}, {Schmidt}, {Schulz von Dratzig}, {Schwering}, {Seo}, {Shan}, {Shi},
  {Siedenburg}, {Son}, {Song}, {Sun}, {Tacconi}, {Tang}, {Tang}, {Tao},
  {Tescaro}, {Ting}, {Ting}, {Tomassetti}, {Torsti}, {T{\"u}rko{\v{g}}lu},
  {Urban}, {Vagelli}, {Valente}, {Vannini}, {Valtonen}, {V{\'a}zquez Acosta},
  {Vecchi}, {Velasco}, {Vialle}, {Vitale}, {Vitillo}, {Wang}, {Wang}, {Wang},
  {Wang}, {Wang}, {Wang}, {Wei}, {Weng}, {Whitman}, {Wienkenh{\"o}ver}, {Wu},
  {Wu}, {Xia}, {Xiong}, {Xu}, {Yan}, {Yang}, {Yang}, {Yang}, {Yi}, {Yu}, {Yu},
  {Zeissler}, {Zhang}, {Zhang}, {Zhang}, {Zhang}, {Zhang}, {Zhang}, {Zheng},
  {Zhu}, {Zhuang}, {Zhukov}, {Zichichi}, {Zimmermann}, {Zuccon}, \& {AMS
  Collaboration}}]{2016PhRvL.117w1102A}
{Aguilar}, M., {Ali Cavasonza}, L., {Ambrosi}, G., {et~al.} 2016{\natexlab{b}},
  \prl, 117, 231102, \dodoi{10.1103/PhysRevLett.117.231102}

\bibitem[{{Aguilar} {et~al.}(2017){Aguilar}, {Ali Cavasonza}, {Alpat},
  {Ambrosi}, {Arruda}, {Attig}, {Aupetit}, {Azzarello}, {Bachlechner}, {Barao},
  {Barrau}, {Barrin}, {Bartoloni}, {Basara}, {Ba{\c{s}}e{\v{g}}mez-du Pree},
  {Battarbee}, {Battiston}, {Becker}, {Behlmann}, {Beischer}, {Berdugo},
  {Bertucci}, {Bindel}, {Bindi}, {de Boer}, {Bollweg}, {Bonnivard}, {Borgia},
  {Boschini}, {Bourquin}, {Bueno}, {Burger}, {Burger}, {Cadoux}, {Cai},
  {Capell}, {Caroff}, {Casaus}, {Castellini}, {Cervelli}, {Chae}, {Chang},
  {Chen}, {Chen}, {Chen}, {Cheng}, {Chou}, {Choumilov}, {Choutko}, {Chung},
  {Clark}, {Clavero}, {Coignet}, {Consolandi}, {Contin}, {Corti}, {Creus},
  {Crispoltoni}, {Cui}, {Dadzie}, {Dai}, {Datta}, {Delgado}, {Della Torre},
  {Demakov}, {Demirk{\"o}z}, {Derome}, {Di Falco}, {Dimiccoli}, {D{\'\i}az},
  {von Doetinchem}, {Dong}, {Donnini}, {Duranti}, {D'Urso}, {Egorov}, {Eline},
  {Eronen}, {Feng}, {Fiandrini}, {Fisher}, {Formato}, {Galaktionov},
  {Gallucci}, {Garc{\'\i}a-L{\'o}pez}, {Gargiulo}, {Gast}, {Gebauer},
  {Gervasi}, {Ghelfi}, {Giovacchini}, {G{\'o}mez-Coral}, {Gong}, {Goy},
  {Grabski}, {Grandi}, {Graziani}, {Guo}, {Haino}, {Han}, {He}, {Heil},
  {Hoffman}, {Hsieh}, {Huang}, {Huang}, {Huh}, {Incagli}, {Ionica}, {Jang},
  {Jia}, {Jinchi}, {Kang}, {Kanishev}, {Khiali}, {Kim}, {Kim}, {Kirn}, {Konak},
  {Kounina}, {Kounine}, {Koutsenko}, {Kulemzin}, {La Vacca}, {Laudi},
  {Laurenti}, {Lazzizzera}, {Lebedev}, {Lee}, {Lee}, {Leluc}, {Li}, {Li}, {Li},
  {Li}, {Li}, {Li}, {Li}, {Lim}, {Lin}, {Lipari}, {Lippert}, {Liu}, {Liu},
  {Lordello}, {Lu}, {Lu}, {Luebelsmeyer}, {Luo}, {Luo}, {Lyu}, {Machate},
  {Ma{\~n}{\'a}}, {Mar{\'\i}n}, {Martin}, {Mart{\'\i}nez}, {Masi}, {Maurin},
  {Menchaca-Rocha}, {Meng}, {Mikuni}, {Mo}, {Mott}, {Nelson}, {Ni}, {Nikonov},
  {Nozzoli}, {Oliva}, {Orcinha}, {Palmonari}, {Palomares}, {Paniccia},
  {Pauluzzi}, {Pensotti}, {Perrina}, {Phan}, {Picot-Clemente}, {Pilo},
  {Pizzolotto}, {Plyaskin}, {Pohl}, {Poireau}, {Quadrani}, {Qi}, {Qin}, {Qu},
  {R{\"a}ih{\"a}}, {Rancoita}, {Rapin}, {Ricol}, {Rosier-Lees}, {Rozhkov},
  {Rozza}, {Sagdeev}, {Schael}, {Schmidt}, {Schulz von Dratzig}, {Schwering},
  {Seo}, {Shan}, {Shi}, {Siedenburg}, {Son}, {Song}, {Tacconi}, {Tang}, {Tang},
  {Tescaro}, {Ting}, {Ting}, {Tomassetti}, {Torsti}, {T{\"u}rko{\v{g}}lu},
  {Urban}, {Vagelli}, {Valente}, {Valtonen}, {V{\'a}zquez Acosta}, {Vecchi},
  {Velasco}, {Vialle}, {Vitale}, {Vitillo}, {Wang}, {Wang}, {Wang}, {Wang},
  {Wang}, {Wang}, {Wei}, {Weng}, {Whitman}, {Wu}, {Wu}, {Xiong}, {Xu}, {Yan},
  {Yang}, {Yang}, {Yang}, {Yi}, {Yu}, {Yu}, {Zannoni}, {Zeissler}, {Zhang},
  {Zhang}, {Zhang}, {Zhang}, {Zhang}, {Zhang}, {Zheng}, {Zhuang}, {Zhukov},
  {Zichichi}, {Zimmermann}, {Zuccon}, \& {AMS
  Collaboration}}]{2017PhRvL.119y1101A}
{Aguilar}, M., {Ali Cavasonza}, L., {Alpat}, B., {et~al.} 2017, \prl, 119,
  251101, \dodoi{10.1103/PhysRevLett.119.251101}

\bibitem[{{Aguilar} {et~al.}(2018{\natexlab{a}}){Aguilar}, {Ali Cavasonza},
  {Ambrosi}, {Arruda}, {Attig}, {Aupetit}, {Azzarello}, {Bachlechner}, {Barao},
  {Barrau}, {Barrin}, {Bartoloni}, {Basara}, {Ba{\c{s}}e{\v{g}}mez-du Pree},
  {Battarbee}, {Battiston}, {Becker}, {Behlmann}, {Beischer}, {Berdugo},
  {Bertucci}, {Bindel}, {Bindi}, {de Boer}, {Bollweg}, {Bonnivard}, {Borgia},
  {Boschini}, {Bourquin}, {Bueno}, {Burger}, {Burger}, {Cadoux}, {Cai},
  {Capell}, {Caroff}, {Casaus}, {Castellini}, {Cervelli}, {Chae}, {Chang},
  {Chen}, {Chen}, {Chen}, {Cheng}, {Chou}, {Choumilov}, {Choutko}, {Chung},
  {Clark}, {Clavero}, {Coignet}, {Consolandi}, {Contin}, {Corti}, {Creus},
  {Crispoltoni}, {Cui}, {Dadzie}, {Dai}, {Datta}, {Delgado}, {Della Torre},
  {Demirk{\"o}z}, {Derome}, {Di Falco}, {Dimiccoli}, {D{\'\i}az}, {von
  Doetinchem}, {Dong}, {Donnini}, {Duranti}, {D'Urso}, {Egorov}, {Eline},
  {Eronen}, {Feng}, {Fiandrini}, {Fisher}, {Formato}, {Galaktionov},
  {Gallucci}, {Garc{\'\i}a-L{\'o}pez}, {Gargiulo}, {Gast}, {Gebauer},
  {Gervasi}, {Ghelfi}, {Giovacchini}, {G{\'o}mez-Coral}, {Gong}, {Goy},
  {Grabski}, {Grandi}, {Graziani}, {Guo}, {Haino}, {Han}, {He}, {Heil},
  {Hsieh}, {Huang}, {Huang}, {Huh}, {Incagli}, {Ionica}, {Jang}, {Jia},
  {Jinchi}, {Kang}, {Kanishev}, {Khiali}, {Kim}, {Kim}, {Kirn}, {Konak},
  {Kounina}, {Kounine}, {Koutsenko}, {Kulemzin}, {La Vacca}, {Laudi},
  {Laurenti}, {Lazzizzera}, {Lebedev}, {Lee}, {Lee}, {Leluc}, {Li}, {Li}, {Li},
  {Li}, {Li}, {Li}, {Li}, {Lim}, {Lin}, {Lipari}, {Lippert}, {Liu}, {Liu},
  {Lordello}, {Lu}, {Lu}, {Luebelsmeyer}, {Luo}, {Luo}, {Lyu}, {Machate},
  {Ma{\~n}{\'a}}, {Mar{\'\i}n}, {Martin}, {Mart{\'\i}nez}, {Masi}, {Maurin},
  {Menchaca-Rocha}, {Meng}, {Mikuni}, {Mo}, {Mott}, {Nelson}, {Ni}, {Nikonov},
  {Nozzoli}, {Oliva}, {Orcinha}, {Palermo}, {Palmonari}, {Palomares},
  {Paniccia}, {Pauluzzi}, {Pensotti}, {Perrina}, {Phan}, {Picot-Clemente},
  {Pilo}, {Pizzolotto}, {Plyaskin}, {Pohl}, {Poireau}, {Quadrani}, {Qi}, {Qin},
  {Qu}, {R{\"a}ih{\"a}}, {Rancoita}, {Rapin}, {Ricol}, {Rosier-Lees},
  {Rozhkov}, {Rozza}, {Sagdeev}, {Schael}, {Schmidt}, {Schulz von Dratzig},
  {Schwering}, {Seo}, {Shan}, {Shi}, {Siedenburg}, {Son}, {Song}, {Tacconi},
  {Tang}, {Tang}, {Tescaro}, {Ting}, {Ting}, {Tomassetti}, {Torsti},
  {T{\"u}rko{\v{g}}lu}, {Urban}, {Vagelli}, {Valente}, {Valtonen}, {V{\'a}zquez
  Acosta}, {Vecchi}, {Velasco}, {Vialle}, {Vitale}, {Wang}, {Wang}, {Wang},
  {Wang}, {Wang}, {Wang}, {Wei}, {Weng}, {Whitman}, {Wu}, {Wu}, {Xiong}, {Xu},
  {Yan}, {Yang}, {Yang}, {Yang}, {Yi}, {Yu}, {Yu}, {Zannoni}, {Zeissler},
  {Zhang}, {Zhang}, {Zhang}, {Zhang}, {Zhang}, {Zhang}, {Zheng}, {Zhuang},
  {Zhukov}, {Zichichi}, {Zimmermann}, {Zuccon}, \& {AMS
  Collaboration}}]{2018PhRvL.120b1101A}
{Aguilar}, M., {Ali Cavasonza}, L., {Ambrosi}, G., {et~al.} 2018{\natexlab{a}},
  \prl, 120, 021101, \dodoi{10.1103/PhysRevLett.120.021101}

\bibitem[{{Aguilar} {et~al.}(2018{\natexlab{b}}){Aguilar}, {Ali Cavasonza},
  {Alpat}, {Ambrosi}, {Arruda}, {Attig}, {Aupetit}, {Azzarello}, {Bachlechner},
  {Barao}, {Barrau}, {Barrin}, {Bartoloni}, {Basara}, {Ba{\c{s}}e{\v{g}}mez-du
  Pree}, {Battarbee}, {Battiston}, {Becker}, {Behlmann}, {Beischer}, {Berdugo},
  {Bertucci}, {Bindel}, {Bindi}, {de Boer}, {Bollweg}, {Bonnivard}, {Borgia},
  {Boschini}, {Bourquin}, {Bueno}, {Burger}, {Burger}, {Cai}, {Capell},
  {Caroff}, {Casaus}, {Castellini}, {Cervelli}, {Chang}, {Chen}, {Chen},
  {Chen}, {Chen}, {Cheng}, {Chou}, {Choumilov}, {Choutko}, {Chung}, {Clark},
  {Clavero}, {Coignet}, {Consolandi}, {Contin}, {Corti}, {Creus},
  {Crispoltoni}, {Cui}, {Dadzie}, {Dai}, {Datta}, {Delgado}, {Della Torre},
  {Demirk{\"o}z}, {Derome}, {Di Falco}, {Dimiccoli}, {D{\'\i}az}, {von
  Doetinchem}, {Dong}, {Donnini}, {Duranti}, {Egorov}, {Eline}, {Eronen},
  {Feng}, {Fiandrini}, {Fisher}, {Formato}, {Galaktionov}, {Gallucci},
  {Garc{\'\i}a-L{\'o}pez}, {Gargiulo}, {Gast}, {Gebauer}, {Gervasi}, {Ghelfi},
  {Giovacchini}, {G{\'o}mez-Coral}, {Gong}, {Goy}, {Grabski}, {Grandi},
  {Graziani}, {Guo}, {Haino}, {Han}, {He}, {Heil}, {Hsieh}, {Huang}, {Huang},
  {Incagli}, {Jia}, {Jinchi}, {Kanishev}, {Khiali}, {Kirn}, {Konak}, {Kounina},
  {Kounine}, {Koutsenko}, {Kulemzin}, {La Vacca}, {Laudi}, {Laurenti},
  {Lazzizzera}, {Lebedev}, {Lee}, {Lee}, {Leluc}, {Li}, {Li}, {Li}, {Li}, {Li},
  {Li}, {Lin}, {Lipari}, {Lippert}, {Liu}, {Liu}, {Liu}, {Lordello}, {Lu},
  {Lu}, {Luebelsmeyer}, {Luo}, {Luo}, {Lyu}, {Machate}, {Ma{\~n}{\'a}},
  {Mar{\'\i}n}, {Martin}, {Mart{\'\i}nez}, {Masi}, {Maurin}, {Menchaca-Rocha},
  {Meng}, {Mikuni}, {Mo}, {Mott}, {Mussolin}, {Nelson}, {Ni}, {Nikonov},
  {Nozzoli}, {Oliva}, {Orcinha}, {Palermo}, {Palmonari}, {Palomares},
  {Paniccia}, {Pauluzzi}, {Pensotti}, {Perrina}, {Phan}, {Picot-Clemente},
  {Pilo}, {Plyaskin}, {Pohl}, {Poireau}, {Quadrani}, {Qi}, {Qin}, {Qu},
  {R{\"a}ih{\"a}}, {Rancoita}, {Rapin}, {Ricol}, {Rosier-Lees}, {Rozhkov},
  {Rozza}, {Sagdeev}, {Schael}, {Schmidt}, {Schulz von Dratzig}, {Schwering},
  {Seo}, {Shan}, {Shi}, {Siedenburg}, {Song}, {Tacconi}, {Tang}, {Tang},
  {Tescaro}, {Tian}, {Ting}, {Ting}, {Tomassetti}, {Torsti}, {Urban},
  {Vagelli}, {Valente}, {Valtonen}, {V{\'a}zquez Acosta}, {Vecchi}, {Velasco},
  {Vialle}, {Wang}, {Wang}, {Wang}, {Wang}, {Wang}, {Wang}, {Wei}, {Wei},
  {Weng}, {Whitman}, {Wu}, {Xiong}, {Xu}, {Yan}, {Yang}, {Yang}, {Yi}, {Yu},
  {Yu}, {Zannoni}, {Zeissler}, {Zhang}, {Zhang}, {Zhang}, {Zhang}, {Zhang},
  {Zhang}, {Zheng}, {Zhuang}, {Zhukov}, {Zichichi}, {Zimmermann}, {Zuccon}, \&
  {AMS Collaboration}}]{2018PhRvL.121e1103A}
{Aguilar}, M., {Ali Cavasonza}, L., {Alpat}, B., {et~al.} 2018{\natexlab{b}},
  \prl, 121, 051103, \dodoi{10.1103/PhysRevLett.121.051103}

\bibitem[{{Aguilar} {et~al.}(2019{\natexlab{a}}){Aguilar}, {Ali Cavasonza},
  {Ambrosi}, {Arruda}, {Attig}, {Azzarello}, {Bachlechner}, {Barao}, {Barrau},
  {Barrin}, {Bartoloni}, {Basara}, {Ba{\textcommabelow s}e{\v{g}}mez-du Pree},
  {Battiston}, {Becker}, {Behlmann}, {Beischer}, {Berdugo}, {Bertucci},
  {Bindi}, {de Boer}, {Bollweg}, {Borgia}, {Boschini}, {Bourquin}, {Bueno},
  {Burger}, {Burger}, {Cai}, {Capell}, {Caroff}, {Casaus}, {Castellini},
  {Cervelli}, {Chang}, {Chen}, {Chen}, {Chen}, {Cheng}, {Chou}, {Choutko},
  {Chung}, {Clark}, {Coignet}, {Consolandi}, {Contin}, {Corti}, {Crispoltoni},
  {Cui}, {Dadzie}, {Dai}, {Datta}, {Delgado}, {Della Torre}, {Demirk{\"o}z},
  {Derome}, {Di Falco}, {Dimiccoli}, {D{\'\i}az}, {von Doetinchem}, {Dong},
  {Donnini}, {Duranti}, {Egorov}, {Eline}, {Eronen}, {Feng}, {Fiandrini},
  {Fisher}, {Formato}, {Galaktionov}, {Garc{\'\i}a-L{\'o}pez}, {Gargiulo},
  {Gast}, {Gebauer}, {Gervasi}, {Giovacchini}, {G{\'o}mez-Coral}, {Gong},
  {Goy}, {Grabski}, {Grandi}, {Graziani}, {Guo}, {Haino}, {Han}, {He}, {Heil},
  {Hsieh}, {Huang}, {Huang}, {Incagli}, {Jia}, {Jinchi}, {Kanishev}, {Khiali},
  {Kirn}, {Konak}, {Kounina}, {Kounine}, {Koutsenko}, {Kulemzin}, {La Vacca},
  {Laudi}, {Laurenti}, {Lazzizzera}, {Lebedev}, {Lee}, {Lee}, {Leluc}, {Li},
  {Li}, {Li}, {Li}, {Light}, {Lin}, {Lippert}, {Liu}, {Liu}, {Liu}, {Lu}, {Lu},
  {Luebelsmeyer}, {Luo}, {Luo}, {Luo}, {Lyu}, {Machate}, {Ma{\~n}{\'a}},
  {Mar{\'\i}n}, {Martin}, {Mart{\'\i}nez}, {Masi}, {Maurin}, {Menchaca-Rocha},
  {Meng}, {Mo}, {Molero}, {Mott}, {Mussolin}, {Nelson}, {Ni}, {Nikonov},
  {Nozzoli}, {Oliva}, {Orcinha}, {Palermo}, {Palmonari}, {Paniccia}, {Pashnin},
  {Pauluzzi}, {Pensotti}, {Perrina}, {Phan}, {Picot-Clemente}, {Plyaskin},
  {Pohl}, {Poireau}, {Popkow}, {Quadrani}, {Qi}, {Qin}, {Qu}, {Rancoita},
  {Rapin}, {Conde}, {Rosier-Lees}, {Rozhkov}, {Rozza}, {Sagdeev}, {Solano},
  {Schael}, {Schmidt}, {Schulz von Dratzig}, {Schwering}, {Seo}, {Shan}, {Shi},
  {Siedenburg}, {Song}, {Sun}, {Tacconi}, {Tang}, {Tang}, {Tian}, {Ting},
  {Ting}, {Tomassetti}, {Torsti}, {Urban}, {Vagelli}, {Valente}, {Valtonen},
  {V{\'a}zquez Acosta}, {Vecchi}, {Velasco}, {Vialle}, {Viz{\'a}n}, {Wang},
  {Wang}, {Wang}, {Wang}, {Wang}, {Wang}, {Wei}, {Weng}, {Wu}, {Xiong}, {Xu},
  {Yan}, {Yang}, {Yi}, {Yu}, {Yu}, {Zannoni}, {Zeissler}, {Zhang}, {Zhang},
  {Zhang}, {Zhang}, {Zhao}, {Zheng}, {Zhuang}, {Zhukov}, {Zichichi},
  {Zimmermann}, {Zuccon}, \& {AMS Collaboration}}]{2019PhRvL.122d1102A}
{Aguilar}, M., {Ali Cavasonza}, L., {Ambrosi}, G., {et~al.} 2019{\natexlab{a}},
  \prl, 122, 041102, \dodoi{10.1103/PhysRevLett.122.041102}

\bibitem[{{Aguilar} {et~al.}(2019{\natexlab{b}}){Aguilar}, {Ali Cavasonza},
  {Alpat}, {Ambrosi}, {Arruda}, {Attig}, {Azzarello}, {Bachlechner}, {Barao},
  {Barrau}, {Barrin}, {Bartoloni}, {Basara}, {Ba{\c{s}}e{\v{g}}mez-du Pree},
  {Battiston}, {Becker}, {Behlmann}, {Beischer}, {Berdugo}, {Bertucci},
  {Bindi}, {de Boer}, {Bollweg}, {Borgia}, {Boschini}, {Bourquin}, {Bueno},
  {Burger}, {Burger}, {Cai}, {Capell}, {Caroff}, {Casaus}, {Castellini},
  {Cervelli}, {Chang}, {Chen}, {Chen}, {Chen}, {Cheng}, {Chou}, {Choutko},
  {Chung}, {Clark}, {Coignet}, {Consolandi}, {Contin}, {Corti}, {Crispoltoni},
  {Cui}, {Dadzie}, {Dai}, {Datta}, {Delgado}, {Della Torre}, {Demirk{\"o}z},
  {Derome}, {Di Falco}, {Di Felice}, {Dimiccoli}, {D{\'\i}az}, {von
  Doetinchem}, {Dong}, {Donnini}, {Duranti}, {Egorov}, {Eline}, {Eronen},
  {Feng}, {Fiandrini}, {Fisher}, {Formato}, {Galaktionov},
  {Garc{\'\i}a-L{\'o}pez}, {Gargiulo}, {Gast}, {Gebauer}, {Gervasi},
  {Giovacchini}, {G{\'o}mez-Coral}, {Gong}, {Goy}, {Grabski}, {Grandi},
  {Graziani}, {Guo}, {Haino}, {Han}, {He}, {Heil}, {Hsieh}, {Huang}, {Huang},
  {Incagli}, {Jia}, {Jinchi}, {Kanishev}, {Khiali}, {Kirn}, {Konak}, {Kounina},
  {Kounine}, {Koutsenko}, {Kulemzin}, {La Vacca}, {Laudi}, {Laurenti},
  {Lazzizzera}, {Lebedev}, {Lee}, {Lee}, {Leluc}, {Li}, {Li}, {Li}, {Li},
  {Light}, {Lin}, {Lippert}, {Liu}, {Liu}, {Liu}, {Lu}, {Lu}, {Luebelsmeyer},
  {Luo}, {Luo}, {Luo}, {Lyu}, {Machate}, {Ma{\~n}{\'a}}, {Mar{\'\i}n},
  {Martin}, {Mart{\'\i}nez}, {Masi}, {Maurin}, {Menchaca-Rocha}, {Meng}, {Mo},
  {Molero}, {Mott}, {Mussolin}, {Nelson}, {Ni}, {Nikonov}, {Nozzoli}, {Oliva},
  {Orcinha}, {Palermo}, {Palmonari}, {Paniccia}, {Pashnin}, {Pauluzzi},
  {Pensotti}, {Perrina}, {Phan}, {Picot-Clemente}, {Plyaskin}, {Pohl},
  {Poireau}, {Popkow}, {Quadrani}, {Qi}, {Qin}, {Qu}, {Rancoita}, {Rapin},
  {Conde}, {Rosier-Lees}, {Rozhkov}, {Rozza}, {Sagdeev}, {Solano}, {Schael},
  {Schmidt}, {von Dratzig}, {Schwering}, {Seo}, {Shan}, {Shi}, {Siedenburg},
  {Song}, {Sun}, {Tacconi}, {Tang}, {Tang}, {Tian}, {Ting}, {Ting},
  {Tomassetti}, {Torsti}, {Urban}, {Vagelli}, {Valente}, {Valtonen}, {Acosta},
  {Vecchi}, {Velasco}, {Vialle}, {Viz{\'a}n}, {Wang}, {Wang}, {Wang}, {Wang},
  {Wang}, {Wang}, {Wei}, {Weng}, {Wu}, {Xiong}, {Xu}, {Yan}, {Yang}, {Yi},
  {Yu}, {Yu}, {Zannoni}, {Zeissler}, {Zhang}, {Zhang}, {Zhang}, {Zhang},
  {Zhao}, {Zheng}, {Zhuang}, {Zhukov}, {Zichichi}, {Zimmermann}, {Zuccon}, \&
  {AMS Collaboration}}]{2019PhRvL.122j1101A}
{Aguilar}, M., {Ali Cavasonza}, L., {Alpat}, B., {et~al.} 2019{\natexlab{b}},
  \prl, 122, 101101, \dodoi{10.1103/PhysRevLett.122.101101}

\bibitem[{{Aguilar} {et~al.}(2020){Aguilar}, {Ali Cavasonza}, {Ambrosi},
  {Arruda}, {Attig}, {Barao}, {Barrin}, {Bartoloni}, {Ba{\c{s}}e{\v{g}}mez-du
  Pree}, {Battiston}, {Becker}, {Behlmann}, {Beischer}, {Berdugo}, {Bertucci},
  {Bindi}, {de Boer}, {Bollweg}, {Borgia}, {Boschini}, {Bourquin}, {Bueno},
  {Burger}, {Burger}, {Burmeister}, {Cai}, {Capell}, {Casaus}, {Castellini},
  {Cervelli}, {Chang}, {Chen}, {Chen}, {Chen}, {Cheng}, {Chou}, {Chouridou},
  {Choutko}, {Chung}, {Clark}, {Coignet}, {Consolandi}, {Contin}, {Corti},
  {Cui}, {Dadzie}, {Dai}, {Delgado}, {Della Torre}, {Demirk{\"o}z}, {Derome},
  {Di Falco}, {Di Felice}, {D{\'\i}az}, {Dimiccoli}, {von Doetinchem}, {Dong},
  {Donnini}, {Duranti}, {Egorov}, {Eline}, {Feng}, {Fiandrini}, {Fisher},
  {Formato}, {Freeman}, {Galaktionov}, {G{\'a}mez}, {Garc{\'\i}a-L{\'o}pez},
  {Gargiulo}, {Gast}, {Gebauer}, {Gervasi}, {Giovacchini}, {G{\'o}mez-Coral},
  {Gong}, {Goy}, {Grabski}, {Grandi}, {Graziani}, {Guo}, {Haino}, {Han},
  {Hashmani}, {He}, {Heber}, {Hsieh}, {Hu}, {Huang}, {Incagli}, {Jang}, {Jia},
  {Jinchi}, {Kanishev}, {Khiali}, {Kim}, {Kirn}, {Konyushikhin}, {Kounina},
  {Kounine}, {Koutsenko}, {Kuhlman}, {Kulemzin}, {La Vacca}, {Laudi},
  {Laurenti}, {Lazzizzera}, {Lebedev}, {Lee}, {Lee}, {Li}, {Li}, {Li}, {Li},
  {Li}, {Li}, {Light}, {Lin}, {Lippert}, {Liu}, {Lu}, {Lu}, {Luebelsmeyer},
  {Luo}, {Lyu}, {Machate}, {Ma{\~n}{\'a}}, {Mar{\'\i}n}, {Marquardt}, {Martin},
  {Mart{\'\i}nez}, {Masi}, {Maurin}, {Menchaca-Rocha}, {Meng}, {Mo}, {Molero},
  {Mott}, {Mussolin}, {Ni}, {Nikonov}, {Nozzoli}, {Oliva}, {Orcinha},
  {Palermo}, {Palmonari}, {Paniccia}, {Pashnin}, {Pauluzzi}, {Pensotti},
  {Phan}, {Piandani}, {Plyaskin}, {Poluianov}, {Qi}, {Qin}, {Qu}, {Quadrani},
  {Rancoita}, {Rapin}, {Reina Conde}, {Rosier-Lees}, {Rozhkov}, {Rozza},
  {Sagdeev}, {Schael}, {Schmidt}, {Schulz von Dratzig}, {Schwering}, {Seo},
  {Shan}, {Shi}, {Siedenburg}, {Solano}, {Sonnabend}, {Song}, {Sun}, {Sun},
  {Tacconi}, {Tang}, {Tang}, {Tian}, {Ting}, {Ting}, {Tomassetti}, {Torsti},
  {T{\"u}ys{\"u}z}, {Urban}, {Usoskin}, {Vagelli}, {Vainio}, {Valente},
  {Valtonen}, {V{\'a}zquez Acosta}, {Vecchi}, {Velasco}, {Vialle}, {Wallmann},
  {Wang}, {Wang}, {Wang}, {Wang}, {Wang}, {Wang}, {Wei}, {Weng}, {Wu}, {Xiong},
  {Xu}, {Yan}, {Yang}, {Yi}, {Yu}, {Yu}, {Zannoni}, {Zhang}, {Zhang}, {Zhang},
  {Zhang}, {Zhang}, {Zhao}, {Zheng}, {Zhuang}, {Zhukov}, {Zichichi},
  {Zimmermann}, {Zuccon}, \& {AMS Collaboration}}]{2020PhRvL.124u1102A}
{Aguilar}, M., {Ali Cavasonza}, L., {Ambrosi}, G., {et~al.} 2020, \prl, 124,
  211102, \dodoi{10.1103/PhysRevLett.124.211102}

\bibitem[{{Aguilar} {et~al.}(2021{\natexlab{a}}){Aguilar}, {Cavasonza},
  {Alpat}, {Ambrosi}, {Arruda}, {Attig}, {Barao}, {Barrin}, {Bartoloni},
  {Ba{\c{s}}e{\v{g}}mez-du Pree}, {Battiston}, {Behlmann}, {Beranek},
  {Berdugo}, {Bertucci}, {Bindi}, {Bollweg}, {Borgia}, {Boschini}, {Bourquin},
  {Bueno}, {Burger}, {Burger}, {Burmeister}, {Cai}, {Capell}, {Casaus},
  {Castellini}, {Cervelli}, {Chang}, {Chen}, {Chen}, {Chen}, {Chen}, {Cheng},
  {Chou}, {Chouridou}, {Choutko}, {Chung}, {Clark}, {Coignet}, {Consolandi},
  {Contin}, {Corti}, {Cui}, {Dadzie}, {Delgado}, {Della Torre}, {Demirk{\"o}z},
  {Derome}, {Di Falco}, {Di Felice}, {D{\'\i}az}, {Dimiccoli}, {von
  Doetinchem}, {Dong}, {Donnini}, {Duranti}, {Egorov}, {Eline}, {Feng},
  {Fiandrini}, {Fisher}, {Formato}, {Freeman}, {G{\'a}mez},
  {Garc{\'\i}a-L{\'o}pez}, {Gargiulo}, {Gast}, {Gervasi}, {Giovacchini},
  {G{\'o}mez-Coral}, {Gong}, {Goy}, {Grabski}, {Grandi}, {Graziani}, {Haino},
  {Han}, {Hashmani}, {He}, {Heber}, {Hsieh}, {Hu}, {Incagli}, {Jang}, {Jia},
  {Jinchi}, {Khiali}, {Kim}, {Kirn}, {Konyushikhin}, {Kounina}, {Kounine},
  {Koutsenko}, {Krasnopevtsev}, {Kuhlman}, {Kulemzin}, {La Vacca}, {Laudi},
  {Laurenti}, {Lazzizzera}, {Lebedev}, {Lee}, {Lee}, {Li}, {Li}, {Li}, {Li},
  {Li}, {Li}, {Liang}, {Light}, {Lin}, {Lippert}, {Liu}, {Liu}, {Lu}, {Lu},
  {Luebelsmeyer}, {Luo}, {Luo}, {Machate}, {Ma{\~n}{\'a}}, {Mar{\'\i}n},
  {Marquardt}, {Martin}, {Mart{\'\i}nez}, {Masi}, {Maurin}, {Medvedeva},
  {Menchaca-Rocha}, {Meng}, {Mikhailov}, {Molero}, {Mott}, {Mussolin},
  {Negrete}, {Nikonov}, {Nozzoli}, {Oliva}, {Orcinha}, {Palermo}, {Palmonari},
  {Paniccia}, {Pashnin}, {Pauluzzi}, {Pensotti}, {Phan}, {Plyaskin}, {Pohl},
  {Poluianov}, {Qin}, {Qu}, {Quadrani}, {Rancoita}, {Rapin}, {Conde}, {Robyn},
  {Rosier-Lees}, {Rozhkov}, {Rozza}, {Sagdeev}, {Schael}, {von Dratzig},
  {Schwering}, {Seo}, {Shakfa}, {Shan}, {Siedenburg}, {Solano}, {Song}, {Song},
  {Sonnabend}, {Strigari}, {Su}, {Sun}, {Sun}, {Tacconi}, {Tang}, {Tang},
  {Tian}, {Ting}, {Ting}, {Tomassetti}, {Torsti}, {T{\"u}ys{\"u}z}, {Urban},
  {Usoskin}, {Vagelli}, {Vainio}, {Valencia-Otero}, {Valente}, {Valtonen},
  {V{\'a}zquez Acosta}, {Vecchi}, {Velasco}, {Vialle}, {Wang}, {Wang}, {Wang},
  {Wang}, {Wang}, {Wang}, {Wang}, {Wang}, {Wang}, {Wei}, {Weng}, {Wu}, {Xiong},
  {Xu}, {Yan}, {Yang}, {Yashin}, {Yi}, {Yu}, {Yu}, {Zannoni}, {Zhang}, {Zhang},
  {Zhang}, {Zhang}, {Zhang}, {Zhao}, {Zheng}, {Zheng}, {Zhuang}, {Zhukov},
  {Zichichi}, {Zuccon}, \& {AMS Collaboration}}]{2021PhRvL.127b1101A}
{Aguilar}, M., {Cavasonza}, L.~A., {Alpat}, B., {et~al.} 2021{\natexlab{a}},
  \prl, 127, 021101, \dodoi{10.1103/PhysRevLett.127.021101}

\bibitem[{{Aguilar} {et~al.}(2021{\natexlab{b}}){Aguilar}, {Cavasonza},
  {Allen}, {Alpat}, {Ambrosi}, {Arruda}, {Attig}, {Barao}, {Barrin},
  {Bartoloni}, {Ba{\c{s}}e{\v{g}}mez-du Pree}, {Battiston}, {Behlmann},
  {Beischer}, {Berdugo}, {Bertucci}, {Bindi}, {de Boer}, {Bollweg}, {Borgia},
  {Boschini}, {Bourquin}, {Bueno}, {Burger}, {Burger}, {Burmeister}, {Cai},
  {Capell}, {Casaus}, {Castellini}, {Cervelli}, {Chang}, {Chen}, {Chen},
  {Chen}, {Chen}, {Cheng}, {Chou}, {Chouridou}, {Choutko}, {Chung}, {Clark},
  {Coignet}, {Consolandi}, {Contin}, {Corti}, {Cui}, {Dadzie}, {Delgado},
  {Della Torre}, {Demirk{\"o}z}, {Derome}, {Di Falco}, {Di Felice},
  {D{\'\i}az}, {Dimiccoli}, {von Doetinchem}, {Dong}, {Donnini}, {Duranti},
  {Egorov}, {Eline}, {Feng}, {Fiandrini}, {Fisher}, {Formato}, {Freeman},
  {Galaktionov}, {G{\'a}mez}, {Garc{\'\i}a-L{\'o}pez}, {Gargiulo}, {Gast},
  {Gervasi}, {Giovacchini}, {G{\'o}mez-Coral}, {Gong}, {Goy}, {Grabski},
  {Grandi}, {Graziani}, {Haino}, {Han}, {Hashmani}, {He}, {Heber}, {Hsieh},
  {Hu}, {Incagli}, {Jang}, {Jia}, {Jinchi}, {Kanishev}, {Khiali}, {Kim},
  {Kirn}, {Konyushikhin}, {Kounina}, {Kounine}, {Koutsenko}, {Kuhlman},
  {Kulemzin}, {La Vacca}, {Laudi}, {Laurenti}, {Lazzizzera}, {Lebedev}, {Lee},
  {Lee}, {Li}, {Li}, {Li}, {Li}, {Li}, {Li}, {Liang}, {Light}, {Lin},
  {Lippert}, {Liu}, {Liu}, {Lu}, {Lu}, {Luebelsmeyer}, {Luo}, {Luo}, {Lyu},
  {Machate}, {Ma{\~n}{\'a}}, {Mar{\'\i}n}, {Marquardt}, {Martin},
  {Mart{\'\i}nez}, {Masi}, {Maurin}, {Menchaca-Rocha}, {Meng}, {Mikhailov},
  {Mo}, {Molero}, {Mott}, {Mussolin}, {Negrete}, {Nikonov}, {Nozzoli}, {Oliva},
  {Orcinha}, {Palermo}, {Palmonari}, {Paniccia}, {Pashnin}, {Pauluzzi},
  {Pensotti}, {Phan}, {Piandani}, {Plyaskin}, {Poluianov}, {Qin}, {Qu},
  {Quadrani}, {Rancoita}, {Rapin}, {Conde}, {Robyn}, {Rosier-Lees}, {Rozhkov},
  {Rozza}, {Sagdeev}, {Schael}, {von Dratzig}, {Schwering}, {Seo}, {Shakfa},
  {Shan}, {Siedenburg}, {Solano}, {Song}, {Song}, {Sonnabend}, {Strigari},
  {Su}, {Sun}, {Sun}, {Tacconi}, {Tang}, {Tang}, {Tian}, {Ting}, {Ting},
  {Tomassetti}, {Torsti}, {T{\"u}ys{\"u}z}, {Urban}, {Usoskin}, {Vagelli},
  {Vainio}, {Valencia-Otero}, {Valente}, {Valtonen}, {V{\'a}zquez Acosta},
  {Vecchi}, {Velasco}, {Vialle}, {Wang}, {Wang}, {Wang}, {Wang}, {Wang},
  {Wang}, {Wang}, {Wang}, {Wang}, {Wei}, {Weng}, {Wu}, {Xiong}, {Xu}, {Yan},
  {Yang}, {Yashin}, {Yi}, {Yu}, {Yu}, {Zannoni}, {Zhang}, {Zhang}, {Zhang},
  {Zhang}, {Zhang}, {Zhao}, {Zheng}, {Zheng}, {Zhuang}, {Zhukov}, {Zichichi},
  {Zimmermann}, {Zuccon}, \& {AMS Collaboration}}]{2021PhRvL.126d1104A}
{Aguilar}, M., {Cavasonza}, L.~A., {Allen}, M.~S., {et~al.} 2021{\natexlab{b}},
  \prl, 126, 041104, \dodoi{10.1103/PhysRevLett.126.041104}

\bibitem[{{Aguilar} {et~al.}(2021{\natexlab{c}}){Aguilar}, {Ali Cavasonza},
  {Ambrosi}, {Arruda}, {Attig}, {Barao}, {Barrin}, {Bartoloni},
  {Ba{\c{s}}e{\u{g}}mez-du Pree}, {Bates}, {Battiston}, {Behlmann}, {Beischer},
  {Berdugo}, {Bertucci}, {Bindi}, {de Boer}, {Bollweg}, {Borgia}, {Boschini},
  {Bourquin}, {Bueno}, {Burger}, {Burger}, {Burmeister}, {Cai}, {Capell},
  {Casaus}, {Castellini}, {Cervelli}, {Chang}, {Chen}, {Chen}, {Chen}, {Cheng},
  {Chou}, {Chouridou}, {Choutko}, {Chung}, {Clark}, {Coignet}, {Consolandi},
  {Contin}, {Corti}, {Cui}, {Dadzie}, {Dai}, {Delgado}, {Della Torre},
  {Demirk{\"o}z}, {Derome}, {Di Falco}, {Di Felice}, {D{\'\i}az}, {Dimiccoli},
  {von Doetinchem}, {Dong}, {Donnini}, {Duranti}, {Egorov}, {Eline}, {Feng},
  {Fiandrini}, {Fisher}, {Formato}, {Freeman}, {Galaktionov}, {G{\'a}mez},
  {Garc{\'\i}a-L{\'o}pez}, {Gargiulo}, {Gast}, {Gebauer}, {Gervasi},
  {Giovacchini}, {G{\'o}mez-Coral}, {Gong}, {Goy}, {Grabski}, {Grandi},
  {Graziani}, {Guo}, {Haino}, {Han}, {Hashmani}, {He}, {Heber}, {Hsieh}, {Hu},
  {Huang}, {Hungerford}, {Incagli}, {Jang}, {Jia}, {Jinchi}, {Kanishev},
  {Khiali}, {Kim}, {Kirn}, {Konyushikhin}, {Kounina}, {Kounine}, {Koutsenko},
  {Kuhlman}, {Kulemzin}, {La Vacca}, {Laudi}, {Laurenti}, {Lazzizzera},
  {Lebedev}, {Lee}, {Lee}, {Leluc}, {Li}, {Li}, {Li}, {Li}, {Li}, {Li},
  {Light}, {Lin}, {Lippert}, {Liu}, {Lu}, {Lu}, {Luebelsmeyer}, {Luo}, {Lyu},
  {Machate}, {Ma{\~n}{\'a}}, {Mar{\'\i}n}, {Marquardt}, {Martin},
  {Mart{\'\i}nez}, {Masi}, {Maurin}, {Menchaca-Rocha}, {Meng}, {Mo}, {Molero},
  {Mott}, {Mussolin}, {Ni}, {Nikonov}, {Nozzoli}, {Oliva}, {Orcinha},
  {Palermo}, {Palmonari}, {Paniccia}, {Pashnin}, {Pauluzzi}, {Pensotti},
  {Phan}, {Plyaskin}, {Pohl}, {Porter}, {Qi}, {Qin}, {Qu}, {Quadrani},
  {Rancoita}, {Rapin}, {Reina Conde}, {Rosier-Lees}, {Rozhkov}, {Rozza},
  {Sagdeev}, {Schael}, {Schmidt}, {Schulz von Dratzig}, {Schwering}, {Seo},
  {Shan}, {Shi}, {Siedenburg}, {Solano}, {Song}, {Sonnabend}, {Sun}, {Sun},
  {Tacconi}, {Tang}, {Tang}, {Tian}, {Ting}, {Ting}, {Tomassetti}, {Torsti},
  {T{\"u}ys{\"u}z}, {Urban}, {Usoskin}, {Vagelli}, {Vainio}, {Valente},
  {Valtonen}, {V{\'a}zquez Acosta}, {Vecchi}, {Velasco}, {Vialle}, {Wang},
  {Wang}, {Wang}, {Wang}, {Wang}, {Wang}, {Wei}, {Weng}, {Wu}, {Xiong}, {Xu},
  {Yan}, {Yang}, {Yi}, {Yu}, {Yu}, {Zannoni}, {Zhang}, {Zhang}, {Zhang},
  {Zhang}, {Zhang}, {Zhao}, {Zheng}, {Zhuang}, {Zhukov}, {Zichichi},
  {Zimmermann}, {Zuccon}, \& {AMS Collaboration}}]{2021PhR...894....1A}
{Aguilar}, M., {Ali Cavasonza}, L., {Ambrosi}, G., {et~al.} 2021{\natexlab{c}},
  \physrep, 894, 1, \dodoi{10.1016/j.physrep.2020.09.003}

\bibitem[{{Aharonian} {et~al.}(2008{\natexlab{a}}){Aharonian}, {Akhperjanian},
  {Barres de Almeida}, {Bazer-Bachi}, {Becherini}, {Behera}, {Benbow},
  {Bernl{\"o}hr}, {Boisson}, {Bochow}, {Borrel}, {Braun}, {Brion}, {Brucker},
  {Brun}, {B{\"u}hler}, {Bulik}, {B{\"u}sching}, {Boutelier}, {Carrigan},
  {Chadwick}, {Charbonnier}, {Chaves}, {Cheesebrough}, {Chounet}, {Clapson},
  {Coignet}, {Costamante}, {Dalton}, {Degrange}, {Deil}, {Dickinson},
  {Djannati-Ata{\"i}}, {Domainko}, {Drury}, {Dubois}, {Dubus}, {Dyks}, {Dyrda},
  {Egberts}, {Emmanoulopoulos}, {Espigat}, {Farnier}, {Feinstein}, {Fiasson},
  {F{\"o}rster}, {Fontaine}, {F{\"u}{\ss}ling}, {Gabici}, {Gallant},
  {G{\'e}rard}, {Giebels}, {Glicenstein}, {Gl{\"u}ck}, {Goret}, {Vivier},
  {V{\"o}lk}, {Volpe}, {Wagner}, {Ward}, {Zdziarski}, \&
  {Zech}}]{2008PhRvL.101z1104A}
{Aharonian}, F., {Akhperjanian}, A.~G., {Barres de Almeida}, U., {et~al.}
  2008{\natexlab{a}}, Physical Review Letters, 101, 261104,
  \dodoi{10.1103/PhysRevLett.101.261104}

\bibitem[{{Aharonian} {et~al.}(2008{\natexlab{b}}){Aharonian}, {Akhperjanian},
  {Bazer-Bachi}, {Behera}, {Beilicke}, {Benbow}, {Berge}, {Bernl{\"o}hr},
  {Boisson}, {Bolz}, {Borrel}, {Braun}, {Brion}, {Brown}, {B{\"u}hler},
  {Bulik}, {B{\"u}sching}, {Boutelier}, {Carrigan}, {Chadwick}, {Chounet},
  {Clapson}, {Coignet}, {Cornils}, {Costamante}, {Degrange}, {Dickinson},
  {Djannati-Ata{\"\i}}, {Domainko}, {O'C. Drury}, {Dubus}, {Dyks}, {Egberts},
  {Emmanoulopoulos}, {Espigat}, {Farnier}, {Feinstein}, {Fiasson},
  {F{\"o}rster}, {Fontaine}, {Fukui}, {Funk}, {Funk}, {F{\"u}{\ss}ling},
  {Gallant}, {Giebels}, {Glicenstein}, {Gl{\"u}ck}, {Goret}, {Hadjichristidis},
  {Hauser}, {Hauser}, {Heinzelmann}, {Henri}, {Hermann}, {Hinton}, {Hoffmann},
  {Hofmann}, {Holleran}, {Hoppe}, {Horns}, {Jacholkowska}, {de Jager},
  {Kendziorra}, {Kerschhaggl}, {Kh{\'e}lifi}, {Komin}, {Kosack}, {Lamanna},
  {Latham}, {Le Gallou}, {Lemi{\`e}re}, {Lemoine-Goumard}, {Lenain}, {Lohse},
  {Martin}, {Martineau-Huynh}, {Marcowith}, {Masterson}, {Maurin}, {McComb},
  {Moderski}, {Moriguchi}, {Moulin}, {de Naurois}, {Nedbal}, {Nolan}, {Olive},
  {Orford}, {Osborne}, {Ostrowski}, {Panter}, {Pedaletti}, {Pelletier},
  {Petrucci}, {Pita}, {P{\"u}hlhofer}, {Punch}, {Ranchon}, {Raubenheimer},
  {Raue}, {Rayner}, {Reimer}, {Renaud}, {Ripken}, {Rob}, {Rolland},
  {Rosier-Lees}, {Rowell}, {Rudak}, {Ruppel}, {Sahakian}, {Santangelo},
  {Saug{\'e}}, {Schlenker}, {Schlickeiser}, {Schr{\"o}der}, {Schwanke},
  {Schwarzburg}, {Schwemmer}, {Shalchi}, {Sol}, {Spangler}, {Stawarz},
  {Steenkamp}, {Stegmann}, {Superina}, {Takeuchi}, {Tam}, {Tavernet},
  {Terrier}, {van Eldik}, {Vasileiadis}, {Venter}, {Vialle}, {Vincent},
  {Vivier}, {V{\"o}lk}, {Volpe}, {Wagner}, \& {Ward}}]{2008A&A...481..401A}
{Aharonian}, F., {Akhperjanian}, A.~G., {Bazer-Bachi}, A.~R., {et~al.}
  2008{\natexlab{b}}, \aap, 481, 401, \dodoi{10.1051/0004-6361:20077765}

\bibitem[{{Aharonian} {et~al.}(2009){Aharonian}, {Akhperjanian}, {Anton},
  {Barres de Almeida}, {Bazer-Bachi}, {Becherini}, {Behera}, {Bernl{\"o}hr},
  {Bochow}, {Boisson}, {Bolmont}, {Borrel}, {Brucker}, {Brun}, {Brun},
  {B{\"u}hler}, {Bulik}, {B{\"u}sching}, {Boutelier}, {Chadwick},
  {Charbonnier}, {Chaves}, {Cheesebrough}, {Chounet}, {Clapson}, {Coignet},
  {Dalton}, {Daniel}, {Davids}, {Degrange}, {Deil}, {Dickinson},
  {Djannati-Ata{\"i}}, {Domainko}, {O'C.~Drury}, {Dubois}, {Dubus}, {Dyks},
  {Dyrda}, {Egberts}, {Emmanoulopoulos}, {Espigat}, {Farnier}, {Feinstein},
  {Fiasson}, {F{\"o}rster}, {Fontaine}, {F{\"u}{\ss}ling}, {Gabici}, {Gallant},
  {G{\'e}rard}, {Gerbig}, {Giebels}, {Glicenstein}, {Gl{\"u}ck}, {Goret},
  {G{\"o}ring}, {Hauser}, {Hauser}, {Heinz}, {Heinzelmann}, {Henri}, {Hermann},
  {Hinton}, {Hoffmann}, {Hofmann}, {Holleran}, {Hoppe}, {Horns},
  {Jacholkowska}, {de Jager}, {Jahn}, {Jung}, {Katarzy{\'n}ski}, {Katz},
  {Kaufmann}, {Kendziorra}, {Kerschhaggl}, {Khangulyan}, {Kh{\'e}lifi},
  {Keogh}, {Klu{\'z}niak}, {Kneiske}, {Komin}, {Kosack}, {Kossakowski},
  {Lamanna}, {Lenain}, {Lohse}, {Marandon}, {Martin}, {Martineau-Huynh},
  {Marcowith}, {Masbou}, {Maurin}, {McComb}, {Medina}, {Moderski}, {Moulin},
  {Naumann-Godo}, {de Naurois}, {Nedbal}, {Nekrassov}, {Nicholas}, {Niemiec},
  {Nolan}, {Ohm}, {Olive}, {de O{\~n}a Wilhelmi}, {Orford}, {Ostrowski},
  {Panter}, {Paz Arribas}, {Pedaletti}, {Pelletier}, {Petrucci}, {Pita},
  {P{\"u}hlhofer}, {Punch}, {Quirrenbach}, {Raubenheimer}, {Raue}, {Rayner},
  {Reimer}, {Renaud}, {Rieger}, {Ripken}, {Rob}, {Rosier-Lees}, {Rowell},
  {Rudak}, {Rulten}, {Ruppel}, {Sahakian}, {Santangelo}, {Schlickeiser},
  {Sch{\"o}ck}, {Schr{\"o}der}, {Schwanke}, {Schwarzburg}, {Schwemmer},
  {Shalchi}, {Sikora}, {Skilton}, {Sol}, {Spangler}, {Stawarz}, {Steenkamp},
  {Stegmann}, {Stinzing}, {Superina}, {Szostek}, {Tam}, {Tavernet}, {Terrier},
  {Tibolla}, {Tluczykont}, {van Eldik}, {Vasileiadis}, {Venter}, {Venter},
  {Vialle}, {Vincent}, {Vivier}, {V{\"o}lk}, {Volpe}, {Wagner}, {Ward},
  {Zdziarski}, \& {Zech}}]{2009A&A...508..561A}
{Aharonian}, F., {Akhperjanian}, A.~G., {Anton}, G., {et~al.} 2009, \aap, 508,
  561, \dodoi{10.1051/0004-6361/200913323}

\bibitem[{{Aharonian} {et~al.}(2021){Aharonian}, {An}, {Axikegu}, {Bai}, {Bao},
  {Bastieri}, {Bi}, {Bi}, {Cai}, {Cai}, {Cao}, {Cao}, {Chang}, {Chang},
  {Chang}, {Chen}, {Chen}, {Chen}, {Chen}, {Chen}, {Chen}, {Chen}, {Chen},
  {Chen}, {Chen}, {Chen}, {Chen}, {Chen}, {Cheng}, {Cheng}, {Cui}, {Cui},
  {Cui}, {Dai}, {Dai}, {Dai}, {Danzengluobu}, {Della Volpe}, {D'Ettorre
  Piazzoli}, {Dong}, {Fan}, {Fan}, {Fan}, {Fang}, {Fang}, {Feng}, {Feng},
  {Feng}, {Feng}, {Gao}, {Gao}, {Gao}, {Gao}, {Ge}, {Geng}, {Gong}, {Gou},
  {Gu}, {Guo}, {Guo}, {Guo}, {Guo}, {Han}, {He}, {He}, {He}, {He}, {He}, {He},
  {Heller}, {Hor}, {Hou}, {Hou}, {Hu}, {Hu}, {Hu}, {Hu}, {Huang}, {Huang},
  {Huang}, {Huang}, {Huang}, {Ji}, {Ji}, {Jia}, {Jiang}, {Jiang}, {Jin},
  {Kuleshov}, {Levochkin}, {Li}, {Li}, {Li}, {Li}, {Li}, {Li}, {Li}, {Li},
  {Li}, {Li}, {Li}, {Li}, {Li}, {Li}, {Li}, {Li}, {Li}, {Liang}, {Liang},
  {Lin}, {Liu}, {Liu}, {Liu}, {Liu}, {Liu}, {Liu}, {Liu}, {Liu}, {Liu}, {Liu},
  {Liu}, {Liu}, {Liu}, {Liu}, {Liu}, {Long}, {Lu}, {Lv}, {Ma}, {Ma}, {Ma},
  {Mao}, {Masood}, {Mitthumsiri}, {Montaruli}, {Nan}, {Pang},
  {Pattarakijwanich}, {Pei}, {Qi}, {Ruffolo}, {Rulev}, {S{\'a}iz}, {Shao},
  {Shchegolev}, {Sheng}, {Shi}, {Song}, {Stenkin}, {Stepanov}, {Sun}, {Sun},
  {Sun}, {Tam}, {Tang}, {Tian}, {Wang}, {Wang}, {Wang}, {Wang}, {Wang}, {Wang},
  {Wang}, {Wang}, {Wang}, {Wang}, {Wang}, {Wang}, {Wang}, {Wang}, {Wang},
  {Wang}, {Wang}, {Wang}, {Wang}, {Wang}, {Wang}, {Wei}, {Wei}, {Wei}, {Wen},
  {Wu}, {Wu}, {Wu}, {Wu}, {Wu}, {Xi}, {Xia}, {Xia}, {Xiang}, {Xiao}, {Xiao},
  {Xin}, {Xin}, {Xing}, {Xu}, {Xu}, {Xue}, {Yan}, {Yang}, {Yang}, {Yang},
  {Yang}, {Yang}, {Yang}, {Yang}, {Yao}, {Yao}, {Ye}, {Yin}, {Yin}, {You},
  {You}, {Yu}, {Yuan}, {Zeng}, {Zeng}, {Zeng}, {Zeng}, {Zha}, {Zhai}, {Zhang},
  {Zhang}, {Zhang}, {Zhang}, {Zhang}, {Zhang}, {Zhang}, {Zhang}, {Zhang},
  {Zhang}, {Zhang}, {Zhang}, {Zhang}, {Zhang}, {Zhang}, {Zhang}, {Zhang},
  {Zhang}, {Zhang}, {Zhao}, {Zhao}, {Zhao}, {Zhao}, {Zhao}, {Zheng}, {Zheng},
  {Zhou}, {Zhou}, {Zhou}, {Zhou}, {Zhou}, {Zhou}, {Zhu}, {Zhu}, {Zhu}, {Zhu},
  {Zuo}, {LHAASO Collaboration}, \& {Huang}}]{2021PhRvL.126x1103A}
{Aharonian}, F., {An}, Q., {Axikegu}, Bai, L.~X., {et~al.} 2021, \prl, 126,
  241103, \dodoi{10.1103/PhysRevLett.126.241103}

\bibitem[{{Ahn} {et~al.}(2010){Ahn}, {Allison}, {Bagliesi}, {Beatty},
  {Bigongiari}, {Childers}, {Conklin}, {Coutu}, {DuVernois}, {Ganel}, {Han},
  {Jeon}, {Kim}, {Lee}, {Lutz}, {Maestro}, {Malinin}, {Marrocchesi}, {Minnick},
  {Mognet}, {Nam}, {Nam}, {Nutter}, {Park}, {Park}, {Seo}, {Sina}, {Wu},
  {Yang}, {Yoon}, {Zei}, \& {Zinn}}]{2010ApJ...714L..89A}
{Ahn}, H.~S., {Allison}, P., {Bagliesi}, M.~G., {et~al.} 2010, \apjl, 714, L89,
  \dodoi{10.1088/2041-8205/714/1/L89}

\bibitem[{{Alekseenko} {et~al.}(2009){Alekseenko}, {Cherniaev}, {Djappuev},
  {Kudjaev}, {Michailova}, {Stenkin}, {Stepanov}, \&
  {Volchenko}}]{2009NuPhS.196..179A}
{Alekseenko}, V.~V., {Cherniaev}, A.~B., {Djappuev}, D.~D., {et~al.} 2009,
  Nuclear Physics B Proceedings Supplements, 196, 179,
  \dodoi{10.1016/j.nuclphysbps.2009.09.032}

\bibitem[{{Alemanno} {et~al.}(2021){Alemanno}, {An}, {Azzarello}, {Barbato},
  {Bernardini}, {Bi}, {Cai}, {Catanzani}, {Chang}, {Chen}, {Chen}, {Chen},
  {Cui}, {Cui}, {Cui}, {Dai}, {D'Amone}, {de Benedittis}, {de Mitri}, {de
  Palma}, {Deliyergiyev}, {di Santo}, {Dong}, {Dong}, {Donvito}, {Droz},
  {Duan}, {Duan}, {D'Urso}, {Fan}, {Fan}, {Fang}, {Fang}, {Feng}, {Feng},
  {Fusco}, {Gao}, {Gargano}, {Gong}, {Gong}, {Guo}, {Guo}, {Guo}, {Han}, {Hu},
  {Huang}, {Huang}, {Huang}, {Ionica}, {Jiang}, {Kong}, {Kotenko}, {Kyratzis},
  {Lei}, {Li}, {Li}, {Li}, {Li}, {Liang}, {Liu}, {Liu}, {Liu}, {Liu}, {Liu},
  {Liu}, {Loparco}, {Luo}, {Ma}, {Ma}, {Ma}, {Ma}, {Marsella}, {Mazziotta},
  {Mo}, {Niu}, {Pan}, {Parenti}, {Peng}, {Peng}, {Perrina}, {Qiao}, {Rao},
  {Ruina}, {Salinas}, {Shang}, {Shen}, {Shen}, {Shen}, {Silveri}, {Song},
  {Stolpovskiy}, {Su}, {Su}, {Sun}, {Surdo}, {Teng}, {Tykhonov}, {Wang},
  {Wang}, {Wang}, {Wang}, {Wang}, {Wang}, {Wang}, {Wang}, {Wang}, {Wei}, {Wei},
  {Wei}, {Wen}, {Wu}, {Wu}, {Wu}, {Wu}, {Wu}, {Xia}, {Xu}, {Xu}, {Xu}, {Xu},
  {Xue}, {Yang}, {Yang}, {Yang}, {Yao}, {Yu}, {Yuan}, {Yuan}, {Yue}, {Zang},
  {Zhang}, {Zhang}, {Zhang}, {Zhang}, {Zhang}, {Zhang}, {Zhang}, {Zhang},
  {Zhang}, {Zhang}, {Zhao}, {Zhao}, {Zhao}, {Zhou}, {Zhu}, \& {Dampe
  Collaboration}}]{2021PhRvL.126t1102A}
{Alemanno}, F., {An}, Q., {Azzarello}, P., {et~al.} 2021, \prl, 126, 201102,
  \dodoi{10.1103/PhysRevLett.126.201102}

\bibitem[{{Alexeyenko} {et~al.}(1981){Alexeyenko}, {Chudakov}, {Gulieva}, \&
  {Sborschikov}}]{1981ICRC....2..146A}
{Alexeyenko}, V.~V., {Chudakov}, A.~E., {Gulieva}, E.~N., \& {Sborschikov},
  V.~G. 1981, International Cosmic Ray Conference, 2, 146

\bibitem[{{Ambrosio} {et~al.}(2003){Ambrosio}, {Antolini}, {Baldini},
  {Barbarino}, {Barish}, {Battistoni}, {Becherini}, {Bellotti}, {Bemporad},
  {Bernardini}, {Bilokon}, {Bower}, {Brigida}, {Bussino}, {Cafagna},
  {Calicchio}, {Campana}, {Carboni}, {Caruso}, {Cecchini}, {Cei}, {Chiarella},
  {Chiarusi}, {Choudhary}, {Coutu}, {Cozzi}, {de Cataldo}, {Dekhissi}, {de
  Marzo}, {de Mitri}, {Derkaoui}, {de Vincenzi}, {di Credico}, {Erriquez},
  {Favuzzi}, {Forti}, {Fusco}, {Giacomelli}, {Giannini}, {Giglietto},
  {Giorgini}, {Grassi}, {Grillo}, {Gustavino}, {Habig}, {Hanson}, {Heinz},
  {Katsavounidis}, {Katsavounidis}, {Kearns}, {Kim}, {Kyriazopoulou},
  {Lamanna}, {Lane}, {Levin}, {Lipari}, {Longley}, {Longo}, {Loparco},
  {Maaroufi}, {Mancarella}, {Mandrioli}, {Margiotta}, {Marini}, {Martello},
  {Marzari-Chiesa}, {Mazziotta}, {Michael}, {Miller}, {Monacelli}, {Montaruli},
  {Monteno}, {Mufson}, {Musser}, {Nicol{\`o}}, {Nolty}, {Orth}, {Osteria},
  {Palamara}, {Patrizii}, {Pazzi}, {Peck}, {Perrone}, {Petrera}, {Popa},
  {Rain{\`o}}, {Reynoldson}, {Ronga}, {Satriano}, {Scapparone}, {Scholberg},
  {Serra}, {Sioli}, {Sirri}, {Sitta}, {Spinelli}, {Spinetti}, {Spurio},
  {Steinberg}, {Stone}, {Sulak}, {Surdo}, {Tarl{\`e}}, {Togo}, {Vakili},
  {Walter}, \& {Webb}}]{2003PhRvD..67d2002A}
{Ambrosio}, M., {Antolini}, R., {Baldini}, A., {et~al.} 2003, \prd, 67, 042002,
  \dodoi{10.1103/PhysRevD.67.042002}

\bibitem[{{Amenomori} {et~al.}(2005){Amenomori}, {Ayabe}, {Cui},
  {Danzengluobu}, {Ding}, {Ding}, {Feng}, {Feng}, {Gao}, {Geng}, {Guo}, {He},
  {He}, {Hibino}, {Hotta}, {Hu}, {Hu}, {Huang}, {Huang}, {Jia}, {Kajino},
  {Kasahara}, {Katayose}, {Kato}, {Kawata}, {Labaciren}, {Le}, {Li}, {Lu},
  {Lu}, {Meng}, {Mizutani}, {Mu}, {Munakata}, {Nagai}, {Nanjo}, {Nishizawa},
  {Ohnishi}, {Ohta}, {Onuma}, {Ouchi}, {Ozawa}, {Ren}, {Saito}, {Sakata},
  {Sasaki}, {Shibata}, {Shiomi}, {Shirai}, {Sugimoto}, {Takita}, {Tan},
  {Tateyama}, {Torii}, {Tsuchiya}, {Udo}, {Utsugi}, {Wang}, {Wang}, {Wang},
  {Wang}, {Wu}, {Xue}, {Yamamoto}, {Yan}, {Yang}, {Yasue}, {Ye}, {Yu}, {Yuan},
  {Yuda}, {Zhang}, {Zhang}, {Zhang}, {Zhang}, {Zhang}, {Zhang}, {Zhaxisangzhu},
  {Zhou}, \& {Tibet As{$\gamma$} Collaboration}}]{2005ApJ...626L..29A}
{Amenomori}, M., {Ayabe}, S., {Cui}, S.~W., {et~al.} 2005, \apjl, 626, L29,
  \dodoi{10.1086/431582}

\bibitem[{{Amenomori} {et~al.}(2006){Amenomori}, {Ayabe}, {Bi}, {Chen}, {Cui},
  {Danzengluobu}, {Ding}, {Ding}, {Feng}, {Feng}, {Feng}, {Gao}, {Geng}, {Guo},
  {He}, {He}, {Hibino}, {Hotta}, {Hu}, {Hu}, {Huang}, {Huang}, {Jia}, {Kajino},
  {Kasahara}, {Katayose}, {Kato}, {Kawata}, {Labaciren}, {Le}, {Li}, {Li},
  {Lou}, {Lu}, {Lu}, {Meng}, {Mizutani}, {Mu}, {Munakata}, {Nagai}, {Nanjo},
  {Nishizawa}, {Ohnishi}, {Ohta}, {Onuma}, {Ouchi}, {Ozawa}, {Ren}, {Saito},
  {Saito}, {Sakata}, {Sako}, {Sasaki}, {Shibata}, {Shiomi}, {Shirai},
  {Sugimoto}, {Takita}, {Tan}, {Tateyama}, {Torii}, {Tsuchiya}, {Udo}, {Wang},
  {Wang}, {Wang}, {Wang}, {Wu}, {Xue}, {Yamamoto}, {Yan}, {Yang}, {Yasue},
  {Ye}, {Yu}, {Yuan}, {Yuda}, {Zhang}, {Zhang}, {Zhang}, {Zhang}, {Zhang},
  {Zhang}, {Zhaxisangzhu}, {Zhou}, \& {Tibet AS{\ensuremath{\gamma}}
  Collaboration}}]{2006Sci...314..439A}
{Amenomori}, M., {Ayabe}, S., {Bi}, X.~J., {et~al.} 2006, Science, 314, 439,
  \dodoi{10.1126/science.1131702}

\bibitem[{{Amenomori} {et~al.}(2008){Amenomori}, {Bi}, {Chen}, {Cui},
  {Danzengluobu}, {Ding}, {Ding}, {Fan}, {Feng}, {Feng}, {Feng}, {Gao}, {Geng},
  {Guo}, {He}, {He}, {Hibino}, {Hotta}, {Hu}, {Hu}, {Huang}, {Huang}, {Jia},
  {Kajino}, {Kasahara}, {Katayose}, {Kato}, {Kawata}, {Labaciren}, {Le}, {Li},
  {Li}, {Lou}, {Lu}, {Lu}, {Meng}, {Mizutani}, {Mu}, {Munakata}, {Nagai},
  {Nanjo}, {Nishizawa}, {Ohnishi}, {Ohta}, {Onuma}, {Ouchi}, {Ozawa}, {Ren},
  {Saito}, {Saito}, {Sakata}, {Sako}, {Shibata}, {Shiomi}, {Shirai},
  {Sugimoto}, {Takita}, {Tan}, {Tateyama}, {Torii}, {Tsuchiya}, {Udo}, {Wang},
  {Wang}, {Wang}, {Wang}, {Wang}, {Wu}, {Xue}, {Yamamoto}, {Yan}, {Yang},
  {Yasue}, {Ye}, {Yu}, {Yuan}, {Yuda}, {Zhang}, {Zhang}, {Zhang}, {Zhang},
  {Zhang}, {Zhang}, {Zhaxisangzhu}, {Zhou}, \& {Tibet AS{\ensuremath{\gamma}}
  Collaboration}}]{2008ApJ...678.1165A}
{Amenomori}, M., {Bi}, X.~J., {Chen}, D., {et~al.} 2008, \apj, 678, 1165,
  \dodoi{10.1086/529514}

\bibitem[{{Amenomori} {et~al.}(2015){Amenomori}, {Bi}, {Chen}, {Chen}, {Chen},
  {Cui}, {Danzengluobu}, {Ding}, {Feng}, {Feng}, {Feng}, {Gou}, {Guo}, {He},
  {He}, {Hibino}, {Hotta}, {Hu}, {Hu}, {Huang}, {Jia}, {Jiang}, {Kajino},
  {Kasahara}, {Katayose}, {Kato}, {Kawata}, {Kozai}, {Labaciren}, {Le}, {Li},
  {Li}, {Li}, {Liu}, {Liu}, {Liu}, {Lu}, {Meng}, {Miyazaki}, {Mizutani},
  {Munakata}, {Nakajima}, {Nakamura}, {Nanjo}, {Nishizawa}, {Niwa}, {Ohnishi},
  {Ohta}, {Ozawa}, {Qian}, {Qu}, {Saito}, {Saito}, {Sakata}, {Sako}, {Shao},
  {Shibata}, {Shiomi}, {Shirai}, {Sugimoto}, {Takita}, {Tan}, {Tateyama},
  {Torii}, {Tsuchiya}, {Udo}, {Wang}, {Wu}, {Xue}, {Yamamoto}, {Yamauchi},
  {Yang}, {Yasue}, {Yuan}, {Yuda}, {Zhai}, {Zhang}, {Zhang}, {Zhang}, {Zhang},
  {Zhang}, {Zhang}, {Zhaxisangzhu}, \& {Zhou}}]{2015ICRC...34..355A}
{Amenomori}, M., {Bi}, X.~J., {Chen}, D., {et~al.} 2015, in International
  Cosmic Ray Conference, Vol.~34, 34th International Cosmic Ray Conference
  (ICRC2015), 355

\bibitem[{{Amenomori} {et~al.}(2017){Amenomori}, {Bi}, {Chen}, {Chen}, {Chen},
  {Cui}, {Danzengluobu}, {Ding}, {Feng}, {Feng}, {Feng}, {Gou}, {Guo}, {He},
  {He}, {Hibino}, {Hotta}, {Hu}, {Hu}, {Huang}, {Jia}, {Jiang}, {Kajino},
  {Kasahara}, {Katayose}, {Kato}, {Kawata}, {Kozai}, {Labaciren}, {Le}, {Li},
  {Li}, {Li}, {Liu}, {Liu}, {Liu}, {Lu}, {Meng}, {Miyazaki}, {Mizutani},
  {Munakata}, {Nakajima}, {Nakamura}, {Nanjo}, {Nishizawa}, {Niwa}, {Ohnishi},
  {Ohta}, {Ozawa}, {Qian}, {Qu}, {Saito}, {Saito}, {Sakata}, {Sako}, {Shao},
  {Shibata}, {Shiomi}, {Shirai}, {Sugimoto}, {Takita}, {Tan}, {Tateyama},
  {Torii}, {Tsuchiya}, {Udo}, {Wang}, {Wu}, {Xue}, {Yamamoto}, {Yamauchi},
  {Yang}, {Yuan}, {Yuda}, {Zhai}, {Zhang}, {Zhang}, {Zhang}, {Zhang}, {Zhang},
  {Zhang}, {Zhaxisangzhu}, {Zhou}, \& {Tibet AS{$\gamma$}
  Collaboration}}]{2017ApJ...836..153A}
---. 2017, \apj, 836, 153, \dodoi{10.3847/1538-4357/836/2/153}

\bibitem[{{Amenomori} {et~al.}(2021){Amenomori}, {Bao}, {Bi}, {Chen}, {Chen},
  {Chen}, {Chen}, {Chen}, {Cirennima}, {Danzengluobu}, {Fang}, {Fang}, {Feng},
  {Feng}, {Feng}, {Gao}, {Gou}, {Guo}, {Guo}, {He}, {He}, {Hibino}, {Hotta},
  {Hu}, {Hu}, {Huang}, {Jia}, {Jiang}, {Jin}, {Kasahara}, {Katayose}, {Kato},
  {Kato}, {Kawata}, {Kihara}, {Ko}, {Kozai}, {Labaciren}, {Li}, {Li}, {Li},
  {Lin}, {Liu}, {Liu}, {Liu}, {Liu}, {Liu}, {Lou}, {Lu}, {Meng}, {Munakata},
  {Nakada}, {Nakamura}, {Nanjo}, {Nishizawa}, {Ohnishi}, {Ohura}, {Ozawa},
  {Qian}, {Qu}, {Saito}, {Sakata}, {Sako}, {Shao}, {Shibata}, {Shiomi},
  {Sugimoto}, {Takano}, {Takita}, {Tan}, {Tateyama}, {Torii}, {Tsuchiya},
  {Udo}, {Wang}, {Wu}, {Xue}, {Yamamoto}, {Yang}, {Yokoe}, {Yuan}, {Zhai},
  {Zhang}, {Zhang}, {Zhang}, {Zhang}, {Zhang}, {Zhang}, {Zhang}, {Zhao},
  {Zhaxisangzhu}, \& {Tibet AS<SUB>{\ensuremath{\gamma}}</SUB>
  Collaboration}}]{2021PhRvL.126n1101A}
{Amenomori}, M., {Bao}, Y.~W., {Bi}, X.~J., {et~al.} 2021, \prl, 126, 141101,
  \dodoi{10.1103/PhysRevLett.126.141101}

\bibitem[{{An} {et~al.}(2019){An}, {Asfandiyarov}, {Azzarello}, {Bernardini},
  {Bi}, {Cai}, {Chang}, {Chen}, {Chen}, {Chen}, {Chen}, {Cui}, {Cui}, {Dai},
  {D'Amone}, {De Benedittis}, {De Mitri}, {Di Santo}, {Ding}, {Dong}, {Dong},
  {Dong}, {Donvito}, {Droz}, {Duan}, {Duan}, {D'Urso}, {Fan}, {Fan}, {Fang},
  {Feng}, {Feng}, {Fusco}, {Gallo}, {Gan}, {Gao}, {Gargano}, {Gong}, {Gong},
  {Guo}, {Guo}, {Guo}, {Han}, {Hu}, {Huang}, {Huang}, {Huang}, {Ionica},
  {Jiang}, {Jin}, {Kong}, {Lei}, {Li}, {Li}, {Li}, {Li}, {Li}, {Liang},
  {Liang}, {Liao}, {Liu}, {Liu}, {Liu}, {Liu}, {Liu}, {Liu}, {Loparco}, {Luo},
  {Ma}, {Ma}, {Ma}, {Ma}, {Ma}, {Marsella}, {Mazziotta}, {Mo}, {Niu}, {Pan},
  {Peng}, {Peng}, {Qiao}, {Rao}, {Salinas}, {Shang}, {Shen}, {Shen}, {Shen},
  {Song}, {Su}, {Su}, {Sun}, {Surdo}, {Teng}, {Tykhonov}, {Vitillo}, {Wang},
  {Wang}, {Wang}, {Wang}, {Wang}, {Wang}, {Wang}, {Wang}, {Wang}, {Wang},
  {Wang}, {Wang}, {Wang}, {Wei}, {Wei}, {Wei}, {Wen}, {Wu}, {Wu}, {Wu}, {Wu},
  {Wu}, {Xi}, {Xia}, {Xu}, {Xu}, {Xu}, {Xu}, {Xue}, {Yang}, {Yang}, {Yang},
  {Yang}, {Yao}, {Yu}, {Yuan}, {Yue}, {Zang}, {Zhang}, {Zhang}, {Zhang},
  {Zhang}, {Zhang}, {Zhang}, {Zhang}, {Zhang}, {Zhang}, {Zhang}, {Zhang},
  {Zhang}, {Zhang}, {Zhao}, {Zhao}, {Zhao}, {Zhou}, {Zhou}, {Zhu}, {Zhu}, \&
  {Zimmer}}]{2019SciA....5.3793A}
{An}, Q., {Asfandiyarov}, R., {Azzarello}, P., {et~al.} 2019, Science Advances,
  5, eaax3793, \dodoi{10.1126/sciadv.aax3793}

\bibitem[{{Andreyev} {et~al.}(1987){Andreyev}, {Chudakov}, {Kozyarivsky},
  {Sidorenko}, {Tulupova}, \& {Voevodsky}}]{1987ICRC....2...22A}
{Andreyev}, Y.~M., {Chudakov}, A.~E., {Kozyarivsky}, V.~A., {et~al.} 1987,
  International Cosmic Ray Conference, 2, 22

\bibitem[{{Atkin} {et~al.}(2017){Atkin}, {Bulatov}, {Dorokhov}, {Gorbunov},
  {Filippov}, {Grebenyuk}, {Karmanov}, {Kovalev}, {Kudryashov}, {Kurganov},
  {Merkin}, {Panov}, {Podorozhny}, {Polkov}, {Porokhovoy}, {Shumikhin},
  {Sveshnikova}, {Tkachenko}, {Tkachev}, {Turundaevskiy}, {Vasiliev}, \&
  {Voronin}}]{2017JCAP...07..020A}
{Atkin}, E., {Bulatov}, V., {Dorokhov}, V., {et~al.} 2017, \jcap, 7, 020,
  \dodoi{10.1088/1475-7516/2017/07/020}

\bibitem[{{Atkin} {et~al.}(2018){Atkin}, {Bulatov}, {Dorokhov}, {Gorbunov},
  {Filippov}, {Grebenyuk}, {Karmanov}, {Kovalev}, {Kudryashov}, {Kurganov},
  {Merkin}, {Panov}, {Podorozhny}, {Polkov}, {Porokhovoy}, {Shumikhin},
  {Tkachenko}, {Tkachev}, {Turundaevskiy}, {Vasiliev}, \&
  {Voronin}}]{2018JETPL.108....5A}
---. 2018, Soviet Journal of Experimental and Theoretical Physics Letters, 108,
  5, \dodoi{10.1134/S0021364018130015}

\bibitem[{{Atoyan} {et~al.}(1995){Atoyan}, {Aharonian}, \&
  {V{\"o}lk}}]{1995PhRvD..52.3265A}
{Atoyan}, A.~M., {Aharonian}, F.~A., \& {V{\"o}lk}, H.~J. 1995, \prd, 52, 3265,
  \dodoi{10.1103/PhysRevD.52.3265}

\bibitem[{{Bartoli} {et~al.}(2015){Bartoli}, {Bernardini}, {Bi}, {Cao},
  {Catalanotti}, {Chen}, {Chen}, {Cui}, {Dai}, {D'Amone}, {Danzengluobu}, {De
  Mitri}, {D'Ettorre Piazzoli}, {Di Girolamo}, {Di Sciascio}, {Feng}, {Feng},
  {Feng}, {Gao}, {Gou}, {Guo}, {He}, {Hu}, {Hu}, {Iacovacci}, {Iuppa}, {Jia},
  {Labaciren}, {Li}, {Liu}, {Liu}, {Liu}, {Lu}, {Ma}, {Ma}, {Mancarella},
  {Mari}, {Marsella}, {Mastroianni}, {Montini}, {Ning}, {Perrone}, {Pistilli},
  {Salvini}, {Santonico}, {Shen}, {Sheng}, {Shi}, {Surdo}, {Tan}, {Vallania},
  {Vernetto}, {Vigorito}, {Wang}, {Wu}, {Wu}, {Xue}, {Yang}, {Yang}, {Yao},
  {Yuan}, {Zha}, {Zhang}, {Zhang}, {Zhang}, {Zhang}, {Zhao}, {Zhaxiciren},
  {Zhaxisangzhu}, {Zhou}, {Zhu}, {Zhu}, \& {ARGO-YBJ
  Collaboration}}]{2015ApJ...809...90B}
{Bartoli}, B., {Bernardini}, P., {Bi}, X.~J., {et~al.} 2015, \apj, 809, 90,
  \dodoi{10.1088/0004-637X/809/1/90}

\bibitem[{{Bercovitch} \& {Agrawal}(1981)}]{1981ICRC...10..246B}
{Bercovitch}, M., \& {Agrawal}, S.~P. 1981, International Cosmic Ray
  Conference, 10, 246

\bibitem[{{Blasi}(2013)}]{2013A&ARv..21...70B}
{Blasi}, P. 2013, \aapr, 21, 70, \dodoi{10.1007/s00159-013-0070-7}

\bibitem[{{Borla Tridon}(2011)}]{2011ICRC....6...47B}
{Borla Tridon}, D. 2011, International Cosmic Ray Conference, 6, 47,
  \dodoi{10.7529/ICRC2011/V06/0680}

\bibitem[{{Case} \& {Bhattacharya}(1996)}]{1996A&AS..120C.437C}
{Case}, G., \& {Bhattacharya}, D. 1996, \aaps, 120, 437

\bibitem[{{Chiavassa} {et~al.}(2015){Chiavassa}, {Apel},
  {Arteaga-Vel{\'a}zquez}, {Bekk}, {Bertaina}, {Bl{\"u}mer}, {Bozdog},
  {Brancus}, {Cantoni}, {Cossavella}, {Daumiller}, {de Souza}, {di Pierro},
  {Doll}, {Engel}, {Fuhrmann}, {Gherghel-Lascu}, {Gils}, {Glasstetter},
  {Grupen}, {Haungs}, {Heck}, {H{\"o}randel}, {Huber}, {Huege}, {Kampert},
  {Kang}, {Klages}, {Link}, {{\L}uczak}, {Mathes}, {Mayer}, {Milke}, {Mitrica},
  {Morello}, {Oehlschl{\"a}ger}, {Ostapchenko}, {Palmieri}, {Pierog}, {Rebel},
  {Roth}, {Schieler}, {Schoo}, {Schr{\"o}der}, {Sima}, {Toma}, {Trinchero},
  {Ulrich}, {Weindl}, {Wochele}, {Zabierowski}, \& {KASCADE-Grande
  Collaboration}}]{2015ICRC...34..281C}
{Chiavassa}, A., {Apel}, W.~D., {Arteaga-Vel{\'a}zquez}, J.~C., {et~al.} 2015,
  in International Cosmic Ray Conference, Vol.~34, 34th International Cosmic
  Ray Conference (ICRC2015), 281

\bibitem[{{Cutler} \& {Groom}(1991)}]{1991ApJ...376..322C}
{Cutler}, D.~J., \& {Groom}, D.~E. 1991, \apj, 376, 322, \dodoi{10.1086/170282}

\bibitem[{{DAMPE Collaboration}(2022)}]{DAMPECOLLABORATION2022}
{DAMPE Collaboration}. 2022, Science Bulletin,
  \dodoi{https://doi.org/10.1016/j.scib.2022.10.002}

\bibitem[{{DAMPE Collaboration} {et~al.}(2017){DAMPE Collaboration}, {Ambrosi},
  {An}, {Asfandiyarov}, {Azzarello}, {Bernardini}, {Bertucci}, {Cai}, {Chang},
  {Chen}, {Chen}, {Chen}, {Chen}, {Cui}, {Cui}, {D'Amone}, {de Benedittis}, {De
  Mitri}, {di Santo}, {Dong}, {Dong}, {Dong}, {Dong}, {Donvito}, {Droz},
  {Duan}, {Duan}, {Duranti}, {D'Urso}, {Fan}, {Fan}, {Fang}, {Feng}, {Feng},
  {Fusco}, {Gallo}, {Gan}, {Gao}, {Gao}, {Gargano}, {Garrappa}, {Gong}, {Gong},
  {Guo}, {Guo}, {Hu}, {Huang}, {Huang}, {Ionica}, {Jiang}, {Jiang}, {Jin},
  {Kong}, {Lei}, {Li}, {Li}, {Li}, {Li}, {Liang}, {Liang}, {Liao}, {Liu},
  {Liu}, {Liu}, {Liu}, {Liu}, {Loparco}, {Ma}, {Ma}, {Ma}, {Ma}, {Ma}, {Ma},
  {Marsella}, {Mazziotta}, {Mo}, {Niu}, {Peng}, {Peng}, {Qiao}, {Rao},
  {Salinas}, {Shang}, {H.~Shen}, {Shen}, {Shen}, {Song}, {Su}, {Su}, {Sun},
  {Surdo}, {Teng}, {Tian}, {Tykhonov}, {Vagelli}, {Vitillo}, {Wang}, {Wang},
  {Wang}, {Wang}, {Wang}, {Wang}, {Wang}, {Wang}, {Wang}, {Wang}, {Wang},
  {Wang}, {Wen}, {Wang}, {Wei}, {Wei}, {Wei}, {Wu}, {Wu}, {Wu}, {Wu}, {Wu},
  {Xi}, {Xia}, {Xin}, {Xu}, {Xu}, {Xu}, {Xue}, {Yang}, {Yang}, {Yang}, {Yang},
  {Yao}, {Yu}, {Yuan}, {Yue}, {Zang}, {Zhang}, {Zhang}, {Zhang}, {Zhang},
  {Zhang}, {Zhang}, {Zhang}, {Zhang}, {Zhang}, {Zhang}, {Zhang}, {Zhang},
  {Zhang}, {Zhang}, {Zhang}, {Zhang}, {Zhang}, {Zhao}, {Zhao}, {Zhao}, {Zhou},
  {Zhou}, {Zhu}, {Zhu}, \& {Zimmer}}]{2017Natur.552...63D}
{DAMPE Collaboration}, {Ambrosi}, G., {An}, Q., {et~al.} 2017, \nat, 552, 63,
  \dodoi{10.1038/nature24475}

\bibitem[{{Evoli} {et~al.}(2017){Evoli}, {Gaggero}, {Vittino}, {Di Bernardo},
  {Di Mauro}, {Ligorini}, {Ullio}, \& {Grasso}}]{2017JCAP...02..015E}
{Evoli}, C., {Gaggero}, D., {Vittino}, A., {et~al.} 2017, \jcap, 2017, 015,
  \dodoi{10.1088/1475-7516/2017/02/015}

\bibitem[{{Faherty} {et~al.}(2007){Faherty}, {Walter}, \&
  {Anderson}}]{2007Ap&SS.308..225F}
{Faherty}, J., {Walter}, F.~M., \& {Anderson}, J. 2007, \apss, 308, 225,
  \dodoi{10.1007/s10509-007-9368-0}

\bibitem[{{Feng} {et~al.}(2016){Feng}, {Tomassetti}, \&
  {Oliva}}]{2016PhRvD..94l3007F}
{Feng}, J., {Tomassetti}, N., \& {Oliva}, A. 2016, \prd, 94, 123007,
  \dodoi{10.1103/PhysRevD.94.123007}

\bibitem[{{Fenton} {et~al.}(1995){Fenton}, {Fenton}, \&
  {Humble}}]{1995ICRC....4..635F}
{Fenton}, K.~B., {Fenton}, A.~G., \& {Humble}, J.~E. 1995, International Cosmic
  Ray Conference, 4, 635

\bibitem[{{Fujita} {et~al.}(2009){Fujita}, {Kohri}, {Yamazaki}, \&
  {Ioka}}]{2009PhRvD..80f3003F}
{Fujita}, Y., {Kohri}, K., {Yamazaki}, R., \& {Ioka}, K. 2009, \prd, 80,
  063003, \dodoi{10.1103/PhysRevD.80.063003}

\bibitem[{{Gleeson} \& {Axford}(1968)}]{1968ApJ...154.1011G}
{Gleeson}, L.~J., \& {Axford}, W.~I. 1968, \apj, 154, 1011,
  \dodoi{10.1086/149822}

\bibitem[{{Gombosi} {et~al.}(1975){Gombosi}, {K{\'o}ta}, {Somogyi}, {Varga},
  {Betev}, {Katsarski}, {Kavlakov}, \& {Khirov}}]{1975ICRC....2..586G}
{Gombosi}, T., {K{\'o}ta}, J., {Somogyi}, A.~J., {et~al.} 1975, International
  Cosmic Ray Conference, 2, 586

\bibitem[{{Grenier} {et~al.}(2015){Grenier}, {Black}, \&
  {Strong}}]{2015ARA&A..53..199G}
{Grenier}, I.~A., {Black}, J.~H., \& {Strong}, A.~W. 2015, \araa, 53, 199,
  \dodoi{10.1146/annurev-astro-082214-122457}

\bibitem[{{Guillian} {et~al.}(2007){Guillian}, {Hosaka}, {Ishihara}, {Kameda},
  {Koshio}, {Minamino}, {Mitsuda}, {Miura}, {Moriyama}, {Nakahata}, {Namba},
  {Obayashi}, {Ogawa}, {Shiozawa}, {Suzuki}, {Takeda}, {Takeuchi}, {Yamada},
  {Higuchi}, {Ishitsuka}, {Kajita}, {Kaneyuki}, {Mitsuka}, {Nakayama},
  {Nishino}, {Okada}, {Okumura}, {Saji}, {Takenaga}, {Desai}, {Kearns},
  {Stone}, {Sulak}, {Wang}, {Goldhaber}, {Casper}, {Gajewski}, {Griskevich},
  {Kropp}, {Liu}, {Mine}, {Smy}, {Sobel}, {Vagins}, {Ganezer}, {Hill}, {Keig},
  {Scholberg}, {Walter}, {Ellsworth}, {Tasaka}, {Kibayashi}, {Learned},
  {Matsuno}, {Messier}, {Hayato}, {Ichikawa}, {Ishida}, {Ishii}, {Iwashita},
  {Kobayashi}, {Nakadaira}, {Nakamura}, {Nitta}, {Oyama}, {Totsuka}, {Suzuki},
  {Hasegawa}, {Kato}, {Maesaka}, {Nakaya}, {Nishikawa}, {Sato}, {Yamamoto},
  {Yokoyama}, {Haines}, {Dazeley}, {Hatakeyama}, {Svoboda}, {Blaufuss},
  {Goodman}, {Sullivan}, {Turcan}, {Habig}, {Fukuda}, {Itow}, {Sakuda},
  {Yoshida}, {Kim}, {Yoo}, {Okazawa}, {Ishizuka}, {Jung}, {Kato}, {Kobayashi},
  {Malek}, {Mauger}, {McGrew}, {Sharkey}, {Yanagisawa}, {Gando}, {Hasegawa},
  {Inoue}, {Shirai}, {Suzuki}, {Nishijima}, {Ishino}, {Watanabe}, {Koshiba},
  {Kielczewska}, {Berns}, {Gran}, {Shiraishi}, {Stachyra}, {Washburn},
  {Wilkes}, \& {Munakata}}]{2007PhRvD..75f2003G}
{Guillian}, G., {Hosaka}, J., {Ishihara}, K., {et~al.} 2007, \prd, 75, 062003,
  \dodoi{10.1103/PhysRevD.75.062003}

\bibitem[{{Guo} {et~al.}(2016){Guo}, {Tian}, \& {Jin}}]{2016ApJ...819...54G}
{Guo}, Y.-Q., {Tian}, Z., \& {Jin}, C. 2016, \apj, 819, 54,
  \dodoi{10.3847/0004-637X/819/1/54}

\bibitem[{{Guo} \& {Yuan}(2018)}]{2018PhRvD..97f3008G}
{Guo}, Y.-Q., \& {Yuan}, Q. 2018, \prd, 97, 063008,
  \dodoi{10.1103/PhysRevD.97.063008}

\bibitem[{{H{\"o}randel}(2003)}]{2003APh....19..193H}
{H{\"o}randel}, J.~R. 2003, Astroparticle Physics, 19, 193,
  \dodoi{10.1016/S0927-6505(02)00198-6}

\bibitem[{{Kohri} {et~al.}(2016){Kohri}, {Ioka}, {Fujita}, \&
  {Yamazaki}}]{2016PTEP.2016b1E01K}
{Kohri}, K., {Ioka}, K., {Fujita}, Y., \& {Yamazaki}, R. 2016, Progress of
  Theoretical and Experimental Physics, 2016, 021E01,
  \dodoi{10.1093/ptep/ptv193}

\bibitem[{{Kyratzis} \& {HERD Collaboration}(2022)}]{2022PhyS...97e4010K}
{Kyratzis}, D., \& {HERD Collaboration}. 2022, \physscr, 97, 054010,
  \dodoi{10.1088/1402-4896/ac63fc}

\bibitem[{{Lee} \& {Ng}(1987)}]{1987ICRC....2...18L}
{Lee}, Y.~W., \& {Ng}, L.~K. 1987, International Cosmic Ray Conference, 2, 18

\bibitem[{{Liu} {et~al.}(2017){Liu}, {Bi}, {Lin}, {Wang}, \&
  {Yin}}]{2017PhRvD..96b3006L}
{Liu}, W., {Bi}, X.-J., {Lin}, S.-J., {Wang}, B.-B., \& {Yin}, P.-F. 2017,
  \prd, 96, 023006, \dodoi{10.1103/PhysRevD.96.023006}

\bibitem[{{Liu} {et~al.}(2019){Liu}, {Guo}, \& {Yuan}}]{2019JCAP...10..010L}
{Liu}, W., {Guo}, Y.-Q., \& {Yuan}, Q. 2019, \jcap, 2019, 010,
  \dodoi{10.1088/1475-7516/2019/10/010}

\bibitem[{{Liu} {et~al.}(2018){Liu}, {Yao}, \& {Guo}}]{2018ApJ...869..176L}
{Liu}, W., {Yao}, Y.-h., \& {Guo}, Y.-Q. 2018, \apj, 869, 176,
  \dodoi{10.3847/1538-4357/aaef39}

\bibitem[{{Luo} {et~al.}(2022){Luo}, {Qiao}, {Liu}, {Cui}, \&
  {Guo}}]{2022ApJ...930...82L}
{Luo}, Q., {Qiao}, B.-q., {Liu}, W., {Cui}, S.-w., \& {Guo}, Y.-q. 2022, \apj,
  930, 82, \dodoi{10.3847/1538-4357/ac6267}

\bibitem[{{Malkov} \& {Moskalenko}(2021)}]{2021ApJ...911..151M}
{Malkov}, M.~A., \& {Moskalenko}, I.~V. 2021, \apj, 911, 151,
  \dodoi{10.3847/1538-4357/abe855}

\bibitem[{{Manchester} {et~al.}(2005){Manchester}, {Hobbs}, {Teoh}, \&
  {Hobbs}}]{2005AJ....129.1993M}
{Manchester}, R.~N., {Hobbs}, G.~B., {Teoh}, A., \& {Hobbs}, M. 2005, \aj, 129,
  1993, \dodoi{10.1086/428488}

\bibitem[{{Mori} {et~al.}(1995){Mori}, {Yasue}, {Munakata}, {Kato}, {Akahane},
  {Koyama}, \& {Kitawada}}]{1995ICRC....4..648M}
{Mori}, S., {Yasue}, S., {Munakata}, K., {et~al.} 1995, International Cosmic
  Ray Conference, 4, 648

\bibitem[{{Munakata} {et~al.}(1995){Munakata}, {Yasue}, {Mori}, {Kato},
  {Koyama}, {Akahane}, {Fujii}, {Ueno}, {Humble}, {Fenton}, {Fenton}, \&
  {Duldig}}]{1995ICRC....4..639M}
{Munakata}, K., {Yasue}, S., {Mori}, S., {et~al.} 1995, International Cosmic
  Ray Conference, 4, 639

\bibitem[{{Munakata} {et~al.}(1997){Munakata}, {Kiuchi}, {Yasue}, {Kato},
  {Mori}, {Hirata}, {Kihara}, {Oyama}, {Mori}, {Fujita}, {Hatakeyama}, {Koga},
  {Maruyama}, {Suzuki}, {Ishizuka}, {Miyano}, {Okazawa}, {Fukuda}, {Hayakawa},
  {Inoue}, {Ishihara}, {Ishino}, {Joukou}, {Kajita}, {Kasuga}, {Koshio},
  {Kumita}, {Matsumoto}, {Nakahata}, {Nakamura}, {Okumura}, {Sakai},
  {Shiozawa}, {Suzuki}, {Suzuki}, {Tomoeda}, {Totsuka}, {Horiuchi},
  {Nishijima}, {Koshiba}, {Suda}, {Suzuki}, {Hara}, {Nagashima}, {Takita},
  {Yamaguchi}, {Hayato}, {Kaneyuki}, {Suzuki}, {Takeuchi}, {Tanimori},
  {Tasaka}, {Ichihara}, {Miyamoto}, \& {Nishikawa}}]{1997PhRvD..56...23M}
{Munakata}, K., {Kiuchi}, T., {Yasue}, S., {et~al.} 1997, \prd, 56, 23,
  \dodoi{10.1103/PhysRevD.56.23}

\bibitem[{{Nagashima} {et~al.}(1989){Nagashima}, {Fujimoto}, {Sakakibara},
  {Fujii}, {Ueno}, {Murakami}, \& {Morishita}}]{1989NCimC..12..695N}
{Nagashima}, K., {Fujimoto}, K., {Sakakibara}, S., {et~al.} 1989, Nuovo Cimento
  C Geophysics Space Physics C, 12, 695, \dodoi{10.1007/BF02511970}

\bibitem[{{Panov} {et~al.}(2007){Panov}, {Adams}, {Ahn}, {Batkov},
  {Bashindzhagyan}, {Watts}, {Wefel}, {Wu}, {Ganel}, {Guzik}, {Gunashingha},
  {Zatsepin}, {Isbert}, {Kim}, {Christl}, {Kouznetsov}, {Panasyuk}, {Seo},
  {Sokolskaya}, {Chang}, {Schmidt}, \& {Fazely}}]{2007BRASP..71..494P}
{Panov}, A.~D., {Adams}, J.~H., J., {Ahn}, H.~S., {et~al.} 2007, Bulletin of
  the Russian Academy of Sciences, Physics, 71, 494,
  \dodoi{10.3103/S1062873807040168}

\bibitem[{{Panov} {et~al.}(2009){Panov}, {Adams}, {Ahn}, {Bashinzhagyan},
  {Watts}, {Wefel}, {Wu}, {Ganel}, {Guzik}, {Zatsepin}, {Isbert}, {Kim},
  {Christl}, {Kouznetsov}, {Panasyuk}, {Seo}, {Sokolskaya}, {Chang}, {Schmidt},
  \& {Fazely}}]{2009BRASP..73..564P}
{Panov}, A.~D., {Adams}, J.~H., {Ahn}, H.~S., {et~al.} 2009, Bulletin of the
  Russian Academy of Sciences, Physics, 73, 564,
  \dodoi{10.3103/S1062873809050098}

\bibitem[{{Qiao} {et~al.}(2019){Qiao}, {Liu}, {Guo}, \&
  {Yuan}}]{2019JCAP...12..007Q}
{Qiao}, B.-Q., {Liu}, W., {Guo}, Y.-Q., \& {Yuan}, Q. 2019, \jcap, 2019, 007,
  \dodoi{10.1088/1475-7516/2019/12/007}

\bibitem[{{Sakakibara} {et~al.}(1973){Sakakibara}, {Ueno}, {Fujimoto}, {Kondo},
  \& {Nagashima}}]{1973ICRC....2.1058S}
{Sakakibara}, S., {Ueno}, H., {Fujimoto}, K., {Kondo}, I., \& {Nagashima}, K.
  1973, International Cosmic Ray Conference, 2, 1058

\bibitem[{{Seo} \& {Ptuskin}(1994)}]{1994ApJ...431..705S}
{Seo}, E.~S., \& {Ptuskin}, V.~S. 1994, \apj, 431, 705, \dodoi{10.1086/174520}

\bibitem[{{Serpico}(2015)}]{2015ICRC...34....9S}
{Serpico}, P. 2015, in International Cosmic Ray Conference, Vol.~34, 34th
  International Cosmic Ray Conference (ICRC2015), 9.
\newblock \doarXiv{1509.04233}

\bibitem[{{Smith} {et~al.}(1994){Smith}, {Cunha}, \&
  {Plez}}]{1994A&A...281L..41S}
{Smith}, V.~V., {Cunha}, K., \& {Plez}, B. 1994, \aap, 281, L41

\bibitem[{{Staszak} \& {VERITAS Collaboration}(2015)}]{2015ICRC...34..868S}
{Staszak}, D., \& {VERITAS Collaboration}. 2015, in International Cosmic Ray
  Conference, Vol.~34, 34th International Cosmic Ray Conference (ICRC2015),
  868.
\newblock \doarXiv{1510.01269}

\bibitem[{{Strong} {et~al.}(2007){Strong}, {Moskalenko}, \&
  {Ptuskin}}]{2007ARNPS..57..285S}
{Strong}, A.~W., {Moskalenko}, I.~V., \& {Ptuskin}, V.~S. 2007, Annual Review
  of Nuclear and Particle Science, 57, 285,
  \dodoi{10.1146/annurev.nucl.57.090506.123011}

\bibitem[{{Sveshnikova} {et~al.}(2013){Sveshnikova}, {Strelnikova}, \&
  {Ptuskin}}]{2013APh....50...33S}
{Sveshnikova}, L.~G., {Strelnikova}, O.~N., \& {Ptuskin}, V.~S. 2013,
  Astroparticle Physics, 50, 33, \dodoi{10.1016/j.astropartphys.2013.08.007}

\bibitem[{{Swinson} \& {Nagashima}(1985)}]{1985P&SS...33.1069S}
{Swinson}, D.~B., \& {Nagashima}, K. 1985, \planss, 33, 1069,
  \dodoi{10.1016/0032-0633(85)90025-X}

\bibitem[{{Thambyahpillai}(1983)}]{1983ICRC....3..383T}
{Thambyahpillai}, T. 1983, International Cosmic Ray Conference, 3, 383

\bibitem[{{Tomassetti}(2012)}]{2012ApJ...752L..13T}
{Tomassetti}, N. 2012, \apjl, 752, L13, \dodoi{10.1088/2041-8205/752/1/L13}

\bibitem[{{Tomassetti}(2015)}]{2015PhRvD..92h1301T}
---. 2015, \prd, 92, 081301, \dodoi{10.1103/PhysRevD.92.081301}

\bibitem[{{Uchiyama} {et~al.}(2012){Uchiyama}, {Funk}, {Katagiri}, {Katsuta},
  {Lemoine-Goumard}, {Tajima}, {Tanaka}, \& {Torres}}]{2012ApJ...749L..35U}
{Uchiyama}, Y., {Funk}, S., {Katagiri}, H., {et~al.} 2012, \apjl, 749, L35,
  \dodoi{10.1088/2041-8205/749/2/L35}

\bibitem[{{Ueno} {et~al.}(1990){Ueno}, {Fujii}, \&
  {Yamada}}]{1990ICRC....6..361U}
{Ueno}, H., {Fujii}, Z., \& {Yamada}, T. 1990, International Cosmic Ray
  Conference, 6, 361

\bibitem[{{Vladimirov} {et~al.}(2012){Vladimirov}, {J{\'o}hannesson},
  {Moskalenko}, \& {Porter}}]{2012ApJ...752...68V}
{Vladimirov}, A.~E., {J{\'o}hannesson}, G., {Moskalenko}, I.~V., \& {Porter},
  T.~A. 2012, \apj, 752, 68, \dodoi{10.1088/0004-637X/752/1/68}

\bibitem[{{Yang} {et~al.}(2016){Yang}, {Aharonian}, \&
  {Evoli}}]{2016PhRvD..93l3007Y}
{Yang}, R., {Aharonian}, F., \& {Evoli}, C. 2016, \prd, 93, 123007,
  \dodoi{10.1103/PhysRevD.93.123007}

\bibitem[{{Yoon} {et~al.}(2017){Yoon}, {Anderson}, {Barrau}, {Conklin},
  {Coutu}, {Derome}, {Han}, {Jeon}, {Kim}, {Kim}, {Lee}, {Lee}, {Lee}, {Lee},
  {Link}, {Menchaca-Rocha}, {Mitchell}, {Mognet}, {Nutter}, {Park},
  {Picot-Clemente}, {Putze}, {Seo}, {Smith}, \& {Wu}}]{2017ApJ...839....5Y}
{Yoon}, Y.~S., {Anderson}, T., {Barrau}, A., {et~al.} 2017, \apj, 839, 5,
  \dodoi{10.3847/1538-4357/aa68e4}

\bibitem[{{Yuan}(2019)}]{2019SCPMA..6249511Y}
{Yuan}, Q. 2019, Science China Physics, Mechanics, and Astronomy, 62, 49511,
  \dodoi{10.1007/s11433-018-9300-0}

\bibitem[{{Yuan} {et~al.}(2017){Yuan}, {Lin}, {Fang}, \&
  {Bi}}]{2017PhRvD..95h3007Y}
{Yuan}, Q., {Lin}, S.-J., {Fang}, K., \& {Bi}, X.-J. 2017, \prd, 95, 083007,
  \dodoi{10.1103/PhysRevD.95.083007}

\bibitem[{{Yue} {et~al.}(2019){Yue}, {Ma}, {Yuan}, {Fan}, {Chen}, {Cui}, {Dai},
  {Dong}, {Huang}, {Jiang}, {Lei}, {Li}, {Liu}, {Liu}, {Liu}, {Luo}, {Pan},
  {Peng}, {Qiao}, {Wei}, {Wu}, {Xu}, {Xu}, {Yuan}, {Zang}, {Zhang}, {Zhang}, \&
  {Zhang}}]{2019FrPhy..1524601Y}
{Yue}, C., {Ma}, P.-X., {Yuan}, Q., {et~al.} 2019, Frontiers of Physics, 15,
  24601, \dodoi{10.1007/s11467-019-0946-8}

\bibitem[{{Zhang} {et~al.}(2021{\natexlab{a}}){Zhang}, {Liu}, {Su}, {Zhu},
  {Xi}, \& {Wang}}]{2021ApJ...923..106Z}
{Zhang}, H.-M., {Liu}, R.-Y., {Su}, Y., {et~al.} 2021{\natexlab{a}}, \apj, 923,
  106, \dodoi{10.3847/1538-4357/ac36c6}

\bibitem[{{Zhang} {et~al.}(2021{\natexlab{b}}){Zhang}, {Qiao}, {Liu}, {Cui},
  {Yuan}, \& {Guo}}]{2021JCAP...05..012Z}
{Zhang}, P.-p., {Qiao}, B.-q., {Liu}, W., {et~al.} 2021{\natexlab{b}}, \jcap,
  2021, 012, \dodoi{10.1088/1475-7516/2021/05/012}

\bibitem[{{Zhang} {et~al.}(2022){Zhang}, {Qiao}, {Yuan}, {Cui}, \&
  {Guo}}]{2022PhRvD.105b3002Z}
{Zhang}, P.-p., {Qiao}, B.-q., {Yuan}, Q., {Cui}, S.-w., \& {Guo}, Y.-q. 2022,
  \prd, 105, 023002, \dodoi{10.1103/PhysRevD.105.023002}

\bibitem[{{Zhao} {et~al.}(2022){Zhao}, {Liu}, {Yuan}, {Hu}, {Bi}, {Wu}, {Zhou},
  \& {Guo}}]{2022ApJ...926...41Z}
{Zhao}, B., {Liu}, W., {Yuan}, Q., {et~al.} 2022, \apj, 926, 41,
  \dodoi{10.3847/1538-4357/ac4416}

\end{thebibliography}
\bibliographystyle{aasjournal}



\end{document}